\documentstyle[12pt, hyperref]{article}

\catcode`\@=11
\def\theequation{\thesubsection.\arabic{equation}}
\@addtoreset{footnote}{section}
\@addtoreset{footnote}{subsection}
\@addtoreset{equation}{subsection}
\catcode`\@=12
\catcode`\@=11

\newcounter{app}

\def\app{\setcounter{equation}{0}
\def\theequation{A\arabic{app}.\arabic{equation}}\par
   \addvspace{4ex}
   \@afterindentfalse
  \secdef\@app\@dapp}
\newcommand\@app{\@startsection {app}{1}{0ex}%
                                   {-3.5ex \@plus -1ex \@minus -.2ex}%
                                   {2.3ex \@plus.2ex}%
                                   {\normalfont\Large\bf}}

\def\@dapp#1{%
{\parindent \z@ \raggedright  \bf #1}\par\nobreak}
\def\l@app#1#2{\ifnum \c@tocdepth >\z@
    \addpenalty\@secpenalty
    \addvspace{1.0em \@plus\p@}%
    \setlength\@tempdima{8.5em}%
    \begingroup
      \parindent \z@ \rightskip \@pnumwidth
      \parfillskip -\@pnumwidth
      \leavevmode \bfseries
      \advance\leftskip\@tempdima
      \hskip -\leftskip
      #1\nobreak\hfil \nobreak\hb@xt@\@pnumwidth{\hss #2}\par
    \endgroup\fi}
\catcode`\@=12

\small  \oddsidemargin -10mm  \evensidemargin 5mm  \topmargin -10mm
\def\be{\begin{equation}}
\def\ee{\end{equation}}
\def\g{\gamma}

\renewcommand{\l}{\langle}
\renewcommand{\r}{\rangle}

\newtheorem{prop}{Proposition}
\newtheorem{remark}{Remark}

\def\bprop{\begin{prop}}
\def\eprop{\end{prop}}
\def\bremark{\begin{remark}}
\def\eremark{\end{remark}}
\begin{document}
\author{A. Yu. Orlov\thanks{Oceanology Institute, Nahimovskii
prospect 36, Moscow, Russia, email address:
orlovs@wave.sio.rssi.ru}}
\title{Soliton Theory, Symmetric Functions and Matrix Integrals}
\date{}
\maketitle

%\vspace{1.5cm}
\begin{abstract}
We consider certain scalar product of symmetric functions which is
parameterized by a function $r$ and an integer $n$. One the one
hand we have a fermionic representation of this scalar product. On
the other hand we get a representation of this product with the
help of multi-integrals. This gives links between a theory of
symmetric functions, soliton theory and models of random matrices
(such as a model of normal matrices).

\end{abstract}
\tableofcontents

\section{Introduction} Our goal is to make links between soliton
theory, theory of symmetric functions and random matrices. We
shall consider a certain class of tau-functions which we call
tau-functions of hypergeometric type. The key point of the paper
is to identify scalar products on the space of symmetric functions
with vacuum expectations of fermionic fields.

We connect different topics: integrals over matrices,
hypergeometric functions, theory of symmetric functions. We also
show that a scalar product of tau-functions is also a
tau-function. The paper is partially based on \cite{nl64} and
\cite{OH}.

The work was reported at NEEDS workshop in Cadiz June 2002 and at
the workshop "Nonlinear physics: theory and experiment" Gallipoli
June-July 2002.

{\bf Symmetric functions}. Symmetric function is a function of
variables ${\bf x} = x_1,\dots ,x_n$ which is invariant under the
action of permutation group $S_n$.  Symmetric polynomials of $n$
variables form a ring denoted by $\Lambda _n$. The number $n$ is
usually irrelevant and may be infinite. The ring of symmetric
functions in infinitely many variables is denoted by $\Lambda$.

Power sums are defined as
\begin{equation}\label{powsum}
p_m=\sum_{i=1}x_i^m,\quad m=1,2,\dots
\end{equation}
Functions $p_{m_1}p_{m_2}\cdots p_{m_k},\quad k=1,2,\dots$ form a
basis in $\Lambda$.

There are different well investigated polynomials of many
variables such as Schur functions, projective Schur functions,
complete symmetric functions \cite{Mac}. Each set form a basis in
$\Lambda$.

There exists a scalar product in $\Lambda $ such that Schur
functions are orthonormal functions:
\begin{equation}\label{shurort}
<s_\lambda ,s_\mu >=\delta_{\lambda,\mu}
\end{equation}

The notion of scalar product is very important in the theory of
symmetric functions. The choice of scalar product gives different
orthogonal and bi-orthogonal systems of polynomials.

{\bf Soliton theory}. KP hierarchy of integrable equations, which
is the most popular example,  consists of semi-infinite set of
nonlinear partial differential equations
\begin{equation}\label{KPh}
\partial_{t_m}u=K_m[u],\quad m=1,2,\dots
\end{equation}
which are commuting flows:
\begin{equation}\label{commute}
\left[\partial_{t_k},\partial_{t_m}\right]u=0
\end{equation}
The first nontrivial one is Kadomtsev-Petviashvili equation
\begin{equation}\label{KP}
\partial_{t_3}u=\frac 14 \partial_{t_1}^3 u+ \frac 34
\partial_{t_1}^{-1}\partial_{t_2}^2u+\frac 34 \partial_{t_1} u^2
\end{equation}
which originally served in plasma physics \cite{KP}. This
equation, which was integrated in \cite{Dr},\cite{ZSh}, now plays
a very important role both, in physics (see \cite{ZM}; see review
in \cite{M} for modern applications) and in mathematics. Each
members of the KP hierarchy is compatible with each other one.

Another very important equation is the equation of two-dimensional
Toda lattice (TL) \cite{AM},\cite{DJKM},\cite{UT}
\begin{equation}\label{Toda}
\partial_{t_1}\partial_{t^*_1}\phi_n=
e^{\phi_{n-1}-\phi_{n}}- e^{\phi_{n}-\phi_{n+1}}
\end{equation}
This equation gives rise to TL hierarchy which contains
derivatives with respect to higher times $t_1,t_2,\dots$ and
$t_1^*,t_2^*,\dots$.

The key point of soliton theory is the notion of tau function,
introduced by Sato (for KP tau-function see \cite{DJKM}). Tau
function is a sort of potential which gives rise both to TL
hierarchy and KP hierarchy. It depends on two semi-infinite sets
of higher times $t_1,t_2,\dots$ and $t_1^*,t_2^*,\dots$, and
discrete variable $n$: $\tau=\tau(n,{\bf t},{\bf t^*})$. More
explicitly we have \cite{DJKM},\cite{JM,D},\cite{UT}:
\begin{equation}\label{tauuphi}
\quad u=2\partial_{t_1}^2\log \tau(n,{\bf t},{\bf t^*}),\quad
\phi_n({\bf t},{\bf t}^*)=-\log \frac{\tau(n+1,{\bf t},{\bf
t}^*)}{\tau(n,{\bf t},{\bf t}^*)}
\end{equation}

In soliton theory so-called Hirota-Miwa  variables ${\bf x},{\bf
y}$, which are related to higher times as
\begin{equation}\label{HMxy}
mt_m=\sum_i x_i^m,\quad mt_m^*=\sum_i y_i^m ,
\end{equation}
are well-known \cite{Miwa}. Tau function is a symmetric function
in Hirota-Miwa variables, higher times $t_m,t_m^*$ are basically
power sums, see (\ref{powsum}), see \cite{KA} as an example.

It is known fact that typical tau function may be presented in the
form of double series in Schur functions \cite{Tinit,TI,TII}:
\begin{equation}\label{tauschurschur}
\tau(n,{\bf t},{\bf t^*})=\sum_{\lambda,\mu}K_{\lambda
\mu}s_\lambda({\bf t})s_\mu ({\bf t^*})
\end{equation}
where coefficients $K_{\lambda \mu }$ are independent of higher
times and solve very special bilinear equations, equivalent to
Hirota bilinear equations.

In the previous paper we use these equations to construct
hypergeometric functions which depend on many variables, these
variables are KP and Toda lattice higher times. Here we shall use
the general approach to integrable hierarchies of Kyoto school
\cite{DJKM}, see
 also \cite{JM,D}. Especially a set
of papers about Toda lattice \cite{UT,Tinit,TI,TII,TT,NTT,T} is
important for us. Let us notice that there is an interesting set
of papers of M.Adler and P. van Moerbeke, where links between
solitons and random matrix theory were worked out from a different
point of view (see \cite{ASM},\cite{AvM} and other papers).

{\bf Random matrices}. In many problems in physics and mathematics
one use probability distribution on different sets of matrices. We
are interested mainly in two matrix models.  The integration
measure $dM_{1,2}$ is different for different ensembles
\begin{equation}\label{RM}
Z=\int e^{V_1(M_1)+V_2(M_2)+U(M_1,M_2)}dM_1dM_2
\end{equation}
where
\begin{equation}\label{VVU}
V_1(M_1)=Tr \sum_{m=1}^\infty M_1^mt_m,\quad V_2(M_2)=Tr
\sum_{m=1}^\infty M_2^mt_m^*
\end{equation}
while $U$ may be different for different ensembles.

The crucial point when evaluating the integral (\ref{RM}) is the
possibility to reduce it to the integral over eigenvalues of
matrices $M_1,M_2$. Then (\ref{RM}) takes the form
\begin{equation}\label{eigen}
Z=C\int e^{\sum_{i=1}^n \sum_{m=1}^\infty \left( x_i^mt_m+
y_i^mt_m^*\right)}e^{U({\bf x},{\bf y})}\prod_{i=1}^n dx_idy_i
\end{equation}

\section{Symmetric functions}

\subsection{Schur functions and scalar product \cite{Mac}}

{\bf Partitions}. Polynomial functions of many variables are
parameterized by partitions. A {\em partition} is any (finite or
infinite) sequence of non-negative integers in decreasing order:
\begin{equation}\label{partition}
\lambda = (\lambda_1, \lambda_2, \dots,\lambda_r,\dots ),\quad
\lambda_1 \ge \lambda_2 \ge \dots \ge\lambda_r \ge \dots
\end{equation}
Non-zero $\lambda_i$ in (\ref{partition}) are called the {\em
parts} of $\lambda$. The number of parts is the {\em length} of
$\lambda$, denoted by $l(\lambda)$. The sum of the parts is the
{\em weight} of $\lambda$, denoted by $|\lambda|$. If
$n=|\lambda|$ we say that $\lambda$ is the {\em partition of} $n$.
The partition of zero is denoted by $0$.

The {\em diagram} of a partition (or {\em Young diagram}) may be
defined as the set of points (or nodes) $(i,j) \in Z^2$ such that
$1\le j \le \lambda_i$. Thus Young diagram is viewed as a subset
of entries in a matrix with $l(\lambda)$ lines and $\lambda_1$
rows. We shall denote the diagram of $\lambda$ by the same symbol
$\lambda$.

For example

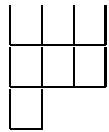
\qquad\qquad\qquad\qquad\qquad\qquad\qquad\begin{tabular}{|c|}
   \\ \hline
\end{tabular}\begin{tabular}{|c|}
   \\ \hline
\end{tabular}\begin{tabular}{|c|}
   \\ \hline
\end{tabular}

\qquad\qquad\qquad\qquad\qquad\qquad\qquad\begin{tabular}{|c|}
   \\ \hline
\end{tabular}\begin{tabular}{|c|}
   \\ \hline
\end{tabular}\begin{tabular}{|c|}
   \\ \hline
\end{tabular}

\qquad\qquad\qquad\qquad\qquad\qquad\qquad\begin{tabular}{|c|}
   \\ \hline
\end{tabular}\\
\\
is the diagram of $(3,3,1)$. The weight of this partition is $7$,
the length is equal to $3$.

The {\em conjugate} of a partition $\lambda$ is the partition
$\lambda '$ whose diagram is the transpose of the diagram
$\lambda$, i.e. the diagram obtained by reflection in the main
diagonal. The partition $(3,2,2)$ is conjugated to $(3,3,1)$, the
corresponding diagram is \\

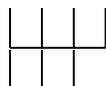
\qquad\qquad\qquad\qquad\qquad\qquad\qquad\begin{tabular}{|c|}
   \\ \hline
\end{tabular}\begin{tabular}{|c|}
   \\ \hline
\end{tabular}\begin{tabular}{|c|}
   \\ \hline
\end{tabular}

\qquad\qquad\qquad\qquad\qquad\qquad\qquad\begin{tabular}{|c|}
   \\ \hline
\end{tabular}\begin{tabular}{|c|}
   \\ \hline
\end{tabular}

\qquad\qquad\qquad\qquad\qquad\qquad\qquad\begin{tabular}{|c|}
   \\ \hline
\end{tabular}\begin{tabular}{|c|}
   \\ \hline
\end{tabular}

\quad

There are different notations for partitions. For instance
sometimes it is convenient to indicate the number of times each
integer occurs as a part:
\begin{equation}\label{mi}
\lambda =(1^{m_1}2^{m_2}\dots r^{m_r} \dots)
\end{equation}
means that exactly $m_i$ of the parts of $\lambda$ are equal to
$i$. For instance partition $(3,3,1)$ is denoted by $(1^13^2)$.

Another notation is due to Frobenius. Suppose that the main
diagonal of the diagram of $\lambda$  consists of $r$ nodes
$(i,i)\quad (1\le i\le r)$. Let $\alpha_i=\lambda_i-i$ be the
number of nodes in the $i$th row of $\lambda$ to the right of
$(i,i)$, for $1\le i\le r$, and let $\beta_i=\lambda_i'-i$ be the
number of nodes in the $i$th column of $\lambda$ below $(i,i)$,
for $1\le i\le r$. We have $\alpha_1>\alpha_2>\cdots >\alpha_r\ge
0$ and $\beta_1>\beta_2>\cdots >\beta_r\ge 0$. Then we denote the
partition $\lambda$ by
\begin{equation}\label{Frob}
\lambda = \left( \alpha_1,\dots ,\alpha_r|\beta_1,\dots
,\beta_r\right)=(\alpha |\beta )
\end{equation}
One may say that Frobenius notation corresponds to hook
decomposition of diagram of $\lambda$, where the biggest hook is
$\left( \alpha_1|\beta_1\right)$ then $\left(
\alpha_2|\beta_2\right)$ and so on up to the smallest one which is
$\left( \alpha_r|\beta_r\right)$. The corners of the hooks are
situated on the main diagonal of the diagram. For instance the
partition $(3,3,1)$ consists of two hooks $(2,2)$ and $(0,1)$:\\

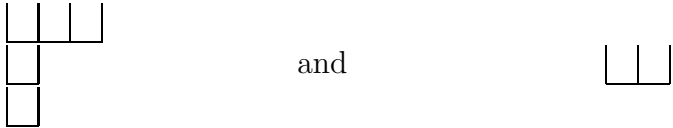
\qquad\qquad\begin{tabular}{|c|}
   \\ \hline
\end{tabular}\begin{tabular}{|c|}
   \\ \hline
\end{tabular}\begin{tabular}{|c|}
   \\ \hline
\end{tabular}

\qquad\qquad\begin{tabular}{|c|}
   \\ \hline
\end{tabular}
\qquad\qquad\qquad\qquad and
\qquad\qquad\qquad\qquad\begin{tabular}{|c|}
   \\ \hline
\end{tabular}\begin{tabular}{|c|}
   \\ \hline
\end{tabular}

\qquad\qquad\begin{tabular}{|c|}
   \\ \hline
\end{tabular}
\\
\\
In Frobenius notation this is $(2,0|2,1)$.

{\bf Schur functions, complete symmetric functions and power
sums}. We consider polynomial symmetric functions of variables
$x_i,i=1,2,\dots,n$ , where the number $n$ is irrelevant and may
be infinity. The collection of variables $x_1,x_2,\dots$ we denote
by ${\bf x}$.

The symmetric polynomials  with rational integer coefficients in
$n$ variables form a ring which is denoted by $\Lambda_n$. A ring
of symmetric functions in countably many variables is denoted by
$\Lambda$ (a precise definition of $\Lambda$ see in \cite{Mac}).

Let us remind some of well-known symmetric functions.

For each $m\ge 1$ the $m$th {\em power sum} is
\begin{equation}\label{powersum}
p_m=\sum_ix_i^m
\end{equation}
The collection of variables $p_1,p_2,\dots$ we denote by ${\bf
p}$.

If we define
\begin{equation}\label{plambda}
p_\lambda=p_{\lambda_1}p_{\lambda_2}\cdots
\end{equation}
for each partition $\lambda=(\lambda_1,\lambda_2, \dots)$, then
the $p_\lambda$ form a basic of symmetric polynomial functions
with rational coefficients.

Having in mind the links with soliton theory it is more convinient
to consider the following "power sums":
\begin{equation}\label{"powersums"}
\gamma_m=\frac 1m \sum_ix_i^m
\end{equation}
which is the same as Hirota-Miwa change of variables (\ref{HMxy}).
The vector
\begin{equation}\label{vectorgamma}
\gamma =(\gamma_1,\gamma_2,\dots )
\end{equation}
is an analog of the higher times variables in KP theory.

{\em Complete symmetric functions} $h_n$ are given by the
generating function
\begin{equation}\label{completesymmetric}
\prod_{i\ge 1}(1-x_iz)^{-1}=\sum_{n\ge 1}h_n({\bf x})z^n
\end{equation}
If we define $h_\lambda=h_{\lambda_1}h_{\lambda_2}\cdots$ for any
$\lambda$, then $h_\lambda$ form a $Z$-basis of $\Lambda$.

The complete symmetric functions are expressed in terms of power
sums with the help of the following generating function:
\begin{equation}\label{ph}
\exp{\sum_{m=1}^\infty \gamma_mz^m}=\sum_{n\ge 1}h_n({\gamma})z^n
\end{equation}

Now suppose that $n$, the number of variables $x_i$, is finite.

Given partition $\lambda$ {\em the Schur function} $s_\lambda$ is
the quotient
\begin{equation}\label{detSc}
s_\lambda({\bf
x})=\frac{\det\left(x_i^{\lambda_i+n-j}\right)_{1\le i,j\le n}}
{\det\left(x_i^{n-j}\right)_{1\le i,j\le n}}
\end{equation}
where denominator is the Vandermond determinant
\begin{equation}\label{Vand}
\det\left(x_i^{n-j}\right)_{1\le i,j\le n}  =\prod_{1\le i<j\le
n}(x_i-x_j)
\end{equation}
Both the numerator and denominator are skew-symmetric, therefore
Schur function in variables $x_1,\dots,x_n$ is symmetric.

One puts $s_0({\bf x})=1$.

{\em The Schur functions $s_\lambda(x_1,\dots,x_n)$, where
$l(\lambda)\le n$, form a $Z$-basis of $\Lambda_n$}.

Each Schur function $s_\lambda$ can be expressed as a polynomial
in the complete symmetric functions $h_n$:
\begin{equation}\label{Schurh}
s_\lambda=\det \left(h_{\lambda_i-i+j}\right)_{1\le i,j\le n}
\end{equation}
where $n\ge l(\lambda)$.

{\bf Orthogonality}. One may define a scalar product on $\Lambda$
by requiring that for any pair of partitions $\lambda$ and $\mu$
we have
\begin{equation}\label{orthp}
<p_\lambda,p_\mu>=\delta_{\lambda\mu}z_{\lambda}
\end{equation}
where $\delta_{\lambda\mu}$ is a Kronecker symbol and
\begin{equation}\label{zlambda}
z_{\lambda}=\prod_{i\ge 1}i^{m_i}\cdot m_i!
\end{equation}
where $m_i=m_i(\lambda)$ is the number of parts of $\lambda$ equal
to $i$.

For any pair of partitions $\lambda$ and $\mu$ we also have
\begin{equation}\label{ss}
<s_\lambda,s_\mu >=\delta_{\lambda ,\mu}
\end{equation}
so that the $s_\lambda$ form an {\em orthonormal} basis of
$\Lambda$. The $s_\lambda$ such that $|\lambda|=n$ form an
orhtonormal basis for all symmetric polynomials of degree $n$.

\subsection{A deformation of the scalar product and series of
hypergeometric type}

{\bf Scalar product $<,>_{r,n}$}. Let us consider a function $r$
which depends on a single variable $n$, the $n$ is integer. Given
partition $\lambda$ let us define
\begin{equation}\label{rlambda}
r_\lambda(x)=\prod_{i,j\in \lambda}r(x+j-i)
\end{equation}
Thus $r_\lambda(n)$ is a product of functions $r$ over all nodes
of Young diagram of the partition $\lambda$ where argument of $r$
is defined by entries $i,j$ of a node. The value of $j-i$ is zero
on the main diagonal; the value $j-i$ is called the content of the
node. For instance, for the partition $(3,3,1)$ the diagram is

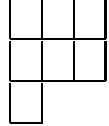
\qquad\qquad\qquad\qquad\qquad\qquad\qquad\begin{tabular}{|c|}
   \\ \hline
\end{tabular}\begin{tabular}{|c|}
   \\ \hline
\end{tabular}\begin{tabular}{|c|}
   \\ \hline
\end{tabular}

\qquad\qquad\qquad\qquad\qquad\qquad\qquad\begin{tabular}{|c|}
   \\ \hline
\end{tabular}\begin{tabular}{|c|}
   \\ \hline
\end{tabular}\begin{tabular}{|c|}
   \\ \hline
\end{tabular}

\qquad\qquad\qquad\qquad\qquad\qquad\qquad\begin{tabular}{|c|}
   \\ \hline
\end{tabular}\\
\\
so our $r_\lambda(x)$ is equal to
$r(x+2)(r(x+1))^2(r(x))^2r(x-1)r(x-2)$.

Given integer $n$ and a function on the lattice $r(n),n\in Z$,
define a scalar product, parameterized by $r,n$ as follows
\begin{equation}\label{scalrn}
<s_\lambda,s_\mu>_{r,n}=r_\lambda(n)\delta_{\lambda\mu}
\end{equation}
where $\delta_{\lambda\mu}$ is Kronecker symbol.

Let $n_i \in Z$ are zeroes of $r$, and $k=\min |n-n_i|$. The
product is non-degenerated on $\Lambda_k$. Really, if $k=n-n_i>0$
due to the definition (\ref{rlambda}) the factor $r_\lambda (n)$
never vanish for the partitions which length is no more then $k$.
In this case the Schur functions of $k$ variables
$\{s_\lambda({\bf x}^k), l(\lambda)\le k\}$  form a basis on
$\Lambda_k$. If $n-n_i=-k<0$ then the factor $r_\lambda(n) $ never
vanish for the partitions $\{\lambda :l(\lambda')\le k\}$ , where
$\lambda'$ is the conjugated partition, then $\{s_\lambda({\bf
x}^k), l(\lambda')\le k\}$ form a basis on $\Lambda_k$.

 If $r$ is non-vanishing function then the
scalar product is non-degenerated on $\Lambda_\infty$.

{\bf A useful formula}. Let ${\gamma}=(\gamma_1,\gamma_2,\dots) $
and $\gamma^*= (\gamma_1^*,\gamma_2^*,\dots) $ be two independent
semi-infinite set of variables. Then we have the following useful
version of Cauchy-Littlewood formula:
\begin{equation}\label{epp}
\exp \sum_{m=1}^\infty m\gamma_m\gamma_m^*=
\sum_{\lambda}s_\lambda(\gamma)s_\lambda(\gamma^*)
\end{equation}
where the sum is going over all partitions including zero
partition and $s_\lambda$ are constructed via (\ref{Schurh}) and
(\ref{ph}).

{\bf Series of hypergeometric type}. Let us consider the following
pair of functions of variables $p_m$
\begin{equation}\label{fpt}
\exp \sum_{m=1}^\infty m\gamma_mt_m,\quad \exp \sum_{m=1}^\infty
m\gamma_mt_m^*
\end{equation}
where $t_m$ and $t_m^*$ are formal parameters. Considering $p_m$
as power sums and using formulae (\ref{epp}) and (\ref{scalrn}),
we evaluate the scalar product of functions (\ref{fpt}):
\begin{equation}\label{exex}
<\exp \sum_{m=1}^\infty m\gamma_mt_m, \exp \sum_{m=1}^\infty
m\gamma_mt_m^*>_{r,n}=\sum_\lambda r_\lambda (n) s_\lambda({\bf
t})s_\lambda({\bf t^*})
\end{equation}
where sum is going over all partitions including zero. The Schur
functions $s_\lambda({\bf t}),s_\lambda({\bf t^*})$ are defined
with the help of
\begin{equation}\label{Schurtt*}
s_\lambda({\bf t})=\det h_{\lambda_i-i+j}({\bf t})_{1\le i,j\le
l(\lambda) },\quad \exp\sum_{m=1}^\infty z^mt_m=\sum_{k=1}^\infty
z^kh_k({\bf t})
\end{equation}

We call the series (\ref{exex}) {\em the series of hypergeometric
type}. It is not necessarily convergent series. If one chooses $r$
as rational function he obtains so-called hypergeometric functions
of matrix argument, or, in other words, hypergeometric functions
related to $GL(N,C)/U(N)$ zonal spherical functions
\cite{GR},\cite{V}. If one takes trigonometric function $r$ we
gets q-deformed versions of these functions, suggested by
I.G.Macdonald and studied by S.Milne in \cite{Milne}.

{\bf Scalar product of series of hypergeometric type}. Considering
the scalar product of functions of $\gamma_m$ and using formulae
(\ref{exex}) and (\ref{scalrn}), we evaluate the scalar product of
series of hypergeometric type (\ref{exex}), and find that the
answer is a series of hypergeometric type again:
\begin{equation}\label{scalprodtautau}
<\tau_{r_1}(k,{\bf t},\gamma),\tau_{r_2}(m,\gamma,{\bf
t^*})>_{r,n}=\tau_{r_3}(0,{\bf t},{\bf t^*})
\end{equation}
where
\begin{equation}\label{r1r2rr3}
r_3(i)=r_1(k+i)r_2(m+i)r_(n+i)
\end{equation}

\section{Solitons, free fermions and fermionic realization of the scalar
product}

In this section we use the language of free fermions as it was
suggested in \cite{JM},\cite{DJKM}.

\subsection{Fermionic operators, Fock spaces, vacuum expectations, tau
functions \cite{DJKM}}

{\bf Free fermions}. We have the algebra of free fermions {\em the
Clifford algebra} ${\bf A}$ over ${\bf C}$ with generators
$\psi_n,\psi_n^* (n \in Z)$, satisfying the defining relations:
\begin{equation} \label{antikom}
[\psi_m,\psi_n]_+=[\psi^*_m,\psi^*_n]_+=0;\qquad [\psi_m,\psi^*_n]_+=
\delta_{mn} .
\end{equation}
Any element of
$W=\left(\oplus_{m \in Z}{\bf C}\psi_m\right)\oplus
\left(\oplus_{m\in Z}{\bf C}\psi_m^*\right)$
 will be referred as a {\em free
fermion}. The Clifford algebra has a standard representation ({\em
Fock representation}) as follows. Put
$W_{an}=\left(\oplus_{m<0}{\bf C}\psi_m\right)\oplus
\left(\oplus_{m\ge 0}{\bf C}\psi_m^*\right)$, and
$W_{cr}=\left(\oplus_{m\ge 0}{\bf C}\psi_m\right)\oplus
\left(\oplus_{m< 0}{\bf C}\psi_m^*\right)$, and consider left
(resp. right) ${\bf A}$-module $F={\bf A}/{\bf A}W_{an}$ (resp
$F^*=W_{cr}{\bf A}{\backslash}{\bf A}$). These are cyclic ${\bf
A}$-modules generated by the vectors $|0\r= 1$ mod ${\bf A}W_{an}$
(resp. by $\l 0|= 1$  mod $W_{cr}{\bf A}$), with the properties
\begin{eqnarray}\label{vak}
\psi_m |0\r=0 \qquad (m<0),\qquad \psi_m^*|0\r =0 \qquad (m \ge 0) , \\
\l 0|\psi_m=0 \qquad (m\ge 0),\qquad \l 0|\psi_m^*=0 \qquad (m<0) .
\end{eqnarray}
Vectors $\l 0|$ and $|0\r$ are refereed as left and right vacuum
vectors. Fermions $w\in W_{an}$ eliminate left vacuum vector,
while fermions $w\in W_{cr}$ eliminate right vacuum vector.

The Fock spaces $F$ and $F^*$ are dual ones, the pairing is
defined with the help of a linear form $\l 0| |0 \r$ on ${\bf A}$
called the {\em vacuum expectation value}. It is given by
\begin{eqnarray}\label{psipsi*vac}
\l 0|1|0 \r=1,\quad \l 0|\psi_m\psi_m^* |0\r=1\quad m<0,
\quad  \l 0|\psi_m^*\psi_m |0\r=1\quad m\ge 0 ,
\end{eqnarray}
\begin{equation}\label{end}
 \l 0|\psi_m\psi_n |0\r=\l 0|\psi^*_m\psi^*_n |0\r=0,\quad \l
0|\psi_m\psi_n^*|0\r=0 \quad m\ne n .
\end{equation}
and by {\bf the Wick rule}, which is
\begin{equation}\label{Wick}
\l 0|w_1 \cdots w_{2n+1}|0 \r =0,\quad \l 0|w_1 \cdots w_{2n} |0\r
=\sum_\sigma sgn\sigma \l 0|w_{\sigma(1)}w_{\sigma(2)}|0\r \cdots
\l 0| w_{\sigma(2n-1)}w_{\sigma(2n)} |0\r ,
\end{equation}
where $w_k \in W$, and $\sigma$ runs over permutations such that
$\sigma(1)<\sigma(2),\dots , \sigma(2n-1)<\sigma(2n)$ and
$\sigma(1)<\sigma(3)<\cdots <\sigma(2n-1)$.

{\bf Lie algebra $\widehat {gl}(\infty)$}. Consider infinite
matrices $(a_{ij})_{i,j\in Z}$ satisfying the condition: there
exists an $N$ such that $a_{ij}=0$ for $|i-j|>N$; these matrices
are called generalized Jacobian matrices. The generalized Jacobian
matrices form Lie algebra, Lie bracket being the usual commutator
of matrices $[A,B]=AB-BA$.

Let us consider a set of linear combinations of quadratic elements
$\sum a_{ij}:\psi_i\psi_j^*:$, where the notation $: :$ means {\em
the normal ordering} which is defined as
$:\psi_i\psi_j^*:=\psi_i\psi_j^*-\l 0|\psi_i\psi_j^*|0\r$.  These
elements together with $1$ span an infinite dimensional Lie
algebra $\widehat{gl}(\infty)$:
\begin{equation}\label{commutator}
[\sum a_{ij}:\psi_i \psi_j^*:,\sum b_{ij} :\psi_i\psi_j^*:]=
\sum c_{ij}:\psi_i \psi_j^*:+c_0 ,
\end{equation}
\begin{equation} c_{ij}=\sum_k a_{ik}b_{kj}-\sum_k b_{ik}a_{kj} ,
\end{equation}
and the last term
\begin{equation}\label{cocycle}
c_0=\sum_{i<0,j\ge 0} a_{ij}b_{ji}-\sum_{i\ge 0,j<0} a_{ij}b_{ji} .
\end{equation}
commutes with each quadratic term.

 We see that Lie algebra of
quadratic elements $\sum a_{ij}:\psi_i \psi_j^*:$ is different of
the algebra of the generalized Jacobian matrices; the difference
is the {\em central extension} $c_0$.

{\bf Bilinear identity}. Now we define the operator $g$ which is
an element of the group corresponding to the Lie algebra
$\widehat{gl}(\infty)$:
\begin{equation}\label{expLie}
g=\exp \sum a_{nm}:\psi_n\psi_m:
\end{equation}
Using (\ref{commutator}) it is possible to derive the following
relation
\begin{equation} g\psi_n=\sum_m \psi_m A_{mn}g,\qquad \psi^*_n g=
g\sum_m A_{nm}\psi^*_m . \label{rot}
\end{equation}
where coefficients $A_{nm}$ are defined by $a_{nm}$.  In turn
(\ref{rot}) yields
\begin{equation} \label{bilinear}
[\sum_{n \in Z}\psi_n \otimes \psi_n^*, g \otimes g ]=0
\end{equation}
The last relation is very important for applications in soliton
theory, and it is equivalent to the so-called Hirota equations.

{\bf Higher times}. Let us introduce
\begin{equation}\label{hamiltonians}
H_n=\sum_{k=-\infty}^\infty \psi_k\psi^*_{k+n},\quad n\neq 0,
\quad
 H({\bf
t})=\sum_{n=1}^{+\infty} t_n H_n ,\quad H^*({\bf
t}^*)=\sum_{n=1}^{+\infty} t_n^* H_{-n}.
\end{equation}
$H_n \in \widehat{gl}(\infty)$, and $H({\bf t}),H^*({\bf t}^*)$
belong to $\widehat{gl}(\infty)$ if one restricts the number of
non-vanishing parameters $t_m,t_m^*$. For $H_n$ we have Heisenberg
algebra commutation relations:
\begin{equation}\label{Heisenberg}
[H_n,H_m]=n\delta_{m+n,0} .
\end{equation}

For future purposes we introduce the following fermions
\begin{equation} \label{fermions}
\psi(z)=\sum_k \psi_k z^k ,\qquad \psi^*(z)=\sum_k \psi^*_k
z^{-k-1}dz
\end{equation}
Using (\ref{antikom}) and (\ref{hamiltonians}) one obtains
\begin{equation}\label{xit}
 e^{H({\bf t})}\psi(z) e^{-H({\bf t})}= \psi(z) e^{\xi({\bf t},z)},
\quad e^{H({\bf t})}\psi^*(z) e^{-H({\bf t})} = \psi^*(z)
e^{-\xi({\bf t},z)} ,
\end{equation}
\begin{equation}\label{xit*}
 e^{-H^*({\bf t}^*)}\psi(z) e^{H^*({\bf t}^*)}=
\psi(z) e^{-\xi({\bf t}^*,z^{-1})}, \quad e^{-H^*({\bf
t}^*)}\psi^*(z) e^{H^*({\bf t}^*)}= \psi^*(z) e^{\xi({\bf
t}^*,z^{-1})} .
\end{equation}
Using
\begin{equation}\label{bilbil}
\sum_{n \in Z}\psi_n \otimes \psi_n^*=\texttt{res}_{z=0} \psi(z)
\otimes
  \psi^*(z)
\end{equation}
and formulae (\ref{xit}),(\ref{xit*}) result in
\begin{equation} \label{bilinearH}
[\sum_{n \in Z}\psi_n \otimes \psi_n^*,  e^{H({\bf t})} \otimes
e^{H({\bf t})} ]=0,\quad [\sum_{n \in Z}\psi_n \otimes \psi_n^*,
 e^{H^*({\bf t}^*)}
\otimes  e^{H^*({\bf t}^*)} ]=0
\end{equation}
Therefore if $g$ solves (\ref{bilinear}) then $e^{H({\bf
t})}ge^{H^*({\bf t}^*)}$ also solves (\ref{bilinear}):
\begin{equation}\label{bilHH}
\left[ \texttt{res}_{z=0} \psi(z) \otimes \psi^*(z),e^{H({\bf
t})}ge^{H^*({\bf t}^*)}\otimes e^{H({\bf t})}ge^{H^*({\bf
t}^*)}\right]=0
\end{equation}

{\bf The KP and TL tau functions} First let us define vacuum
vectors labelled by the integer $M$:
\begin{eqnarray}
\l n|=\l 0|\Psi^{*}_{n},\qquad |n\r=\Psi_{n}|0\r ,
\end{eqnarray}
\begin{eqnarray}
\Psi_{n}=\psi_{n-1}\cdots\psi_1\psi_0 \quad n>0,
\qquad \Psi_{n}=\psi^{*}_{n}\cdots\psi^{*}_{-2}\psi^{*}_{-1}\quad n<0 ,
 \nonumber\\
\Psi^{*}_{n}=\psi^{*}_{0}\psi^{*}_1\cdots\psi^{*}_{n-1} \quad n>0,
\qquad \Psi^{*}_{n}=\psi_{-1}\psi_{-2}\cdots\psi_{n}\quad n<0 .
\end{eqnarray}

Given $g$, which satisfy bilinear identity (\ref{bilinear}), one
constructs KP and TL tau-functions (\ref{tauuphi}) as follows:
\begin{equation}\label{taucorKP}
\tau_{KP}(n,{\bf t})= \langle n|e^{H({\bf t})}g|n \rangle ,
\end{equation}
\begin{equation}\label{taucor}
\tau_{TL}(n,{\bf t},{\bf t}^*)= \langle n|e^{H({\bf
t})}ge^{H^*({\bf t}^*)}|n \rangle .
\end{equation}
If one fixes the variables $n,{\bf t^*}$, then $\tau_{TL}$ may be
recognized as $\tau_{KP}$.

The times ${\bf t}=(t_1,t_2,\dots)$ and  ${\bf t}^*=(t_1^*,t_2^*,\dots)$
 are called higher Toda lattice times \cite{JM,UT} (the
first set ${\bf t}$ is in the same time the set of higher KP
times. The first times of this set $t_1,t_2,t_3$ are independent
variables for KP equation (\ref{KP}), which is the first
nontrivial equation in the KP hierarchy).

{\bf Fermions and Schur functions}. In the KP theory it is
suitable to use the definition of
 the  Schur function $s_\lambda$ in terms of higher times
\begin{equation}\label{Schurht}
s_{{\lambda}}({\bf t})=\det(h_{n_i-i+j}({\bf t}))_{1\le i,j\le r}
,
\end{equation}
where $h_m({\bf t})$ is the elementary Schur polynomial defined by
the Taylor's expansion:
\begin{equation}\label{elSchur}
e^{\xi({\bf t},z)}=\exp(\sum_{k=1}^{+\infty}t_kz^k)=
\sum_{n=0}^{+\infty}z^n h_n({\bf t}) .
\end{equation}

{\bf Lemma 1} \cite{DJKM} \\
For $-j_1<\cdots <-j_k<0\le i_s<\cdots <i_1$, $s-k\ge 0$ the next formula is valid:
\begin{equation}\label{glemma}
\l s-k|e^{H({\bf t})}\psi^*_{-j_1}\cdots
\psi^*_{-j_k}\psi_{i_s}\cdots\psi_{i_1}|0\r= (-1)^{j_1+\cdots
+j_k+(k-s)(k-s+1)/2}s_{{\lambda}}({\bf t}) ,
\end{equation}
where the partition ${\lambda}=(n_1,\dots , n_{s-k},
n_{s-k+1},\dots , n_{s-k+j_1})$ is defined by the pair of
partitions:
\begin{eqnarray}
(n_1, \dots , n_{s-k})=(i_1-(s-k)+1, i_2-(s-k)+2, \dots , i_{s-k}) ,\\
(n_{s-k+1},\dots , n_{s-k+j_1})=(i_{s-k+1},\dots , i_s|j_1-1,\dots , j_k-1) .
\end{eqnarray}
The proof is achieved by direct calculation.
Here $(\dots | \dots)$ is another notation for a partition due to Frobenius
(see \cite{Mac}). \\

\subsection{Partitions and free fermions}

{\bf Lemma 2} Given partition $\lambda=(i_1,\dots,i_s|j_1-1,\dots
, j_s-1 )$ let us introduce the notation
\begin{equation}\label{|lambda>}
|\lambda \r=(-1)^{j_1+\cdots +j_s}\psi_{-j_1}^*\cdots
\psi_{-j_s}^*\psi_{i_s}\cdots \psi_{i_1} |0\r
\end{equation}
\begin{equation}\label{<lambda |}
\l \lambda| =(-1)^{j_1+\cdots +j_s} \l
0|\psi^*_{i_1}\cdots\psi^*_{i_s}\psi_{-j_s} \cdots\psi_{-j_1}
\end{equation}
Then
\begin{equation}\label{lambdamu}
\l \lambda |\mu \r=\delta_{\lambda,\mu}
\end{equation}
where $\delta$ is Kronecker symbol, and
\begin{equation}\label{s(H^*)|0>}
s_\lambda({\bf H^*})|0\r= |\lambda \r
\end{equation}
\begin{equation}\label{<0|s(H)}
\l 0|s_\lambda({\bf H})= \l \lambda |
\end{equation}
here $s_\lambda$ is defined according to (\ref{Schurtt*}) where
instead of ${\bf t}=(t_1,t_2,\dots )$ we have
\begin{equation}\label{bf H*}
{\bf H^*}=(H_{-1},\frac{H_{-2}}{2},\dots,\frac{H_{-m}}{m},\dots)
\end{equation}
\begin{equation}\label{bf H}
{\bf H}=(H_1,\frac{H_2}{2},\dots,\frac{H_m}{m},\dots)
\end{equation}

The proof of (\ref{s(H^*)|0>}) and of (\ref{<0|s(H)}) follows from
using Lemma 1 and the development
\begin{equation}\label{expHt}
e^{\sum_{m=1}^\infty H_mt_m}=\sum_{\lambda }s_\lambda({\bf
H})s_\lambda({\bf t})
\end{equation}

{\bf Lemma 3}
\begin{equation}\label{s(A)|0>}
s_\lambda(-{\bf A})|0\r=r_\lambda(0)|\lambda \r
\end{equation}
where $s_\lambda$ is defined according to (\ref{Schurtt*}) where
instead of ${\bf t}=(t_1,t_2,\dots )$ we have
\begin{equation}\label{bf A}
{\bf A}=(A_1,\frac{A_2}{2},\dots,\frac{A_m}{m},\dots)
\end{equation}
\begin{equation}\label{A_k}
A_k= \sum_{n=-\infty}^\infty \psi^*_{n-k} \psi_n r(n)r(n-1)\cdots
r(n-k+1), \quad k=1,2,\dots  .
\end{equation}

The proof follows from the following. First the components of
vector ${\bf A}$ mutually commute: $[A_m,A_k]=0$. Then we apply
the development
\begin{equation}\label{expAt}
e^{\sum_{m=1}^\infty A_mt_m}=\sum_{\lambda }s_\lambda(-{\bf
A})s_\lambda({\bf t})
\end{equation}
and apply the both sides to the right vacuum vector:
\begin{equation}\label{expAvac}
e^{\sum_{m=1}^\infty A_mt_m}|0\r=\sum_{\lambda }s_\lambda({\bf
t})s_\lambda(-{\bf A})|0\r
\end{equation}
Then we develop $e^{\sum_{m=1}^\infty A_mt_m}|0\r$ using Taylor
expansion of the exponential function and also use the explicit
form of $(\ref{A_k})$, Lemma 1, the definition of $|\lambda \r$
and also (\ref{lambdamu}).

\subsection{Vacuum expectation value as a scalar product. Symmetric
function theory consideration.}

Let us remember the definition of scalar product (\ref{orthp}) on
the space of symmetric functions. It is known that one may present
it directly in terms of the variables $\gamma$ and the derivatives
with respect to these variables:
\begin{equation}\label{derivatives}
{\tilde \partial}=(\partial_{\gamma_1},\frac 12
\partial_{\gamma_2},\dots,\frac 1n
\partial_{\gamma_n},\dots)
\end{equation}

Then it is known, that due to (\ref{orthp}) the scalar product of
polynomial functions, say, $f$ and $g$, may be defined as
\begin{equation}\label{scalarproduct}
\l f,g\r  =\left(f({\tilde \partial})\cdot
g(\gamma\right)|_{\gamma=0}
\end{equation}
For instance one checks that (\ref{scalarproduct}) results in true
answer:
\begin{equation}\label{producttt}
\l \gamma_n,\gamma_m\r  =\frac 1n \delta_{n,m}, \quad \l p_n,p_m\r
=n \delta_{n,m}
\end{equation}

The following statement is important

\bprop Let scalar product is defined as in (\ref{scalarproduct}).
Then
\begin{equation}\label{vacscal}
\l f,g\r  =\l 0|f({\bf { H}})g({\bf H^*})|0\r
\end{equation}
where
\begin{equation}\label{tildeHH*}
{\bf {
H}}=\left(H_1,\frac{H_2}{2},\dots,\frac{H_n}{n},\dots\right),\quad
{\bf
H^*}=\left(H_{-1},\frac{H_{-2}}{2},\dots,\frac{H_{-n}}{n},\dots\right)
\end{equation}
\eprop

 It follows from the fact that higher times and derivatives with respect
 to higher times gives a realization of Heisenberg algebra $
H_kH_m-H_mH_k=k\delta_{k+m,0}$:
\begin{equation}
\partial_{\gamma_k}\cdot m\gamma_m-m\gamma_m\cdot
\partial_{\gamma_k}=k\delta_{k+m,0}
\end{equation}
and from the comparison of
\begin{equation}\label{Hvacuum}
\partial_{\gamma_m}\cdot 1=0,\qquad Free \quad term \quad of \quad
\partial_{\gamma_m}=0
\end{equation}
with
\begin{equation}\label{Hvacuum}
H_m|n\r=0,m>0,\qquad \l n|H_m=0,m<0
\end{equation}

~

Now let us consider deformed scalar product (\ref{scalrn})
\begin{equation}\label{rortSchur}
\l s_\mu,s_\lambda\r  _{r,n}=r_\lambda(n) \delta_{\mu,\lambda}
\end{equation}
Each polynomial function is a linear combination of Schur
functions. We have the following realization of the deformed
scalar product

\bprop
\begin{equation}\label{vacscal_rn}
\l f,g\r  _{r,n}=\l n|f({\bf { H}})g(-{\bf A})|n\r
\end{equation}
where
\begin{equation}\label{tildeHH*}
{\bf {
H}}=\left(H_1,\frac{H_2}{2},\dots,\frac{H_m}{m},\dots\right),\quad
{\bf A}=\left(A_1,\frac{A_2}{2},\dots,\frac{A_m}{m},\dots\right)
\end{equation}
$H_k$ and $A_k$ are defined respectively by (\ref{hamiltonians})
and (\ref{A_k}).
 \eprop

\subsection{KP tau-function $\tau_r(n,{\bf t},{\bf t^*} )$}

For the collection of independent variables ${\bf t^*}
=(t_1^*,t_2^*,\dots)$ we denote
\begin{equation}\label{Abeta}
A({\bf t^*})=\sum_{m=1}^\infty t_m^*A_m .
\end{equation}
where $A_m$ were defined above in (\ref{A_k}).

Using the explicit form of $A_m$ (\ref{A_k}) and anti-commutation
relations (\ref{antikom}) we obtain
\begin{equation}\label{psixir}
e^{A({\bf t^*})}\psi(z)e^{-A({\bf t}^*)}= e^{-\xi_r({\bf
t}^*,z^{-1})}\cdot \psi(z)
\end{equation}
\begin{equation}\label{psi*xir}
e^{A({\bf t}^*)}\psi^*(z)e^{-A({\bf t}^*)}= e^{\xi_{r'}({\bf
t}^*,z^{-1})}\cdot \psi^*(z)
\end{equation}
where operators
\begin{equation}\label{xir}
\xi_r({\bf t}^*,z^{-1})=\sum_{m=1}^{+\infty}t_m\left(\frac 1z
r(D)\right)^m, \quad D=z\frac{d}{dz},\quad r'(D)=r(-D)
\end{equation}
act on all functions of $z$ on the right hand side according to
the rule which was described in the beginning of the subsection
3.4, namely $r(D)\cdot z^n=r(n)z^n$. The exponents in
(\ref{psixir}),(\ref{psi*xir}),(\ref{xir}) are considered as their
Taylor series.

Therefore we have

{\bf Lemma} The fermionic operator $e^{A^({\bf t^*})}$ solves
bilinear identity (\ref{bilinear}):
\begin{equation}\label{bilAA}
\left[ \texttt{res}_{z=0} \psi(z) \otimes \psi^*(z),e^{A^({\bf
t^*})}\otimes e^{A({\bf t^*})}\right]=0
\end{equation}
Of cause this Lemma is also a result of a general treating in
\cite{DJKM}.

Now it is natural to consider the following tau function:
\begin{equation}\label{tauhyp1}
\tau_r(n,{\bf t},{\bf t^*} ):=\langle n|e^{H({\bf t})} e^{-A( {\bf
t^* })} |n\rangle .
\end{equation}

Let us use the fermionic realization of scalar product
(\ref{vacscal_rn}). We immediately obtain

\bprop
\begin{equation}\label{tauscal}
\tau_r(n,{\bf t},{\bf t^*})=\l e^{\sum_{m=1}^\infty
mt_m\gamma_m},e^{\sum_{m=1}^\infty mt_m^*\gamma_m}\r  _{r,n}
\end{equation}
\eprop

Due to (\ref{exex}) we get

\bprop
\begin{equation}\label{tauhyp}
\tau_r(n,{\bf t},{\bf t^*} ) = \sum_{{\lambda}} r_{
\lambda}(n)s_{\lambda }({\bf t})s_{\lambda }({\bf t^* }) .
\end{equation}
\eprop

We shall not consider the problem of convergence of this series.
The variables $M,{\bf t}$ play the role of KP higher times, ${\bf
t^*}$ is a collection of group times for a commuting subalgebra of
additional symmetries of KP (see \cite{OW,D',ASM} and Remark 7 in
\cite{O}). From different point of view (\ref{tauhyp}) is a
tau-function of two-dimensional Toda lattice \cite{UT} with two
sets of continuous variables ${\bf t},{\bf t^*}$ and one discrete
variable $M$. Formula (\ref{tauhyp}) is symmetric with respect to
${\bf t} \leftrightarrow {\bf t^*}$. \\
For given $r$ we define the function $r'$:
\begin{equation}\label{r'}
r'(n):=r(-n) .
\end{equation}

Tau function $\tau_r$ has properties
\begin{equation}\label{tausim}
\tau_r(n,{\bf t},{\bf t^*} )=\tau_r(n,{\bf t^*},{\bf t}),\quad
\tau_{r'}(-n,-{\bf t},-{\bf t^*}) =\tau_r(n,{\bf t},{\bf t^*} ).
\end{equation}
Also $\tau_r(n,{\bf t},{\bf t^*} )$ does not change if $t_m \to
a^mt_m,t_m^* \to a^{-m}t_m^*,m=1,2,\dots ,$, and it does change if
$t_m \to a^mt_m,m=1,2,\dots, r(n) \to a^{-1} r(n)$. (\ref{tausim})
follows from the relations
\begin{equation}
 r'_{\lambda}(n)=r_{{\lambda}'}(-n),\quad
s_{\lambda}({\bf t})=(-)^{|{\lambda}|}s_{{\lambda}'}(-{\bf t}).
\end{equation}

Let us notice that tau-function (\ref{tauhyp}) can be viewed as a
result of action of additional symmetries
\cite{O},\cite{D},\cite{OW} on the vacuum tau-function .

\subsection{Algebra of $\Psi DO$ on a circle and a development
of $e^{A({\bf t^*})},e^{{\tilde A}({\bf t})}$ }

Given functions $r,\tilde{r}$, the operators
\begin{equation}\label{A_m}
A_m= -\sum_{n=-\infty}^\infty r(n)\cdots
r(n-m+1)\psi_n\psi^*_{n-m}
 , \quad m=1,2,\dots  ,
\end{equation}
\begin{equation}\label{tA_k}
\tilde{A}_m=\sum_{n=-\infty}^\infty \tilde{r}(n+1)\cdots
\tilde{r}(n+m) \psi_n\psi_{n+m}^*, \quad m=1,2,\dots
\end{equation}
belong to the Lie algebra of pseudo-differential operators on a
circle with central extension. These operators may be also
rewritten in the following form
\begin{equation}\label{tdopsim}
{A}_m=\frac{1}{2\pi \sqrt{-1}} \oint :\psi^*(z) \left(\frac 1z
{r}(D)\right)^m \cdot\psi(z):
\end{equation}
\begin{equation}\label{tdopsim'}
\tilde{A}_m=-\frac{1}{2\pi \sqrt{-1}} \oint :\psi^*(z)
\left(\tilde{r}(D)z\right)^m \cdot\psi(z):
\end{equation}
where $D=z\frac{d}{dz}$, pseudo-differential operators
$r(D),\tilde{r}(D)$ act on functions of $z$ to the right. This
action is given as follows: $r(D)\cdot z^n=r(n)z^n$, where
$z^n,n\in Z$ is the basis of holomorphic functions in the
punctured disk $1>|z|>0$. Central extensions of the Lie algebra of
generators (\ref{tdopsim})
 (remember the subsection 3.1) are described
 by the formulae
\begin{equation}\label{ce}
\omega_n (A_m,A_k)=\delta_{mk}\tilde{r}(n+m-1)\cdots\tilde{r}(n)
r(n)\cdots r(n-m+1),\quad \omega_n-\omega_{n'}\sim 0
\end{equation}
where $n$ is an integer. These extension which differs by the
choice of $n$ are cohomological ones. The choice corresponds to
the choice of normal ordering $::$, which may chosen in a
different way as $:a:\quad :=a-\l n|a|n\r$, $n$ is an integer. (We
choose it by $n=0$ see subsection 3.1).
 When functions $r,\tilde{r}$ are polynomials the operators
$\left(\frac 1z r(D)\right)^n,\left(\tilde{r}(D)z\right)^n$ belong
to the $W_\infty$ algebra, while the fermionic operators
(\ref{A_k}),(\ref{tA_k}) belong to the algebra with central
extension denoted by ${\widehat W}_\infty$ \cite{KR}.

We are interested in calculating the vacuum expectation value of
different products of operators of type $e^{\sum A_m t_m^*}$ and
$e^{\sum{\tilde A}_m t_m}$.

{\bf Lemma} Let partitions ${\lambda}=(i_1,\dots ,i_s|j_1-1,\dots
,j_s-1)$ and ${\mu}=(\tilde{i}_1,\dots
,\tilde{i}_r|\tilde{j}_1-1,\dots ,\tilde{j}_r-1)$ satisfy the
relation ${\lambda}\supseteq{\mu}$. The following  is valid:
\begin{eqnarray}\label{e^Aexp}
\l
0|\psi^*_{\tilde{i}_1}\cdots\psi^*_{\tilde{i}_r}\psi_{-\tilde{j}_r}\cdots
\psi_{-\tilde{j}_1}e^{A({\bf t^*})}\psi^*_{-j_1}\cdots
\psi^*_{-j_s}\psi_{i_s}\cdots\psi_{i_1}|0\r=\nonumber\\
=(-1)^{\tilde{j}_1+\cdots+\tilde{j}_r+j_1+\cdots+j_s}s_{\lambda/\mu}
({\bf t^*})r_{\lambda/\mu}(0)
\end{eqnarray}
\begin{eqnarray}\label{e^tAexp}
\l 0|\psi^*_{i_1}\cdots\psi^*_{i_s}\psi_{-j_s}\cdots \psi_{-j_1}
e^{{\tilde A}({\bf t})} \psi^*_{-{\tilde j}_1}\cdots
\psi^*_{-{\tilde j}_r}\psi_{{\tilde i}_r}\cdots\psi_{{\tilde
i}_1}|0\r
=\nonumber\\
=(-1)^{\tilde{j}_1+\cdots+\tilde{j}_r+j_1+\cdots+j_s}s_{\lambda/\mu}
({\bf t})\tilde{r}_{\lambda/\mu}(0)
\end{eqnarray}
where $s_{\lambda/\mu}({\bf t})$ is a skew Schur function
\cite{Mac}:
\begin{equation}\label{skewScur}
s_{\lambda/\mu}({\bf t})=\det\left(h_{\lambda_i-\mu_j-i+j}({\bf
t})\right),
\end{equation}
and
\begin{equation}\label{skewr}
{r}_{\lambda/\mu}(n):=\prod_{i,j\in \lambda/\mu}r(n+j-i),\quad
{\tilde r}_{\lambda/\mu}(n):=\prod_{i,j\in \lambda/\mu}{\tilde
r}(n+j-i)
\end{equation}
The  proof is achieved by a development of
$,e^A=1+A+\cdots,e^{\tilde A}=1+{\tilde A}+\cdots$ and the direct
evaluation of vacuum expectations (\ref{e^Aexp}),(\ref{e^tAexp}),
and a determinant formula for the skew Schur function of Example
22 in Sec 5
of \cite{Mac}.\\

Let us introduce vectors
\begin{equation}\label{rlambdanfermion}
|\lambda,n \r=(-1)^{j_1+\cdots+j_s}\psi^*_{-j_1+n}\cdots
\psi^*_{-j_s+n}\psi_{i_s+n}\cdots\psi_{i_1+n}|n\r
\end{equation}
\begin{equation}\label{llambdanfermion}
\l\lambda,n |= (-1)^{j_1+\cdots+j_s}\l
n|\psi^*_{i_1+n}\cdots\psi^*_{i_s+n}\psi_{-j_s+n}\cdots
\psi_{-j_1+n}
\end{equation}
As we see
\begin{equation}\label{ortlambdan}
\l \lambda,n|\mu,m\r=\delta_{mn}\delta_{\lambda\mu}
\end{equation}

Using Lemma 1 and the development (\ref{expAt}) we get

{\bf Lemma}
\begin{equation}\label{lambdan}
|\lambda,n \r= s_\lambda({\bf H^*})|n\r ,\quad \l \lambda,n |= \l
n|s_\lambda({\bf { H}})
\end{equation}
and
\begin{equation}\label{lambdaSchurr}
s_\lambda(-{\bf A})|n\r=r_\lambda (n) s_\lambda({\bf
H^*})|n\r,\quad \l n|s_\lambda({\bf {\tilde A}})= {\tilde
r}_\lambda (n)\l n|s_\lambda({\bf H})
\end{equation}
\begin{equation}\label{tildeAvac}
{\bf
H}=\left(H_1,\frac{H_2}{2},\dots,\frac{H_m}{m},\dots\right),\quad
{\bf {\tilde A}}=\left(\tilde A_1,\frac{\tilde
A_2}{2},\dots,\frac{\tilde A_m}{m},\dots \right)
\end{equation}
where $A_m,{\tilde A}_m$ are given by (\ref{A_k}), (\ref{tA_k}).

There are the following developments:

\bprop
\begin{equation}\label{AskewSchurdec}
 e^{-A({\bf t^*})}=\sum_{n \in Z}\sum_{\lambda \supseteq \mu}
|\mu,n\r s_{\lambda/\mu}({\bf t^*})r_{\lambda/\mu}(n)\l \lambda
,n|
\end{equation}
\begin{equation}\label{tildeAskewSchurdec}
 e^{{\tilde A}({\bf t})}=\sum_{n \in Z}\sum_{\lambda \supseteq \mu}
|\lambda,n\r s_{\lambda/\mu}({\bf t}){\tilde r}_{\lambda/\mu}(n)\l
\mu ,n|
\end{equation}
where $s_{\lambda/\mu}({\bf t})$ is the skew Schur function
(\ref{skewScur}) and
${r}_{\lambda/\mu}(n),\tilde{r}_{\lambda/\mu}(n)$ are defined in
(\ref{skewr}).

\eprop

The proof follows from the relation \cite{Mac}
\begin{equation}\label{skewort}
\l s_{\lambda/\mu},s_\nu \r  =\l s_\lambda,s_\mu s_\nu \r
\end{equation}

\section{Integral representation of the scalar product and
random matrices}

In case the function $r$ has zero the scalar product
(\ref{rortSchurmulti}) is degenerate one. For simplicity let us
take $r(0)=0,r(k)\neq 0, k \le n$. The scalar product
(\ref{rortSchurmulti}) non generate on the subspace of symmetric
functions spanned by Schur functions $\{ s_\lambda ,l(\lambda)\le
n \}$, $l(\lambda)$ is the length of partition $\lambda$.

We found the integral representation of this scalar product.

To get this representation one should suggest that there exist a
function $\mu_r(xy)$ with the following properties
\begin{equation}\label{stringfr}
\left(y-\frac 1x r(-D_x)\right)\mu_r(xy)=0,\quad D_x=x\frac{d}{dx}
\end{equation}
\begin{equation}\label{stringfr*}
\left(x-\frac 1y r(-D_y)\right)\mu_r(xy)=0,\quad D_y=y\frac{d}{dy}
\end{equation}
and the integral
\begin{equation}\label{normirovka}
\int_\Gamma \mu_r (xy) dxdy=A
\end{equation}
is finite and does not vanish.

The integration domain $\Gamma$ should be chosen in such a way
that the operator $\frac 1x r(D),D=x\frac{d}{dx}$ is conjugated to
the operator $\frac 1x r(-D),D=x\frac{d}{dx}$, and also
\begin{equation}\label{domainGamma}
\frac 1A \int_\Gamma \mu_r (xy)x^n dxdy=\frac 1A \int \int_\Gamma
\mu_r(xy) {y}^n dxdy=\delta_{n,0}
\end{equation}
At last with the help of (\ref{stringfr}) we get
\begin{equation}\label{intzz^*}
\frac 1A  \int_\Gamma \mu_r (xy)x^n{y}^m
dxdy=\delta_{n,m}r(1)r(2)\cdots r(n)
\end{equation}
Now one can evaluate the integral
\begin{equation}\label{intSchur}
I_{\lambda,\nu}=A^{-n}\int \cdots \int \Delta({\bf x})\Delta({\bf
y})s_\lambda({\bf x})s_\nu({\bf y}) \prod_{k=1}^n
\mu_r(x_k{y}_k)dx_kdy_k
\end{equation}
where ${\bf x}=(x_1,\dots,x_n),{\bf y}=(y_1,\dots,y_n)$
\begin{equation}\label{vandzz^*}
\Delta({\bf z})=\prod_{i<j}^n(z_i-z_j)
\end{equation}
Using the definition of the Schur functions (\ref{detSc}) and the
definition of $r_\lambda(n)$ (\ref{rlambda}) we finely get the
relation
\begin{equation}\label{Idelta}
I_{\lambda\nu}=r_\lambda(n)\delta_{\nu,\lambda}
\end{equation}
and we obtain

\bprop
\begin{equation}\label{fgscalint}
\l f,g \r_{r,n} =A^{-n}\int_\Gamma \cdots \int_\Gamma \Delta({\bf
x})\Delta({\bf y})f({\bf x})g ({\bf y}) \prod_{k=1}^n
\mu_r(x_k{y}_k)dx_kdy_k
\end{equation}
\eprop

{\bf In case ${\bf r(0)=0}$} the series
\begin{equation}\label{tau1/r}
\mu_r(xy)=
1+\frac{xy}{r(-1)}+\frac{\left(xy\right)^2}{r(-1)r(-2)}+ \cdots
\end{equation}
is the formal solution of (\ref{stringfr}),(\ref{stringfr*}). This
series is equal to the tau-function of hypergeometric type related
to $1/r'$ , where
\begin{equation}\label{defr'}
  r'(n):= r(-n)
\end{equation}

\begin{equation}\label{mutau1/r}
\mu_r(xy)=\tau_{1/{r'}}\left(1,{\bf t}(x),{\bf
t^*}(y)\right),\quad kt_k=x^k,\quad kt_k^*={y}^k
\end{equation}
Let us suggest that the series (\ref{tau1/r}) is convergent (this
fact depends on the choice of the function $r$) , and the integral
(\ref{normirovka}) exists, then one may choose this $\mu_r$ for
(\ref{fgscalint}).

Let us remember that if
\begin{equation}\label{tt*xy}
{\bf t}({\bf x})=(t_1({\bf x}),t_2({\bf x}),\dots ),\quad {\bf
t^*}({\bf y})=(t_1^*({\bf y}),t_2^*({\bf y}),\dots ),\quad
mt_m=\sum_{i=1}^n x_i^m,\quad mt_m^*=\sum_{i=1}^n y_i^m
\end{equation}
we have the following determinant representation \cite{pa24} ,
\cite{pd22} (see an Appendix below
(\ref{det2}),
(\ref{vand}),
(\ref{vandtilde+}))
\begin{equation}\label{tau1taun}
{\Delta({\bf x})\Delta({\bf y})}\tau_{r}\left(n,{\bf t}({\bf
x}),{\bf t^*}({\bf y})\right)= c_n{\det
\left(\tau_{r}\left(1,x_i,y_k\right)\right)_{i,k=1}^n},\quad
c_n=\prod_{k=0}^{n-1}\left( r(k)\right)^{k-n}
\end{equation}
where ${\bf x}=(x_1,\dots,x_n),{\bf y}=(y_1,\dots,y_n)$ which is
true on the level of formal series.

$\tau_{1/{r'}}\left(n,{\bf t}({\bf x}),{\bf t^*}({\bf y})\right)$
is the tau-function (\ref{tauhyp}),
 where the function $r$ is chosen as $1/{r'}$
\begin{equation}\label{tau1/r'}
\tau_{1/{r'}}\left(n,{\bf t}({\bf x}),{\bf t^*}({\bf y})\right)=
1+\frac{\left(\sum_{i=1}^n x_i\right)\left(\sum_{i=1}^n
y_i\right)}{r(-1)}+\cdots
\end{equation}
Due to determinant representation (\ref{tau1taun}) we see that
this series is a convergent one, since we consider the series
(\ref{mutau1/r}) to be convergent.

\bprop
Let the tau function $\tau_{1/r'}$ of (\ref{tauhyp}), where
$r'(n)\equiv r(-n)$, is a convergent series, and let $r(0)=0$.
Then we have the following realization of scalar product
(\ref{scalrn})
\begin{equation}\label{inttau1/r}
\l f,g \r_{r,n} =A^{-n}\int_\Gamma \cdots \int_\Gamma f({\bf x})g
({\bf y})\tau_{1/{r'}}\left(n,{\bf t}({\bf x}),{\bf t^*}({\bf
y})\right) \left(\Delta({\bf x})\right)^2\left(\Delta({\bf
y})\right)^2\prod_{k=1}^n dx_kdy_k
\end{equation}
\eprop

The proof follows from the fact that the function
$\tau_{1/{r'}}\left(n,{\bf t}({\bf x}),{\bf t^*}({\bf y})\right)$
is a symmetric function of variables $x_1,\dots,x_n$. Then by
changing notations of variables $x_i$ inside of the integral we
obtain the Preposition.

Then it follows from (\ref{inttau1/r}),(\ref{exex}) that for the
same conditions the following duality relation is true
\begin{equation}\label{duality}
\tau_r\left(n,{\bf t},{\bf t^*})\right)=
\end{equation}
\begin{equation}
A^{-n}c_n^{-1}\int_\Gamma \cdots \int_\Gamma
e^{\sum_{i=1}^n\sum_{m=1}^\infty (t_mx_i^m + t_m^*y_i^m)}
\tau_{1/{r'}}\left(n,{\bf t}({\bf x}),{\bf t^*}({\bf y})\right)
\left(\Delta({\bf x})\right)^2\left(\Delta({\bf
y})\right)^2\prod_{k=1}^n dx_kdy_k
\end{equation}

Also due to (\ref{scalprodtautau}),(\ref{r1r2rr3}) we have the
following
\begin{equation}\label{tau3tau2tau1tau}
\tau_{r_3}\left(0,{\bf t},{\bf t^*}\right)=
\end{equation}
\begin{equation}
A^{-n}c_n^{-1}\int_\Gamma \cdots \int_\Gamma
\tau_{r_1}\left(k,{\bf t},{\bf t^*}({\bf
x})\right)\tau_{1/{r'}}\left(n,{\bf t}({\bf x}),{\bf t^*}({\bf
y})\right) \tau_{r_2}\left(m,{\bf t}({\bf y}),{\bf
t^*}\right)\Delta^2({\bf x})\Delta^2({\bf y})\prod_{k=1}^n
dx_kdy_k
\end{equation}
where
\begin{equation}
r_3(i)=r_1(k+i)r_2(m+i)r_(n+i)
\end{equation}

\subsection{Standard scalar product}

Let us suggest a realization of the standard scalar product
(\ref{scalarproduct}). This scalar product is a particular case of
(\ref{rortSchur}).
 In this case $r\equiv 1$. This function has no zeroes, therefore
 the consideration of the previous subsection is not valid.
 We realize (\ref{scalarproduct}) as follows
\begin{equation}\label{standscalreal}
<f,g>=\int f({\bf t})e^{-\sum_{m=1}^{\infty}m|t_m|^2}g({\bf {\bar
t}})\prod_{m=1}^{\infty}\frac{md^2t_m}{\pi}
\end{equation}
where ${\bar t}_m$ is the complex conjugated of $t_m$. The formula
for the scalar product of two tau-functions is similar to
(\ref{tau3tau2tau1tau}):
\begin{equation}\label{tautaustand}
<\tau_{r_1}(k),\tau_{r_2}(m)>=\tau_{r_3}(0)=
\end{equation}
where due to (\ref{tauhyp}),(\ref{scalarproduct})
$r_3(i)=r_1(k+i)r_2(m+i)$, and
\begin{equation}
=\int \tau_{r_1}\left(k,{\bf t},{\gamma}\right)
e^{-\sum_{k=1}^{\infty}k|\gamma_k|^2}\tau_{r_2}(m,{\bar \gamma},
{\bf t^*})\prod_{i=1}^{\infty}\frac{kd^2\gamma_k}{\pi}
\end{equation}

\subsection{Normal matrix model}

Let us consider a model of normal matrices. A matrix is called
normal if it commutes with Hermitian conjugated matrix.

Let $M$ be $n$ by $n$ normal matrix. The following integral is
called the model of normal matrices:
\begin{equation}\label{normint}
I^{NM}(n,{\bf t},{\bf t^*};{\bf u})=\int dMdM^+
e^{U(MM^+)+V_1(M)+V_2(M^+)}=
\end{equation}
\begin{equation}\label{normint-}
C\int \cdots \int |\Delta({\bf z})|^2 e^{\sum_{i=1}^nU(z_i{\bar
z}_i)+V_1(z_i)+V_2({\bar z}_i)}\prod_{i=1}^n dz_id{\bar z}_i
\end{equation}
where  $dM=\prod_{i<k} d\Re M_{ik} d\Im M_{ik}\prod_{i=1}^N
dM_{ii}$, $C$ is a number which appears due to the angle
integration, and $U$ is defined by ${\bf u}=(u_1,u_2,\dots)$:
\begin{equation}\label{UV}
U(MM^+)=\sum_{n=1}^\infty u_n Tr (MM^+)^n,\quad
V_1(M)=\sum_{m=1}^\infty t_m Tr M^m,\quad V_2(M)=\sum_{m=1}^\infty
t_m^* Tr \left(M^+\right)^m
\end{equation}
\begin{equation}\label{Delt}
\Delta({\bf z})=\prod_{i<k}^n(z_i-z_k)
\end{equation}
and $\Delta(z)=1$ for $n=1$.

As we see the integral (\ref{normint-}) has a form of
(\ref{fgscalint}), where ${\bf x}={\bf z},{\bf y}={\bf {\bar z}}$
and
\begin{equation}\label{fgmu}
f({\bf z})=e^{\sum_{i=1}^n\sum_{m=1}^\infty t_mz_i^m},\quad g({\bf
{\bar z}})=e^{\sum_{i=1}^n\sum_{m=1}^\infty t_m^*{\bar
z}_i^m},\quad \mu_r(z_i{\bar z}_i)= e^{U(z_i{\bar z}_i)}
\end{equation}
Comparing $e^U$ with $\mu_r$ of (\ref{tau1/r}) we obtain that the
set of variables ${\bf u}=(u_1,u_2,\dots)$ is related to values of
$r(n),n=-1,-2,\dots$ as follows
\begin{equation}\label{ur}
h_m({\bf u})=\frac {1}{r(-1)}\cdots \frac {1}{r(-m)},\quad
r(-m)=\frac{h_{m-1}({\bf u})}{h_m({\bf u})},\quad
e^{\sum_{k=1}^\infty u_k z^k}=\sum_{m=0}^\infty h_m({\bf u})z^m
\end{equation}

Finely  formula (\ref{exex}), where $p_m=\sum_{i=1}^nz_i^m$,
yields the asymptotic series for (\ref{normint}):
\begin{equation}\label{INMpert}
I^{NM}(n,{\bf t},{\bf t^*};{\bf
u})=<e^{\sum_{i=1}^n\sum_{m=1}^\infty t_mz_i^m}
,e^{\sum_{i=1}^n\sum_{m=1}^\infty t_m^*{ z}_i^m}
>_{r,n}=\sum_{\lambda}r_\lambda(n) s_\lambda({\bf
t})s_\lambda({\bf t^*})
\end{equation}

\subsection{Two-matrix model }
Details of this and of the next section one can find in \cite{OH}.
Let us evaluate the following integral over $n$ by $n$  matrices
$M_1$ and $M_2$, where $M_1$ is a Hermitian matrix and $M_2$ is an
anti-Hermitian one
\begin{equation}\label{HTMMmeasure}
I^{2MM}(n,{\bf t},{\bf t^*})=\int
e^{V_1(M_1)+V_2(M_2)}e^{-\texttt{Tr} M_1M_2}dM_1dM_2
\end{equation}
where
\begin{equation}\label{HTMMV}
V_1(M_1)=\sum_{m=1}^\infty t_m\textrm{Tr} M_1
^m,\quad
V_2(M_2)=\sum_{m=1}^\infty t_m^*\textrm{Tr} M_2^m
\end{equation}
It is well-known \cite{HCIZ},\cite{Mehta} that this integral
reduces to the integral over eigenvalues $x_i$ and $y_i$  of
matrices $M_1$ and $M_2$ respectively.

To do it first we diagonalize the matrices:
$M_1=U_1XU_1^{-1},M_2=\sqrt{-1}U_2YU_2^{-1}$, where $U_1,U_2$ are
unitary matrices and $X,Y$ are diagonal matrices with real
entries. Then one calculates the Jacobians of the change of
variables $M_1 \rightarrow (X,U_1),M_2 \rightarrow (Y,U_2)$ :
$dM_1=\left(\Delta({\bf x})\right)^2\prod_{k=1}^n
dx_k\prod_{i<j}d\Re(U_1)_{ij}d\Im(U_1)_{ij}$ and $
dM_2=\left(\Delta({\bf y})\right)^2\prod_{k=1}^n
dy_k\prod_{i<j}d\Re(U_2)_{ij}d\Im(U_2)_{ij}$. As a result we get
\begin{equation}\label{2MMIZ}
I^{HTMM}(n,{\bf t},{\bf t^*})=\int
e^{\sum_{i=1}^n\sum_{m=1}^\infty t_mx_i^m}
e^{\sum_{i=1}^n\sum_{m=1}^\infty t_my_i^m} I({\bf x},{\bf y})
\Delta^2({\bf x})\Delta^2({\bf y})\prod_{k=1}^n dx_kdy_k \int
d_*U_1
\end{equation}
where
\begin{equation}\label{IZ}
I({\bf x},{\bf y})= \int e^{-\texttt{Tr} XUYU^{-1}}dUU^{-1}
\end{equation}
is the so-called {\bf Harish-Chandra-Itzykson-Zuber integral}.
This integral was calculated \cite{HCIZ}, \cite{Mehta} :
\begin{equation}\label{IZdet}
I({\bf x},{\bf y})=\frac{\det
\left(e^{-x_iy_k\sqrt{-1}}\right)_{i,k=1}^n}{\Delta({\bf
x})\Delta({\bf y})}
\end{equation}
Due to (\ref{tau1taun}) formula (\ref{IZdet}) is a manifest of the
fact that the Harish-Chandra-Itzykson-Zuber integral actually is a
tau-function (see a different approach to this fact in \cite{ZJ}).
This is the tau-function $\tau_r$ which corresponds to
$r(n)=-1/n$, in other words we can write
\begin{equation}\label{IZ=tau}
\frac{I({\bf x},{\bf
y})}{0!\cdots(n-1)!}=\tau_{1/{r'}}\left(n,{\bf t}({\bf x}),{\bf
t^*}({\bf y})\right),\quad 1/{r'(n)}=-\frac 1n
\end{equation}
For $n=1$, $x_1=x,y_1=y$ we have
\begin{equation}\label{tau(1)1/r}
\tau_{1/{r'}}\left(1,{\bf t}({\bf x}),{\bf t^*}({\bf
y})\right)=e^{-xy\sqrt{-1}}
\end{equation}
Due to the equality
\begin{equation}\label{integrexp}
\int_{-\infty}^\infty\int_{-\infty}^\infty
e^{-xy\sqrt{-1}}dxdy=2\pi
\end{equation}
the condition (\ref{normirovka} ) is fulfilled.

Therefore due to (\ref{inttau1/r}) we obtain that the model of two
Hermitian random matrices is the series of hypergeometric type
(\ref{exex}), where $p_m=\sum_{i=1}^nz_i^m$ and $r(n)=n$:
\begin{equation}\label{2MM=scalprod}
I^{2MM}(n,{\bf t},{\bf t^*})=\frac{C_n}{(2\pi)^n}
<e^{\sum_{i=1}^n\sum_{m=1}^\infty t_mz_i^m}
,e^{\sum_{i=1}^n\sum_{m=1}^\infty t_m^*{ z}_i^m}
>_{r,n}=\frac{C_n}{(2\pi)^n}\sum_{\lambda}(n)_\lambda s_\lambda({\bf
t})s_\lambda({\bf t^*})
\end{equation}
where we remember that according to (\ref{rlambda})
\begin{equation}\label{(n)lambda}
(n)_\lambda=\prod_{i,j\in
\lambda}(n+j-i)=\frac{\Gamma(n+1+\lambda_1)\Gamma(n+\lambda_2)\cdots
\Gamma(\lambda_n)}{\Gamma(n+1)\Gamma(n)\cdots \Gamma(1)}
\end{equation}
and $C_n$ incorporates the volume of unitary group.

The same result one can obtain with the help of (\ref{fgscalint}),
using the known result that integral (\ref{2MMIZ}) is equal to
\begin{equation}
C_n\int_{-\infty}^\infty \cdots \int_{-\infty}^\infty
e^{\sum_{i=1}^n\sum_{m=1}^\infty t_mx_i^m}
e^{\sum_{i=1}^n\sum_{m=1}^\infty t_my_i^m}e^{-\sum_{i=1}^n
x_iy_i\sqrt{-1}} {\Delta({\bf x})\Delta({\bf y})}\prod_{i=1}^n
dx_idy_i
\end{equation}
where $x_i$ and $y_i$ are eigenvalues of $M_1$ and $M_2$
respectively.

This integral has a form of (\ref{fgmu}) if one takes the
integrals over variables $x_i$ be along the real axe, while the
integrals over variables $y_i$ be along the imaginary axe. Then we
take $r(n)=n$ so that
\begin{equation}\label{2MMmuint}
\mu_r(xy)=\tau_{1/{r'}}=1-xy+\frac{x^2y^2}{2}+\cdots=e^{-xy}
\end{equation}
\begin{equation}\label{2MMnorm}
\int_{\Re}\int_{\Im} e^{-xy}dxdy=2\pi
\end{equation}

Thus we have the following asymptotic perturbation series
\cite{OH}
\begin{equation}\label{2MMss}
\frac{I^{2MM}(n,{\bf t},{\bf t^*})}{I^{2MM}(n,0,0)}
=\sum_{\lambda} (n)_{\lambda}s_{\lambda}({\bf t})s_{\lambda}({\bf
t}^*)
\end{equation}
In case all higher coupling constants are equal to 0 (namely, in
the case of Gauss matrix integral) this series can be easily
evaluated as
\begin{equation}\label{gauss2MM}
I^{2MM}(1,t_1,t_2,0,0,\dots;t_1^*,t_2^*,0,0,\dots)=\sum_{m=0}^\infty
m!h_m({\bf t})h_m({\bf t^*})= \frac {\exp
\frac{t_1t_1^*+t_2\left(t_1^*\right)^2+t_2^*(t_1)^2}
{1-4t_2t_2^*}}{\sqrt{1-4t_2t_2^*}}
\end{equation}
and for $n>1$ one can use the formula (for example see \cite{OH})
\begin{equation}\label{taun=dettau1}
\tau_r\left(n,{\bf t},{\bf t^*}\right)=\frac{1}{(n-1)!\cdots
0!}\det \left(
\frac{\partial^{k+m}}{\partial_{t_1}^k\partial_{t_1^*}^m}\tau_r\left(1,{\bf
t},{\bf t^*}\right)\right)_{k,m=0}^{n-1}
\end{equation}
 which is true for the case $r(0)=0$.

\subsection{Hermitian one matrix model}

Let us evaluate simplest perturbation terms for one matrix model
using the series (\ref{gauss2MM}) \cite{OH}. Let us take all
$t_k=0$ except $t_4$, and all $t_k^*=0$ except $t_2^*$ \cite{BEH}.
If we put
\begin{equation}\label{tg}
g_4=-4N^{-1}t_4,\quad g=(2Nt_2^*)^{-1}
\end{equation}

we obtain that
\begin{equation}\label{1MM}
\sum_{\lambda} (N)_{\lambda}s_{\lambda}({\bf t})s_{\lambda}({\bf
t}^*)=\int dM e^{-NTr(\frac g2 M^2+g_4 M^4)}
\end{equation}

We chose the normalization of matrix integral in such a way that
it is equal to $1$ when $g_4=0$.

{\bf Remark}. As it is well-known, see for instance
\cite{Eynardpreprint}, according to Feynman rules for one matrix
model we have (a) to each propagator (double line) is associated a
factor $1/(Ng)$ (which is $2t_2^*$ in our notations) (b) to each
four leg vertex is associated a factor $(-Ng_4)$ (which is $4t_4$
in our notations) (c) to each closed single line is associated a
factor $N$. Therefore one may say that the factors $(N)_{\lambda}$
in (\ref{1MM}) is responsible for closed lines, the factors
$s_{\lambda}({\bf t}=0,t_2,0,\dots)$ are responsible for
propagators and the factors $s_{\lambda}({\bf t}^*)$ are
responsible for vertices. As we see we get the following structure
for the Feynman diagrams, containing $k=|{\lambda}|/4$ vertices
and $2k=|{\lambda}|/2$ propagators:
\begin{equation}\label{numberN}
(Ng_4)^{|{\lambda}|/4}(Ng)^{-|{\lambda}|/2}\sum_{|{\lambda}|=4k}
(N)_{\lambda}a_{\lambda}b_{\lambda}
\end{equation}
where the number $|{\lambda}|$ is the weight of partition
${\lambda}$, and $a_{\lambda},b_{\lambda}$ are numbers:
\begin{equation}\label{ab}
a_{\lambda}=s_{\lambda}(0,0,0,1,0,\dots),\quad
b_{\lambda}=s_{\lambda}(0,1,0,\dots)
\end{equation}

Since only $t_4$ is non vanishing the first non vanishing Schur
functions correspond to ${\lambda}$ of weight $4$, which are
$(4),(3,1),(2,2),(2,1,1),(1,1,1,1)$. Let us notice that two last
partition are conjugated to the two first, and the partition
$(2,2)$ is self conjugated.

To evaluate l.h.s. of (\ref{1MM}) we use (\ref{rlambda}) ,
(\ref{Schurh}) and take into account the property for conjugated
partition ${\lambda}'$ we have
\begin{equation}\label{sym}
s_{{\lambda}'}(-{\bf t}) =(-)^{|{\lambda|}}s_{{\lambda}}({\bf
t}),\quad r'_{{\lambda }'}(-n)=r_{\lambda}(n)
\end{equation}
where $|{\lambda}|$ is the number of boxes in the Young diagram
(or the same - the weight of the partition), and for $r'$ see
(\ref{r'}).

The partition ${\lambda}=(4)$ gives
\begin{equation}\label{4}
N(N+1)(N+2)(N+3)t_4\frac{(t_2^*)^2}{2!}=
(N^4+6N^3+11N^2+6N)\left(-\frac{Ng_4}{4}\right)
\left(\frac{1}{8N^2g^2}\right)
\end{equation}

Due to (\ref{sym}) , (\ref{tg})  the conjugated partition
$(1,1,1,1)$ gives the same answer but $N \rightarrow -N$. (Thus
one needs to keep only even powers of $N$).

The partition $(3,1)$ gives
\begin{equation}\label{3,1}
N(N+1)(N+2)(N-1)(-t_4)\frac{(-t_2^*)^2}{2!}=
(N^4+2N^3-N^2-2N)\left(-\frac{Ng_4}{4}\right)
\left(\frac{1}{8N^2g^2}\right)
\end{equation}

Again due to the contribution of the conjugated partition
$(2,1,1)$ one needs to keep only even powers of $N$.

First Schur function in the l.h.s of (\ref{1MM}) is equal to zero
(due to (\ref{Schurh}) and since only $t_4$ is non vanishing).

Finely we find first perturbation terms as

\begin{equation}\label{2term}
1-\left(\frac{N^2}{2}+\frac 14 \right)\frac{g_4}{g^2}+\cdots
\end{equation}

This answer coincides with the well-known answer one gets by
Feynman diagram method, see \cite{Eynardpreprint}

For the next order one takes partitions of weight $8$, as a
consequence that it is only $t_4$ that is non vanishing. One can
get that non vanishing contribution is only due to self conjugated
partitions $(3,3,2)$, $(4,2,1,1)$, and to partitions
$(8);(7,1),(4,4);(6,1,1), (4,3,1);(5,1,1,1)$ and conjugated to
these ones. The enumerated partitions yields respectively
$h_4^2,h_4^2$, then $h_8;-h_8,h_4^2$; $-h_8,-h_4^2;-h_8$ for
$s_{\lambda}({\bf t}^*)$. And it yields respectively
$h_4^2-h_4h_2^2,h_4^2-h_4h_2^2$, then
$h_8;-h_8,h_4^2;-h_8-h_6h_2,-h_4^2+h_6h_2;-h_8+h_6h_2$ for
$s_{\lambda}({\bf t})$. For partitions $(3,3,2)$, $(4,2,1,1)$, and
for partitions $(8); (7,1),(4,4);(6,1,1), (4,3,1);(5,1,1,1)$ we
get  respectively
\begin{equation}\label{332}
N(N+1)(N+2)(N-1)(N)(N+1)(N-2)(N-1) \left(t_4^2\right)
\left(\left(\frac{(t_2^*)^2}{2!}\right)^2-\frac{(t_2^*)^2}{2!}(t_2^*)^2
\right)
\end{equation}
\begin{equation}\label{4211}
N(N+1)(N+2)(N+3)(N-1)(N)(N-2)(N-3)(t_4^2)
\left(\left(\frac{(t_2^*)^2}{2!}\right)^2-
\frac{(t_2^*)^2}{2!}(t^*_2)^2\right)
\end{equation}
and
\begin{equation}\label{8}
N(N+1)(N+2)(N+3)(N+4)(N+5)(N+6)(N+7)
\left(\frac{t_4^2}{2!}\right)\left(\frac{(t_2^*)^4}{4!}\right)
\end{equation}
\begin{equation}\label{71}
N(N+1)(N+2)(N+3)(N+4)(N+5)(N+6)(N-1)
\left(-\frac{t_4^2}{2!}\right)\left(-\frac{(t_2^*)^4}{4!}\right)
\end{equation}
\begin{equation}\label{44}
N(N+1)(N+2)(N+3)(N-1)(N)(N+1)(N+2)
\left(t_4^2\right)\left(\frac{(t_2^*)^2}{2!}\right)^2
\end{equation}
\begin{equation}\label{611}
N(N+1)(N+2)(N+3)(N+4)(N+5)(N-1)(N-2)
\left(\frac{t_4^2}{2!}\right)\left(\frac{(t_2^*)^4}{4!}-
t_2^*\frac{(t^*_2)^3}{3!}\right)
\end{equation}
\begin{equation}\label{431}
N(N+1)(N+2)(N+3)(N-1)(N)(N+1)(N-2)
\left(-t_4^2\right)\left(-\left(\frac{(t_2^*)^2}{2!}\right)^2+
t_2^*\frac{(t^*_2)^3}{3!}\right)
\end{equation}
\begin{equation}\label{5111}
N(N+1)(N+2)(N+3)(N+4)(N-1)(N-2)(N-3)
\left(-\frac{t_4^2}{2!}\right)\left(-\frac{(t_2^*)^4}{4!}
+t_2^*\frac{(t^*_2)^3}{3!}\right)
\end{equation}
At the highest order at $N$ which is $N^6$ one gets zero.

\begin{equation}\label{sum}
(32N^6+320N^4+488N^2)t_4^2(t_2^*)^4=
(32N^6+320N^4+488N^2)\left(\frac{Ng_4}{4}\right)^2
\left(\frac{2}{Ng}\right)^4=
\end{equation}
$$
\frac{g_4^2}{g^4}(32N^4+320N^2+488)
$$

\section{Appendices A}

\subsection{Baker-Akhiezer functions and bilinear identities
\cite{DJKM}}

Vertex operators $V_\infty({\bf t},z)$, $V_\infty^*({\bf t},z)$
and $V_0({\bf t}^*,z)$, $V_0^*({\bf t}^*,z)$ act on the space
$C[t_1,t_2,\dots ]$ of polynomials in infinitely many variables,
and are defined by the formulae:
\begin{equation}\label{vertex}
V_\infty ({\bf t},z)=z^Me^{\xi({\bf
t},z)}e^{-\xi(\tilde{\partial},z^{-1})},\quad V^*_\infty ({\bf
t},z)=z^{-M}e^{-\xi({\bf t},z)}e^{\xi(\tilde{\partial},z^{-1})} ,
\end{equation}
\begin{equation}\label{vertex0}
 V_0({\bf t}^*,z)=z^{M}e^{-\xi({\bf t}^*,z^{-1})}e^{\xi(\tilde{\partial}^*,z)},\quad
V^*_0({\bf t}^*,z)=z^{-M}e^{\xi({\bf
t}^*,z^{-1})}e^{-\xi(\tilde{\partial}^*,z)} ,
\end{equation}
where $\tilde{\partial}=(\frac{\partial}{\partial t_1},
\frac{1}{2}\frac{\partial} {\partial
t_2},\frac{1}{3}\frac{\partial}{\partial t_3},\dots)$,
$\tilde{\partial}^*=(\frac{\partial}{\partial t_1^*},
\frac{1}{2}\frac{\partial} {\partial
t_2^*},\frac{1}{3}\frac{\partial}{\partial t_3^*},\dots)$.

We have the rules of the bosonization:
\begin{eqnarray} \label{bos1}
\l M+1|e^{H({\bf t})}\psi(z)=V_\infty ({\bf t},z)\l M|e^{H({\bf
t})},\quad
 \l M-1|e^{H({\bf t})}\psi^*(z)=
 V^*_\infty ({\bf t},z)\l M |e^{H({\bf t})}\frac{dz}{z} ,
\end{eqnarray}
\begin{eqnarray} \label{bos2}
 \psi (z)e^{H^*({\bf t}^*)}|M\r=
 V_0({\bf t}^*,z)e^{H^*({\bf t}^*)}|M-1\r ,
 \psi^*(z)e^{H^*({\bf t}^*)}|M\r=V^*_0({\bf t}^*,z) e^{H^*({\bf
t}^*)}|M+1\r\frac{dz}{z}
\end{eqnarray}

The Baker-Akhiezer functions and conjugated Baker-Akhiezer
functions are:
\begin{eqnarray}\label{baker}
w_\infty (M,{\bf t},{\bf t}^*,z)=\frac{V_\infty({\bf
t},z)\tau}{\tau},\quad
w^*_\infty (M,{\bf t},{\bf t}^*,z)=\frac{V^*_\infty({\bf t},z)\tau}{\tau} ,\\
w_0(M,{\bf t},{\bf t}^*,z)=\frac{V_0({\bf
t}^*,z)\tau(M+1)}{\tau(M)},\quad
 w^*_0(M,{\bf t},{\bf t}^*,z)=\frac{V^*_0({\bf t}^*
,z)\tau(M-1)}{\tau(M)} ,
\end{eqnarray}
where
\begin{equation}\label{taufunction}
\tau(M)=\tau(M,{\bf t},{\bf t}^*)= \langle M|e^{H({\bf
t})}ge^{H^*({\bf t}^*)}|M\rangle .
\end{equation}
Both KP and TL hierarchies are described by the bilinear identity:
\begin{equation}\label{bi}
\oint w_\infty (M,{\bf t},{\bf t}^*,z) w^*_\infty (M',{\bf
t}',{{\bf t}'}^*,z)dz= \oint w_0 (M,{\bf t},{\bf t}^*,z^{-1})
w^*_0 (M',{\bf t}',{{\bf t}'}^*,z^{-1})z^{-2}dz ,
\end{equation}
which holds for any ${\bf t},{\bf t}^*,{\bf t}',{{\bf t}'}^* $
for any integers $M,M'$.\\

The Schur functions $s_{{\lambda}}({\bf t})$ are well-known
examples of tau-functions which correspond to rational solutions
of the KP hierarchy. It is known that not any linear combination
of Schur functions turns to be a KP tau-function, in order to find
these combinations one should solve bilinear difference equation,
see \cite{JM}, which is actually a version of discrete Hirota
equation. Below we shall present KP tau-functions which are
infinite series of Schur polynomials, and which turn to be known
hypergeometric functions (\ref{vh}),(\ref{hZ}).

\subsection{$H_0({\bf T})$, fermions $\psi({\bf T},z),
\psi^*({\bf T},z)$, bosonization rules \cite{nl64}}

Throughout this section we assume that $r,{\tilde r}\neq 0$. \\
Put
\begin{equation}\label{rT}
r(n)=e^{T_{n-1}-T_n},\quad {\tilde r}(n)=e^{{\tilde
T}_{n-1}-{\tilde T}_n}
\end{equation}
New variables $T_n,n \in Z$ (and also ${\tilde T}_n$) are defined
up to a constant independent of $n$. \\
We define a generator $H_0({\bf T})\in \widehat{gl}(\infty)$ :
\begin{equation}\label{H0+}
H_0({\bf T}):= \sum_{n=-\infty}^{\infty} T_n :\psi^*_n \psi_n: ,
\end{equation}
which produces the following transformation:
\begin{equation}\label{prop41}
e^{\mp H_0({\bf T})}\psi_n e^{\pm H_0({\bf T})}=e^{\pm T_n}\psi_n,
\quad e^{\mp H_0({\bf T})}\psi^*_n e^{\pm H_0({\bf T})}=e^{\mp
T_n}\psi^*_n
\end{equation}
This is the transformation which sends $A_m,{\tilde A}_m$ of
(\ref{A_k}), (\ref{tA_k}) to $H_{-m},H_m$ respectively. Therefore
\begin{equation}\label{prop43}
e^{H_0({\tilde {\bf T}})}\tilde{A}({\bf t}) e^{-H_0({\tilde {\bf
T}})}= H({\bf t}),\quad e^{-H_0({\bf T})}A({\bf t}^*) e^{H_0({\bf
T})}=-H^*({\bf t}^*) .
\end{equation}

 It is convenient to consider the following fermionic operators:
\begin{eqnarray}\label{kruchferm}
\psi({\bf T},z)=e^{H_0({\bf T})}\psi(z)e^{-H_0({\bf T})}
=\sum_{n=-\infty}^{n=+\infty} e^{-T_n}z^n\psi_n ,\\
\label{kruchferm*}\psi^*({\bf T},z)=e^{H_0({\bf
T})}\sum_{n=-\infty}^{n=+\infty} z^{-n-1}\psi^*_n e^{-H_0({\bf
T})}= \sum_{n=-\infty}^{n=+\infty} e^{T_n}z^{-n-1}\psi^*_n .
\end{eqnarray}

Now we consider Hirota-Miwa change of variables:  ${\bf t}=\pm
{\bf t}({\bf x}^N)$ and ${{\bf t^*}=\pm \bf t^*}({\bf y}^N)$.

Considering (\ref{bos1}),(\ref{bos2}) and using
(\ref{prop43}),(\ref{kruchferm}),(\ref{kruchferm*})
(\ref{bos1}),(\ref{bos2}) we get the following bosonization rules
\begin{eqnarray}\label{mm*}
e^{-A({\bf t^*}({\bf y}^N))}|M\r= \frac{\psi({\bf
T},y_1)\cdots\psi({\bf T},y_N)|M-N\r } {\Delta^{+}(M, N,{\bf
T},{\bf y}^N)} ,\\ \label{mm*l-} e^{-A(-{\bf t^*}({\bf
y}^N))}|M\r= \frac{\psi^*({\bf T},y_1)\cdots\psi^*({\bf
T},y_N)|M+N\r} {\Delta^{-}(M, N,{\bf T},{\bf y}^N)} ,\\
\label{mm*r+} \l M|e^{{\tilde{A}}({\bf t}({\bf x}^N))} = \frac{\l
M-N|\psi^*(-{\bf \tilde{T}},\frac{1}{x_N})\cdots \psi^*(-{\bf
\tilde{T}},\frac{1}{x_1})} {{\tilde \Delta}^{+}(M, N,{\bf
\tilde{T}},{\bf x}^N)},\\ \label{mm*r-} \l M|e^{\tilde{A}(-{\bf
t}({\bf x}^N))} =\frac{ \l M+N|\psi (-{\bf
\tilde{T}},\frac{1}{x_N})\cdots\psi (-{\bf
\tilde{T}},\frac{1}{x_1})} {{\tilde \Delta}^{-}(M, N,{\bf
\tilde{T}},{\bf x}^N)} .
\end{eqnarray}
 The coefficients are
\begin{eqnarray}\label{vand}
\Delta^\pm (M, N,{\bf T},{\bf y}^N)=
\frac{\prod_{i<j}^N(y_i-y_j)}{(y_1\cdots y_N)^{N\mp M}}\times
\frac{\tau(M,{\bf 0,T,0})}{\tau(M \mp N,{\bf 0,T,0})},\\
 \label{vandtilde+}
{\tilde \Delta}^\pm(M, N,{\bf \tilde{T}},{\bf x}^N)=
\frac{\prod_{i<j}^N (x_i-x_j)}{(x_1\cdots x_N)^{N-1\mp M}}\times
\frac{\tau(M,{\bf 0,\tilde{T},0})}{\tau(M\mp N,{\bf
0,\tilde{T},0})},
\end{eqnarray}
where in case $N=1$ the Vandermond products
$\prod_{i<j}^N(y_i-y_j), \prod_{i<j}^N (x_i-x_j)$ are replaced by
$1$. The origin of the notation $\tau(M,{\bf 0,T,0})$ is explained
latter, see (\ref{H0+}),(\ref{H0-}), and used for
\begin{equation}\label{0T01}
\tau (n,{\bf 0},{\bf T},{\bf 0} ) =e^{T_{n-1}+\cdots
+T_1+T_0}=e^{nT_0}\prod_{k=0}^{n-1}\left(r(k)\right)^{k-n}, \quad
n>0,
\end{equation}
\begin{equation}
\tau (0,{\bf 0},{\bf T},{\bf 0} )=1, \quad n=0 ,
\end{equation}
\begin{equation}\label{0T02}
\tau (n,{\bf 0},{\bf T},{\bf 0} ) = e^{-T_{n}-\cdots
-T_{-2}-T_{-1}}=e^{nT_0}\prod_{k=0}^{-n-1}\left(r(k)\right)^{k+n},
\quad n<0 .
\end{equation}

 Therefore in Hirota-Miwa
variables one can rewrite the vacuum expectation of the exponents
of (\ref{A_m}) and (\ref{tA_k}):
\begin{eqnarray}\label{fermicor}
\l M|e^{{\tilde{A}}({\bf t}({\bf x}^N))}
e^{-A({\bf t^*}({\bf y}^N))}|M\r=\nonumber\\
\frac{ \l M-N|\psi^*(-{\bf \tilde{T}},\frac{1}{x_N})\dots
\psi^*(-{\bf \tilde{T}},\frac{1}{x_1}) \psi({\bf
T},y_1)\cdots\psi({\bf T},y_N)|M-N\r} {{\tilde \Delta}^{+}(M,
N,{\bf \tilde{T}},{\bf x}^N)
 \Delta^{+}(M, N,{\bf T},{\bf y}^N)},
\end{eqnarray}
\begin{eqnarray}\label{fermicor-}
\l M|e^{\tilde{A}(-{\bf t}({\bf x}^N))}
e^{-A(-{\bf t^*}({\bf y}^N))}|M\r=\nonumber\\
\frac{ \l M+N|\psi (-{\bf \tilde{T}},\frac{1}{x_N})\cdots\psi
(-{\bf \tilde{T}},\frac{1}{x_1}) \psi^*({\bf
T},y_1)\cdots\psi^*({\bf T},y_N)|M+N\r} {{\tilde \Delta}^{-}(M,
N,{\bf \tilde{T}},{\bf x}^N) \Delta^{-}(M, N,{\bf T},{\bf y}^N)}.
\end{eqnarray}
This representation is suitable for an application of Wick rule
(\ref{Wick}) , see (\ref{det2}) below.

\subsection{Toda lattice tau-function
$\tau(n,{\bf t},{\bf T},{\bf t}^*)$ \cite{nl64}}

Here we consider {\em Toda lattice} (TL)
\cite{M},\cite{DJKM},\cite{UT}. Our notations $M,{\bf t},{\bf
t}^*$ correspond to the notations $s,x,-y$ respectively in
\cite{UT}.

Let us consider a special type TL tau-function (\ref{taucor}),
which depends on the three sets of variables ${\bf t},{\bf T},{\bf
t}^*$ and on $M \in Z$:
\begin{equation}\label{tau3}
\tau(M,{\bf t},{\bf T},{\bf t}^*)= \langle M|e^{H({\bf t})}
\exp\left(\sum_{-\infty}^{\infty}T_n:\psi_n^*\psi_n:\right)
e^{H^*({\bf t}^*)}|M\rangle ,
\end{equation}
where $:\psi_n^*\psi_n:=\psi^*_n\psi_n -\l 0| \psi_n^*\psi_n |0\r
$. Since the operator $\sum_{-\infty}^{\infty}:\psi_n^*\psi_n:$
commutes with all elements of the $\widehat{gl} (\infty)$ algebra,
one can put $T_{-1}=0$ in (\ref{tau3}). With respect to the KP and
the TL dynamics the variables $T_n$ have a meaning of integrals of
motion.

As we shall see the hypergeometric functions
(\ref{ordinary}),(\ref{qordinary}),(\ref{vh}),(\ref{hZ1}) listed
in the Appendix are ratios of tau-functions (\ref{tau3}) evaluated
at special values of times $M,{\bf t},{\bf T},{\bf t^*}$. It is
true only in the case when all parameters $a_k$ of the
hypergeometric functions are non integers. For the case when at
least one of the indices $a_k$ is an integer, we will need a
tau-function of an open Toda chain which will be considered in the
Appendix.

Tau-function (\ref{tau3}) is linear in each $e^{T_n}$. It is
described by the  Proposition

\bprop
\begin{equation}\label{tauhyp'}
\frac{\tau (M,{\bf t},{\bf T},{\bf t}^* )} {\tau (M,{\bf 0},{\bf
T},{\bf 0} )} =1+ \sum_{{\lambda}\neq {\bf 0}}
e^{(T_{M-1}-T_{n_1+M-1})+(T_{M-2}-T_{n_2+M-2})+\cdots +
(T_{M-l}-T_{n_l+M-l}) }s_{\lambda }({\bf t}) s_{\lambda }({\bf  t
}^*) .
\end{equation}
The sum is going over all different partitions
\begin{equation}\label{part}
{\lambda}=(n_1,n_2,\dots,n_l),\quad l=1,2,3,... ,
\end{equation}
excluding the partition ${\bf 0}$.\\
\eprop Let $r \neq 0$,$\tilde{r} \neq 0$. Then we put
\begin{equation}\label{rT}
 r(n)=e^{T_{n-1}-T_{n}} ,\quad \tilde{r}(n)=
e^{\tilde{T}_{n-1}-\tilde{T}_{n}}.
\end{equation}
to show the equivalence of (\ref{tauhyp}) and (\ref{tauhyp'}) in
this case. We have
\begin{equation}\label{H0+}
\tau (n,{\bf 0},{\bf T},{\bf 0} ) =e^{-T_{n-1}-\cdots
-T_1-T_0}=e^{-nT_0}\prod_{k=0}^{n-1}\left(r(k)\right)^{n-k}, \quad
n
> 0,
\end{equation}
\begin{equation}
\tau (0,{\bf 0},{\bf T},{\bf 0} )=1, \quad n=0 ,
\end{equation}
\begin{equation}\label{H0-}
\tau (n,{\bf 0},{\bf T},{\bf 0} ) = e^{T_{n}+\cdots
+T_{-2}+T_{-1}}=e^{-nT_0}\prod_{k=0}^{-n-1}\left(r(k)\right)^{-k-n},
\quad n<0 .
\end{equation}

\bprop Let $\tau(n,{\bf t},{\bf T},{\bf t}^*)$ is the Toda lattice
tau function (\ref{tau3}).  Then
\begin{equation}\label{tauTLAAtauKP}
\frac{\tau(n,{\bf t},\tilde{{\bf T}}+{\bf T},{\bf t}^*)}
{\tau(n,{{\bf 0}},\tilde{{\bf T}}+{\bf T},{\bf 0})}= \l n |
e^{\tilde{A} ({\bf t})} e^{-A({\bf t}^*)} | n \r = \tau
_{\tilde{r}r}(n,{\bf t},{\bf t}^*).
\end{equation}
where
\begin{equation}\label{AtA}
A ({\bf t}^*)=\sum_{k=1}^\infty A_kt^*_k,\quad {\tilde A}({\bf
t})=\sum_{k=1}^\infty {\tilde A}_kt_k,
\end{equation}
$A_k,{\tilde A}_k$ are defined by
(\ref{tdopsim}),(\ref{tdopsim'}). The functions $\tilde{r},r$ are
related to $\tilde{\bf T},{\bf T}$ by (\ref{rT}), the product
$r\tilde{r}$ has no zeroes at integer values of it's argument.

 This proposition
follows from formulas(\ref{prop41})-(\ref{prop43}). \eprop

We shall consider $\tilde{r}=1$.

The notation $\tau_r(M,{\bf t},{\bf t}^*)$  will be used only for
the KP tau-function (\ref{tauhyp1}). Notation $\tau(M,{\bf t},{\bf
T},{\bf t}^*)$ is used for Toda lattice tau-function.  These
tau-functions are related via (\ref{teplitsGauss}). KP
tau-function $\tau(n):=\tau_r(n,{\bf t},{\bf t}^*)$ obeys the
following Hirota equation:
\begin{equation}\label{hirota}
\tau(n)\partial_{t^*_1}
\partial_{t_1}\tau(n)-
\partial_{t_1}\tau(n)
\partial_{t^*_1}\tau(n)=r(n)\tau(n-1)\tau(n+1).
\end{equation}

The equation
\begin{equation}\label{rToda}
\partial_{t_1}\partial_{t^*_1}\phi_n=
r(n)e^{\phi_{n-1}-\phi_{n}}- r(n+1)e^{\phi_{n}-\phi_{n+1}}
\end{equation}
which is similar to the Toda lattice equation holds for
\begin{equation}\label{phi_n}
\phi_n({\bf t},{\bf t}^*)=-\log \frac{\tau_r(n+1,{\bf t},{\bf
t}^*)}{\tau_r(n,{\bf t},{\bf t}^*)}
\end{equation}

Equations (\ref{rToda}) and (\ref{hirota}) are still true in case
$r$ has zeroes.\\
If the function $r$ has no integer zeroes, using the change of
variables
\begin{equation}\label{phivarphi}
 \varphi_n=-\phi_n - T_n ,
\end{equation}
we obtain Toda lattice equation in the standard form \cite{UT}:
\begin{equation}\label{Toda}
\partial_{t_1}\partial_{t^*_1}\varphi_n=e^{\varphi_{n+1}-\varphi_{n}}-
e^{\varphi_{n}-\varphi_{n-1}}.
\end{equation}
As we see from (\ref{phivarphi}) the variables $T_n$ might have
the meaning of asymptotic values of the fields $\phi_n$ for the
class of tau-functions (\ref{tau3}) which is characterized by the
property $\varphi_n \to 0$ as $t_1 \to 0$.

\subsection{Expressions and linear equations for
tau-functions I (Hirota-Miwa variables)}

It is well-known fact that tau-functions solves bilinear
equations. The class of tau-functions under our consideration
solves also linear equations.

The relations of this subsection correspond to the case when at
least one set of variables (either ${\bf t}$ or ${\bf t}^*$ or
both ones) is substituted with Hirota-Miwa variables.

 In cases ${\bf t}={\bf
t}({\bf x}^N)$ (using (\ref{mm*r+}) and (\ref{psi*xir})) and ${\bf
t}=-{\bf t}({\bf x}^N)$ (using (\ref{mm*r-}) and (\ref{psixir}))
one gets the following representations
\begin{equation}\label{tau+xi}
 \tau_r(M,{\bf
t}({\bf x}^N),{\bf t}^*)= \Delta^{-1}\left(e^{\xi_{r'}({\bf
t}^*,x_1)}\cdots e^{\xi_{r'}({\bf
t}^*,x_N)}\cdot\Delta\right),\quad \Delta= \frac{\prod_{i<j}
(x_i-x_j)}{(x_1\cdots x_N)^{N-1-M}}
\end{equation}
\begin{equation}\label{tau-xi}
\tau_r(M,-{\bf t}({\bf x}^N),{\bf
t}^*)= \Delta^{-1}\left(e^{-\xi_r({\bf t}^*,x_1)}\cdots
e^{-\xi_r({\bf t}^*,x_N)}\cdot\Delta \right),\quad \Delta=
\frac{\prod_{i<j} (x_i-x_j)}{(x_1\cdots x_N)^{N-1+M}}
\end{equation}
 where the exponents are formal series
$e^\xi=1+\xi+\cdots$, and $r(D_i)\prod_k x_k^{a_k} =r(a_i) \prod_k
x_k^{a_k} $. Let us write down that according to
(\ref{xir}),(\ref{r'})
\begin{equation}\label{xirx}
\xi_{r'}({\bf
t}^*,x_i)=\sum_{m=1}^{+\infty}t_m^*\left(x_ir(D_i)\right)^m, \quad
\xi_{r}({\bf
t}^*,x)=\sum_{m=1}^{+\infty}t_m^*\left(x_ir(-D_i)\right)^m, \quad
D_i=x_i\frac{\partial}{\partial x_i}
\end{equation}
Let us note
that $[\xi_{r'}({\bf t}^*,x_n),\xi_{r'}({\bf t}^*,x_m)]=0=
[\xi_{r}({\bf t}^*,x_n),\xi_{r}({\bf t}^*,x_m)]$ for all $n,m$.

(\ref{tau-xi}) may be also obtained from (\ref{tau+xi})
with the help of (\ref{tausim}).

From (\ref{tau+xi}) it follows for $m=1,2,\dots $
that
\begin{equation}
\left(\frac{\partial}{\partial
t^*_m}-\sum_{i=1}^N (x_ir(D_{x_i}))^m \right) \left(\Delta
\tau_r(M,{\bf t}({\bf x}^N),{\bf t}^*)\right)=0, \quad \Delta=
\frac{\prod_{i<j} (x_i-x_j)}{(x_1\cdots x_N)^{N-1-M}}.
\end{equation}

In case $ {\bf t}={\bf t}({\bf x}^N), {\bf t}^*={\bf t}^*({\bf
y}^{N^*})$ we get
\begin{eqnarray}\label{repr}
\tau_r(M, {\bf t}({\bf x}^N), {\bf t}^*({\bf y}^{N^*}))
=\sum_{{\lambda}\in P}r_{ \bf n }(M) s_{\lambda }({\bf
x}^N)s_{\lambda}({\bf y}^{N^*})\\ \label{reprx}
=\frac{1}{\Delta(x)} \prod_{i=1}^N\prod_{j=1}^{N^*}
(1-y_jx_ir(D_{x_i}))^{-1}\cdot \Delta(x)\\ \label{repry}
=\frac{1}{\Delta(y)} \prod_{i=1}^{N}\prod_{j=1}^{N^*}
(1-x_iy_jr(D_{y_j}))^{-1}\cdot \Delta(y)
\end{eqnarray}
where
\begin{equation}\label{reprNN*}
\Delta(x)=
\frac{\prod_{i<j}^N (x_i-x_j)}{(x_1\cdots x_N)^{N-1-M}},\quad
\Delta(y)=
\frac{\prod_{i<j}^{N^*} (y_i-y_j)}{(y_1\cdots y_N)^{N^*-1-M}}
\end{equation}

Looking at (\ref{reprx}) ,(\ref{repry}) one derives the following
system of linear equations
\begin{equation}\label{linuryx}
\left(D_{y_j}-\sum_{i=1}^N\frac{1}{1-y_jx_ir(D_{x_i})}
+N\right)\left( \Delta(x) \tau_r(M, {\bf t}({\bf x}^N), {\bf
t}^*({\bf y}^{N^*}))\right)=0,\quad j=1,\dots,N^*
\end{equation}
\begin{equation}\label{linurxy}
\left(D_{x_i}-\sum_{j=1}^{N^*}\frac{1}{1-x_iy_jr(D_{y_j})} +N^*
\right)\left( \Delta(y) \tau_r(M, {\bf t}({\bf x}^N), {\bf
t}^*({\bf y}^{N^*}))\right)=0,\quad i=1,\dots,N
\end{equation}

In some examples below we shall take specify  ${\bf t}^*$ as
follows.

(1) Let us take
\begin{equation}\label{t+xNt*1t*infty}
mt_m=\sum_{k=1}^N x_k^m,\quad {\bf t}^*=(1,0,0,\dots)
\end{equation}
Then one consider the equation (\ref{linur+}) with
$k=1$ (taking $t_1^*=1$):
\begin{equation}\label{reprt1*=1}
\tau_r(M, {\bf t}({\bf x}^N), {\bf t}^*)= \frac{1}{\Delta}
e^{x_1r(D_1)+\cdots+x_Nr(D_N)}\cdot \Delta,
\end{equation}
where
\begin{equation}
\Delta = \frac{\prod_{i<j}
(x_i-x_j)}{(x_1\cdots x_N)^{N-M-1}}
\end{equation}
and $D_i=x_i\partial_{x_i}$. We get
\begin{equation}\label{linurt1*=1}
\left(\sum_{i=1}^N D_i -\frac{1}{{\Delta}}\left(\sum_{i=1}^N x_i
r(D_i)\right) {\Delta} \right) \tau_r(M, {\bf t}({\bf x}^N), {\bf
t}^*)=0
\end{equation}

(2) Let us take
\begin{equation}\label{t+xNzt*inftyq}
mt_m=\sum_{k=1}^N x_k^m,\quad mt^*_m= \frac{z^m}{1-q^m},\quad
m=1,2,\dots
\end{equation}
 Let us note that the tau-function $\tau_r({\bf
t},{\bf t}^*)$ will not change if instead of (\ref{t+xNzt*inftyq})
one take
\begin{equation}\label{zt+xNt*inftyq}
 mt_m=\sum_{k=1}^N
(zx_k)^m,\quad mt^*_m= \frac{1}{1-q^m},\quad m=1,2,\dots
\end{equation}
This gives the representation as follows
\begin{equation}\label{reprt*q}
\tau_r(M, {\bf t}({\bf x}^N), {\bf t}^*)=
\frac{1}{\Delta(x)}
\prod_{i=1}^N\frac{1}{\left(x_ir(D_{x_i});q\right)_\infty}\cdot
\Delta(x)
\end{equation}
We see that
\begin{equation}
\left(q^{\frac{N^2-N}{2}-NM}q^{D_1+\cdots
D_N}- \prod_{i=1}^N(1-x_ir(D_i))\right) \left(\Delta(x)\tau_r(M,
{\bf t}({\bf x}^N), {\bf t}^*)\right)=0
\end{equation}

We shall write down more linear equations, which follow from the
explicit fermionic representation of the tau-function
(\ref{tauhyp1}) via the bosonization formulae (\ref{fermicor})
 and (\ref{fermicor-}) These equations may be also viewed as the
constraint which result in the string equations (\ref{string1})
and (\ref{string2}).

For the variables ${\bf x}^N$ in case ${\bf t}={\bf t}({\bf
x}^N)$, one can use the relation $\l M|A_m=0$, and makes profit of
the relation $A_m=e^{H_0}H_{-m}e^{-H_0}$ inside the fermionic
vacuum expectation value (\ref{fermicor}). This way we get the
partial differential equations for the tau-function
(\ref{tauhyp}):
\begin{equation}\label{linur+}
\frac{\partial \tau_r(M, {\bf t}({\bf x}^N),
{\bf t}^*)} {\partial t^*_m}= \frac{1}{\Delta }\left(\sum_{i=1}^N
(x_ir(D_{x_i}))^m\right) \Delta \tau_r(M, {\bf t}({\bf x}^N), {\bf
t}^*),\quad m=1,2,3,\dots ,
\end{equation}
where $\Delta$ is
proportional to ${\tilde{\Delta}}^+$ of (\ref{vandtilde+}):
\begin{equation}\label{tildeDelta+}
\Delta= \frac{\prod_{i<j}
(x_i-x_j)}{(x_1\cdots x_N)^{N-1-M}}.
\end{equation}

In variables ${\bf y}^{(\infty)}$, ${\bf t^*}= -{\bf t^*}({\bf
y}^{(\infty)})$, we can rewrite (\ref{linur+}):
\begin{eqnarray}
(-1)^{k}\sum_{i=1}^{+\infty}\frac{e_{k-1}\left(\frac{1}{y_1},\dots,
\frac{1}{y_{i-1}},\frac{1}{y_{i+1}},\dots\right)} {\prod_{j\neq
i}(1-\frac{y_i}{y_j})}\frac{\partial \tau_r(M,-{\bf t}({\bf x}^N),
-{\bf t^*}({\bf y}^{(\infty)}))}
{\partial y_i}=\nonumber\\
\frac{1}{{\tilde \Delta}}\left(\sum_{i=1}^N
(x_ir(-D_{x_i}))^k\right) {\tilde \Delta} \tau_r(M, -{\bf t}({\bf
x}^N), -{\bf t^*}({\bf y}^{(\infty)})),
\end{eqnarray}
where $e_k({\bf y})$ is a symmetric function defined through the
relation
$\prod_{i=1}^{+\infty}(1+ty_i)=\sum_{k=0}^{+\infty}t^ke_k({\bf
y})$.

Also we have
\begin{eqnarray}\label{linur2}
 \left(\sum_{k=1}^{M+N-1} k -\sum_{i=1}^N D_{x_i}\right)
{\tilde \Delta}^{-}\tau(M, -{\bf t}({\bf x}^N),{\bf T},
-{\bf t}^{*}({\bf y}^{(N')}))\Delta^{-}=\nonumber\\
 \left(\sum_{k=1}^{M+N'-1} k -\sum_{i=1}^{N'}\left(\frac{1}{y_i}D_{y_i}y_i
\right)\right) {\tilde \Delta}^{-}\tau(M, -{\bf t}({\bf x}^N),{\bf
T}, -{\bf t}^{*}({\bf y}^{(N')}))\Delta^{-},
\end{eqnarray}
where ${\tilde \Delta}^{-}={\tilde \Delta}^{-}(M, N,{\bf 0},{\bf
x}^N)$ and $\Delta^{-}=\Delta^{-}(M, N,{\bf 0},{\bf y}^{(N')})$.
This formula is obtained by the insertion of the fermionic
operator $ res_z:\psi^*(z)z\frac{d}{dz}\psi(z):$ inside the
fermionic vacuum expectation.

These formulae can be also written in terms of higher KP and TL
times, with the help of vertex operator action, see the {\em
Subsection ``Vertex operator action''}. Then the relations
(\ref{linur+}) are the infinitesimal version of
(\ref{stringvert1}), while the relation (\ref{linur2}) is the
infinitesimal version of (\ref{stringvert2}).

In some examples below we shall take specify  ${\bf t}^*$ as
follows. (1) Let us take
\begin{equation}\label{t+xNt*1t*infty}
mt_m=\sum_{k=1}^N x_k^m,\quad {\bf t}^*=(t_1^*,0,0,\dots)
\end{equation}
One notes that that the tau-function $\tau_r({\bf
t},{\bf t}^*)$ will not change if instead of (\ref{t+xNzt*inftyq})
one take
\begin{equation}\label{t*1t+xNt*infty}
 mt_m=\sum_{k=1}^N
(t_1x_k)^m,\quad {\bf t}^*=(1,0,0,\dots)
\end{equation}
Then one
consider the equation (\ref{linur+}) with $k=1$, taking $t_1=1$.
We put $D_i=x_i\partial_{x_i}$ and get
\begin{equation}\label{linurtauzon}
\left(\sum_{i=1}^N D_i
-\frac{1}{{\Delta}}\left(\sum_{i=1}^N x_i r(D_i)\right) {\Delta}
\right)\tau_r({\bf t}({\bf x}^N),{\bf t}^*)=0
\end{equation}
where
$r$ is given by (\ref{op1}), and
\begin{equation}
\Delta =
\frac{\prod_{i<j} (x_i-x_j)}{(x_1\cdots x_N)^{N-M-1}}
\end{equation}

(2) Let us take
\begin{equation}\label{t+xNzt*inftyq}
mt_m=\sum_{k=1}^N x_k^m,\quad mt^*_m= \frac{z^m}{1-q^m},\quad
m=1,2,\dots
\end{equation}
Let us note that the tau-function $\tau_r({\bf
t},{\bf t}^*)$ will not change if instead of (\ref{t+xNzt*inftyq})
one take
\begin{equation}\label{zt+xNt*inftyq}
 mt_m=\sum_{k=1}^N
(zx_k)^m,\quad mt^*_m= \frac{1}{1-q^m},\quad m=1,2,\dots
\end{equation}
Then we apply the operator $(1-q^D_z),D_z=z\frac {d}{dz}$ to
$e^{-A({\bf t}^*)}|0\r $ ( ${\bf t}^* $ of (\ref{t+xNzt*inftyq} )
) and get $e^{ -A({\bf t}^*)}
(1-\sum_{n=0}^{+\infty}\psi^*_{-n}\psi_0)|0\r=e^{ -A({\bf t}^*)}
(\sum_{n=1}^{+\infty}\psi^*_{-n}\psi_0)|0\r=e^{-A({\bf t}^*)}
(A_1+\dots|0\r$, where terms denoted by dots vanish inside the
vacuum expectation $\l 0| e^{H({\bf t}({\bf x}^N)}e^{-A({\bf
t}^*)}A_1|0\r$. Then one consider the equation (\ref{linur+}) with
$k=1$, taking $z=1$. We put $D_i=x_i\partial_{x_i}$ and because
get
\begin{equation}\label{linurtauzonq}
\left(\sum_{i=1}^N (1-q^{D_1+\cdots +D_N})
-\frac{1}{{\Delta}}\left(\prod_{i=1}^N (1-x_i r(D_i))\right)
{\Delta} \right)\tau_r({\bf t}({\bf x}^N),{\bf t}^*)=0
\end{equation}
where $r$ is given by (\ref{op1}),
and
\begin{equation}
\Delta = \frac{\prod_{i<j}
(x_i-x_j)}{(x_1\cdots x_N)^{N-M-1}}
\end{equation}
The simplest
way to derive linear equation (\ref{linur+}),
(\ref{linurtauzon}),(\ref{linurtauzonq}) is to use the
representations (\ref{tau+xi}),(\ref{tau-xi}).

\subsection{The vertex operator action.
Linear equations for  tau-functions II } Now we present relations
between hypergeometric functions which follow from the soliton
theory, for instance see \cite{ASM},\cite{D'}. (In \cite{nl64}
there were few misprints in the formulae presented below). Let us
introduce the operators which act on functions of ${\bf t}$
variables:
\begin{equation}\label{Omega}
\Omega_r^{(\infty)}({\bf t},{\bf t}^*):=- \frac{1}{2\pi \sqrt{-1}}
\lim _{\epsilon \to 0} \oint V^*_\infty ({\bf t},z+\epsilon)
\xi_r({\bf t}^* ,z^{-1})V_\infty ({\bf t},z) \frac{dz}{z} ,
\end{equation}
\begin{equation}
\Omega_r^{(0)}({\bf t}^*,{\bf t}):= \frac{1}{2\pi \sqrt{-1}} \lim
_{\epsilon \to 0} \oint  V_0^* ({\bf t}^*,z+\epsilon)
\xi_r^{(0)}({\bf t},z)V_0 ({\bf t}^*,z) \frac{dz}{z} ,
\end{equation}
where $V_\infty({\bf t},z),V^*_\infty({\bf t},z), V_0({\bf
t}^*,z),V_0^*({\bf t}^*,z)$ are defined by (\ref{vertex}), and
\begin{equation}
 \xi_r({\bf t}^* ,z^{-1})=\sum_{m=1}^{+\infty}t^*_m
\left(\frac 1z r(D)\right)^m,\quad \xi_r^{(0)}({\bf
t},z)=\sum_{m=1}^{+\infty}t_m \left(r(D)z\right)^m
\end{equation}
For instance
\begin{equation}
 r=1:\qquad \Omega_r^{(\infty)}({\bf t},{\bf t}^*)=
 \Omega_r^{(0)}({\bf t}^*,{\bf t})
=\sum_{n>0} nt_nt^*_n .
\end{equation}

Also we consider
\begin{equation}
 Z_{nn}({\bf t})=-
\frac{1}{4\pi^2} \oint \frac {z^n}{{z^*}^n}V^*_\infty ({\bf
t},z^*)V_\infty ({\bf t},z) \frac{dzdz^*}{zz^*} ,\quad
Z_{nn}^*({\bf t}^*) =-\frac{1}{4\pi^2} \oint \frac
{z^n}{{z^*}^n}V^*_0 ({\bf t}^*,z^*)V_0 ({\bf t}^*,z)
\frac{dzdz^*}{zz^*} .
\end{equation}

The bosonization formulae (\ref{bos1}),(\ref{bos2})  result in

\bprop We have shift argument formulae for the tau function
(\ref{tauhyp}):
\begin{equation}\label{stringvert1}
e^{\Omega_r^{(\infty)}({\bf t},\gamma)}\cdot
\tau_r(M,{\bf t},{\bf
t}^*)= \tau_r(M,{\bf t},{\bf t}^* +\g),\quad
e^{\Omega_r^{(0)}({\bf t}^*,\gamma)}\cdot
\tau_r(M,{\bf t},{\bf
t}^*)= \tau_r(M,{\bf t}+\g,{\bf t}^*) .
\end{equation}
Also we have
\begin{equation}\label{stringvert2}
e^{\sum_{-\infty}^{\infty}\g_n Z_{nn}({\bf t})}
\cdot \tau (M,{\bf
t},{\bf T},{\bf t}^*)= e^{\sum_{-\infty}^{\infty}
\g_nZ_{nn}^*({\bf
t}^*)}\cdot \tau (M,{\bf t},{\bf T},{\bf t}^*)=
 \tau(M,{\bf t},{\bf T}+\g,{\bf t}^*),
\end{equation}
 For instance
\begin{equation}
 e^{\Omega_r^{(\infty)}({\bf t},{\bf
t}^*)}\cdot 1=e^{\Omega_r^{(0)}({\bf t}^*,{\bf t})}\cdot
1=\tau_r(M,{\bf t},{\bf t}^*)
\end{equation}
Also
\begin{equation}
e^{\sum_{-\infty}^{\infty}T_n Z_{nn}} \cdot \exp
\left(\sum_{n=1}^\infty nt_nt^*_n\right)=
e^{\sum_{-\infty}^{\infty}T_nZ_{nn}^*}\cdot \exp
\left(\sum_{n=1}^\infty nt_nt^*_n\right)=
\tau (M,{\bf t},{\bf
T},{\bf t}^*).
\end{equation}
\eprop

\subsection{Determinant formulae I (Hirota-Miwa variables)
 \cite{nl64}}

In the case ${\bf t}={\bf t}({\bf x}^N)$ one can apply Wick
theorem \cite{JM} to obtain a determinant formulae. \bprop{A
generalization of Milne's determinant formula}
\begin{equation}\label{det1}
\tau_r(M,{\bf t}({\bf x}^N),{{\bf t}^*})=\frac{
\det\left(x_i^{N-k}\tau_r(M-k+1,{\bf t}(x_i),{{\bf t}^*})
\right)_{i,k=1}^{N}}{ \det\left(x_i^{N-k}\right)_{i,k=1}^N} .
\end{equation}
{\bf Proof}\\
\begin{eqnarray}
\tau_r(M,{\bf t}({\bf x}^N),{{\bf t}^*})=
\l M|e^{H({\bf t}({\bf x}^N)}e^{-A({{\bf t}^*})}|M\r=\\
\label{miwatau}
\frac{x_1^{N-M-1}\cdots x_N^{N-M-1}}{\prod_{i<j} (x_i-x_j)}\l M|\psi_{M-1}
\cdots \psi_{M-N}\psi^*(\frac{1}{x_N})\cdots\psi^*(\frac{1}{x_1})
e^{-A({{\bf t}^*})}|M\r=\\
\frac{(x_1\cdots x_N)^{N-M-1}}{\prod_{i<j} (x_i-x_j)} \det\left(\l
M|\psi_{M-k}\psi^*(\frac{1}{x_i})e^{-A({{\bf
t}^*})}|M\r\right)_{i,k=1}^N=\\ \frac{
\det\left(x_i^{N-k}\tau_r(M-k+1,{\bf t}(x_i),{{\bf t}^*})
\right)_{i,k=1}^{N}}{ \det\left(x_i^{N-k}\right)_{i,k=1}^N}
\end{eqnarray}
Last equality follows from:
\begin{eqnarray}
\l M|\psi_{M-k}\psi^*(\frac{1}{x_i})e^{-A({{\bf t}^*})}|M\r=\\
=\l M|\psi_{M-1}\cdots \psi_{M-k+1}\psi_{M-k}\psi^*(\frac{1}{x_i})
e^{-A({{\bf t}^*})}\psi^*_{M-k+1}\cdots \psi^*_{M-1}|M\r+\\
+\sum_{j=1}^{k-1} a^k_j({\bf t}^*)\l M|\psi_{M-1}\cdots
\psi_{M-k+j}\psi^*(\frac{1}{x_i})
e^{-A({{\bf t}^*})}\psi^*_{M-k+j+1}\cdots \psi^*_{M-1}|M\r=\\
=\l M-k+1|\psi_{M-k}\psi^*(\frac{1}{x_i})e^{-A({{\bf t}^*})}|M-k+1\r+\\
+\sum_{j=1}^{k-1} a^k_j({\bf t}^*)\l M-k+1+j|
\psi_{M-k+j}\psi^*(\frac{1}{x_i})e^{-A({{\bf t}^*})}|M-k+1+j\r=\\
=x_i^{M-k+1}\tau_r(M-k+1,{\bf t}(x_i),{\bf t}^*)+\sum_{j=1}^{k-1}
a^k_j({\bf t}^*) x_i^{M-k+1+j} \tau_r(M-k+1+j, {\bf t}(x_i), {\bf
t}^*)
\end{eqnarray}
Where the functions $a^k_j({\bf t}^*)$ must be derived as the
results of action of operator $e^{-A({\bf t}^*)}$ on the fermions
$\psi_{M-1},\dots , \psi_{M-k}$. Thus we have:
\begin{eqnarray}
x_i^{N-M-1}\l M|\psi_{M-k}\psi^*(\frac{1}{x_i})e^{-A({{\bf t}^*})}|M\r=\\
=x_i^{N-k}\tau_r(M-k+1,{\bf t}(x_i),{\bf t}^*)+\sum_{l=1}^{k-1}
a^k_{k-l}({\bf t}^*) x_i^{N-l} \tau_r(M-l+1,{\bf t}(x_i), {\bf
t}^*)
\end{eqnarray}
\eprop

\quad

 In the case ${\bf t}={\bf t}({\bf x}^N),{\bf t^*}={\bf t^*}({\bf
y}^N)$ we apply to (\ref{fermicor}) the Wick theorem to obtain a
different Proposition \bprop
 For
$r\neq 0$  we take a tau function $\tau_r(M,{\bf t}({\bf
x}^N),{\bf t^*}({\bf y}^N))$ and  apply Wick's theorem. We get the
determinant formula:
\begin{equation}\label{det2}
 \tau_r(M,{\bf
t}({\bf x}^N),{\bf t^*}({\bf y}^N))=
\frac{c_{M}}{c_{M-N}}\frac{\det
\left(\tau_r(M-N+1,x_i,y_j)\right)_{i,j=1}^N} {\Delta({\bf
x}^N)\Delta({\bf y}^N)},
\end{equation}
where
\begin{equation}\label{cn}
c_n=\prod_{k=0}^{n-1}\left( r(k)\right)^{k-n}
\end{equation}
\begin{equation}\label{tau(1)}
\tau_r(M-N+1,x_i,y_j)=1+r(M-N+1)x_iy_j+r(M-N+1)r(M-N+2)x_i^2y_j^2+\cdots
\end{equation}
\eprop

\subsection{Determinant formulae II}

 There is also determinant formulae for tau functions in
 variables ${\bf t},{\bf t^*}$. These one we get in case the
 function $r$ has zeroes. If $r(M_0)=0$ and $n>0$ then
\begin{equation}\label{detzeroesr}
\tau_r(M_0+n,{\bf t},{\bf t^*})=c_n\det \left(
\frac{\partial^{a+b}}{\partial t_1^a\partial
{t_1^*}^b}\tau_r(M_0+1,{\bf t},{\bf t^*})\right)_{a,b=0}^{n-1}
\end{equation}
\begin{equation}\label{cn}
c_n=\prod_{k=0}^{n-1}\left( r(k)\right)^{k-n}
\end{equation}

\subsection{Integral representations I (Hirota-Miwa variables)
 \cite{nl64}}
In the case ${\bf t}={\bf t}({\bf x}^N),{\bf t^*}={\bf t^*}({\bf
y}^N)$ we get an integral representation formulae. For the
fermions (\ref{kruchferm}) we easily get the
relations:
\begin{equation}
\int \psi({\bf T},\alpha z)d\mu(\alpha)=
\psi\left({\bf T}+{\bf T}(\mu),z\right),\quad \int
\psi^*\left(-{\bf T},\frac {1}{\alpha z}\right)d{\tilde
\mu}(\alpha)= \psi^*\left(-{\bf T}-{\bf T}({\tilde \mu}),\frac 1z
\right)
\end{equation}
where $\mu,{\tilde \mu}$ are some
integration measures, and shifts of times $T_n$ are defined in
terms of the moments:
\begin{equation}
\int \alpha^n
d\mu(\alpha)=e^{-T_n(\mu)},\quad \int {\alpha}^n  d{\tilde
\mu}(\alpha)= e^{-T_n({\tilde \mu})}.
\end{equation}
Therefore thanks to the bosonization formulae (\ref{fermicor}) we
have the relations for the tau-function; below ${\bf t}^*$ is
defined via Hirota-Miwa variables.
 \bprop{Integral representation
formula holds}
\begin{eqnarray}\label{intrepr}
\int{\tilde \Delta}_{\bf \tilde{T}}({\tilde \alpha}{\bf x}^N)
\frac{\tau\left(M,{\bf t}(\tilde{\alpha}{\bf x}^N), {\bf
T+\tilde{T}},{\bf t}({\bf \alpha y}_{(N)})\right)}
{\tau\left(M,{\bf 0}, {\bf T+\tilde{T}},{\bf 0}\right)}
\Delta_{\bf T}({\bf \alpha y}_{(N)}) \prod_{i=1}^N d{\tilde
\mu}({\tilde \alpha}_i)\prod_{i=1}^N d\mu(\alpha_i)
\nonumber\\
={\tilde \Delta}_{\bf \tilde{T}+{\bf \tilde{T}(\tilde{\mu})}}({\bf
x}^N) \frac{\tau\left(M,{\bf t}({\bf x}^N),{\bf T}+{\bf
\tilde{T}}+ {{\bf \tilde{T}}}({\tilde \mu})+{\bf T}(\mu),{\bf
t}({\bf y}^N) \right)}{\tau\left(M,{\bf 0},{\bf T}+{\bf
\tilde{T}}+ {{\bf \tilde{T}}}({\tilde \mu})+{\bf T}(\mu),{\bf
0}\right)} \Delta_{{\bf T}+{\bf T}(\mu)}({\bf y}^N).
\end{eqnarray}
where $\Delta_{{\bf T}}({\bf \alpha y}_{(N)})= \Delta^+ (M, N,
{\bf T}, {\bf \alpha y}_{(N)})$, ${\tilde \Delta}_{{\bf
\tilde{T}}}({\tilde \alpha}{\bf x}^N)=
{\tilde \Delta}^+(M, N, {\bf \tilde{T}},{\tilde \alpha}{\bf x}^N)$,\\
${\bf \alpha y}_{(N)} =\left(\alpha_1 y_1,\alpha_2 y_2,\dots,
\alpha_N y_N\right)$ and $\tilde {\bf \alpha}{\bf x}^N
=\left({\tilde \alpha}_1 x_1, {\tilde \alpha}_2 x_2,\dots,{\tilde
\alpha}_N x_N\right)$. In particular
\begin{eqnarray}\label{leftintegral}
\int \frac{\tau\left(M,{\bf t},{\bf T},{\bf t}({\bf \alpha y}_{(N)})\right)}
{\tau\left(M,{\bf 0},{\bf T},{\bf 0}\right)}
\Delta_{{\bf T}}({\bf \alpha y}_{(N)})
\prod_{i=1}^N d\mu(\alpha_i)=\nonumber\\
\frac{\tau\left(M,{\bf t},{\bf T}+{\bf T}(\mu),{\bf t}({\bf
y}^N)\right)} {\tau\left(M,{\bf 0},{\bf T}+{\bf T}(\mu),{\bf
0}\right)} \Delta_{{\bf T}+{\bf T}(\mu)}({\bf y}^N).
\end{eqnarray}
\eprop
Remember that arbitrary linear combination of tau-functions
is not a tau-function.
Formulae (\ref{intrepr}) and also (\ref{leftintegral}) give
the integral representations for the tau-function (\ref{tauhyp1}).
It may help to express a tau-function with the help
of a more simple one.

We shall use the following integrals
\begin{equation}\label{Hankel*}
\frac{1}{2\pi\sqrt{-1}}\int_{C}\psi^*(-{\bf T},\frac{\alpha}{ x})
e^{\alpha}{\alpha}^{-b}
d\alpha = \psi^*(-{\bf T}-{\bf T}^b,\frac {1}{x}),
\end{equation}
\begin{equation}\label{Gamma*}
\int_0^\infty \psi^*(-{\bf T},\frac{1}{\alpha x})
 e^{-\alpha}{\alpha}^{a-1}d\alpha
=\psi^*(-{\bf T}-{\bf T}^a,\frac{1}{x}),
\end{equation}
\begin{equation}\label{Beta*}
\frac{1}{\Gamma(b-a)}
\int_0^1\psi^*(-{\bf T},\frac{1}{\alpha x}) {\alpha}^{a-1}(1-\alpha )^{b-a-1}
d\alpha = \psi^*(-{\bf T}-{\bf T}^c,\frac{1}{x}),
\end{equation}
where $C$ in (\ref{Hankel*}) starts at $-\infty$ on the real axis,
circles the origin in the
counterclockwise direction and returns to the starting point, and
\begin{equation}\label{Tabc}
T_n^b=\ln \Gamma(b+n+1), \quad T_n^a=-\ln \Gamma(a+n+1),\quad
T_n^c=\ln \frac {\Gamma(b+n+1)}{\Gamma(a+n+1)\Gamma(b-a)}.
\end{equation}
Let us remind that q-versions of exponential function, of gamma function
and of beta function are known as follows \cite{KV}
\begin{equation}\label{qexp}
\exp_q(x)=\frac{1}{(x(1-q);q)_\infty},\quad
\Gamma_q(x)=(1-q)^{1-a}\frac{(q;q)_\infty}{(q^x;q)_\infty},\quad
B_q(x,y)=\frac{\Gamma_q(x)\Gamma_q(y)}{\Gamma_q(x+y)}
\end{equation}
Gamma and beta functions have the $q$-integral representations
\begin{equation}\label{qintGammaBeta}
\Gamma_q(x)=-q^{-x}\int_0^\infty
\frac{\alpha ^{x-1}} {\exp_q(\alpha)} d_q\alpha,\quad
B_q(x,y)=\int_0^1 \alpha^{x-1}\frac{(\alpha q;q)_\infty}
{(\alpha q^y;q)_\infty} d_q\alpha
\end{equation}
where $q$-integral are defined as
\begin{equation}
 \int_0^\infty f(x)d_qx=(1-q)\sum_{j=-\infty}^{j=\infty}q^jf(q^j),\quad
\int_0^c f(x)d_qx=c(1-q)\sum_{j=0}^{j=\infty}q^jf(cq^j)
\end{equation}
The following  $q$-integrals will be of use:
\begin{equation}\label{qHankel*}
\frac{q^{-a}}{q-1}
\int_0^\infty \psi^*(-{\bf T},\frac{q}{\alpha (1-q)x})
\frac{\alpha ^{a-1}} {\exp_q(\alpha)} d_q\alpha
=\psi^*\left(-{\bf T}-{\bf T}(a,q),\frac{1}{x}\right),
\end{equation}
\begin{equation}\label{qBeta*}
\frac {1}{\Gamma_q(b-a)}
\int_0^1\psi^*(-{\bf T},\frac{1}{\alpha x})
 {\alpha}^{a-1}\frac{(\alpha q;q)_\infty }
{(\alpha q^{b-a};q)_\infty } d_q\alpha = \psi^*(-{\bf T}-{\bf
T}(a,b,q),\frac{1}{x}).
\end{equation}
where, as one can check
using (\ref{qintGammaBeta}),
\begin{equation}
T_n(a,q)=\ln
\frac{1}{(1-q)^{n}\Gamma_q(a+n+1)},\quad T_n(a,b,q)= \ln  \frac
{\Gamma_q(b+n+1)}{\Gamma_q(a+n+1)}.
\end{equation}
In the same way one can consider Hirota-Miwa variables
(\ref{HMxy}). In the {\em Examples} below we shall present
hypergeometric functions listed in the {\em Subsections 1.2 and
1.3} as tau-functions of the type (\ref{fermicor}). Then we are
able to write down integration formulae, namely
(\ref{leftintegral}), which express ${}_{p+1}\Phi_{s}$ and
${}_{p+1}\Phi_{s+1}$ in terms of ${}_p\Phi_s$ with the help of
 (\ref{qHankel*}), (\ref{qBeta*}) and (\ref{intrepr}).
In \cite{Milne} different integral representation formula was presented,
which was based on the $q$-analog of Selberg's integral of Askey and
Kadell.
By taking the limit $q \to 1$ one can
consider functions  ${}_{p}F_s$.
Using (\ref{Tabc}), one can express
${}_{p+1}{\mathcal{F}}_{s}$, ${}_{p+1}{\mathcal{F}}_{s+1}$ and
${}_{p}{\mathcal{F}}_{s+1}$
 as integrals of
${}_p{\mathcal{F}}_s$ with the help of (\ref{Gamma*}), (\ref{Beta*}) and
(\ref{Hankel*}) respectively.

\subsection{Integral representations II}

Now let us consider the case when $\tau_r$ depends on ${\bf
t},{\bf t^*}$. Using (\ref{tauscal}) and (\ref{fgscalint}) one
gets the integral representation for $\tau_r$ in case the function
$r$ has zero. Let us again choose $r(0)=0$. The  have the
representation described above

\bprop

\begin{equation}\label{intSchur}
\tau_r(M,{\bf t},{\bf t^*}) =\int \cdots \int
\Delta(z)\Delta(z^*)\prod_{k=1}^M e^{\sum_{n=1}^\infty
\left(z^n_kt_n+{z^*_k}^nt^*_n\right)}\mu_r(z_k{z^*}_k)dz_kdz^*_k
\end{equation}
where
\begin{equation}\label{vandzz^*}
\Delta(z)=\prod_{i<j}^M(z_i-z_j),\quad
\Delta(z^*)=\prod_{i<j}^M(z_i^*-z_j^*)
\end{equation}
and $\mu_r$ is defined by (\ref{tau1/r} ),(\ref{normirovka} ).

\eprop

 Now it is interesting to notice that if the series
$\tau_{1/r}$ (which defines $\mu_r$, see (\ref{tau1/r})) is a
convergent series then the series $\tau_r$ may be divergent one.
In this case one can consider the integral $I_r(M,{\bf t},{\bf
t^*})$ is an analog of Borel sum for the divergent series
$\tau_r$, see \cite{OH}.

\subsection{Examples of hypergeometric series \cite{nl64}}
The main point of this subsection is the observation that if
$r(D)$ is a rational function of $D$ then $\tau_r$ is a
hypergeometric series. If  $r(D)$ is a rational function of $q^D$
we obtain $q$-deformed hypergeometric series.
Now let us consider various $r(D)$.\\

{\bf Example 1} Let all parameters $b_k$ be non integers.
\begin{equation}
{}_pr_s(D)=\frac{(D+a_1)(D+a_2)\cdots
(D+a_p)}{(D+b_1)(D+b_2)\cdots (D+b_s)} . \label{op1}
\end{equation}

 For the vacuum expectation value (\ref{tauhyp}) we have:
\begin{equation}
{}^p\tau_r^s(M,{\bf t},{\bf t}^*)=
\sum_{{\lambda}}s_{{\lambda}}({\bf t})s_{{\lambda}}({\bf t}^*)
\frac{(a_1+M)_{{\lambda}}\cdots (a_p+M)_{{\lambda}}}
{(b_1+M)_{{\lambda}}\cdots (b_s+M)_{{\lambda}}} . \label{tau}
\end{equation}

 If in formula (\ref{tau}) we put
\begin{equation}\label{t^*}
t^*_1=1 , \qquad t^*_i=0,\quad i>1 ,
\end{equation}
then $s_{{\lambda}}({\bf t}^*)=H^{-1}_{{\lambda}}$, and we obtain
the hypergeometric function related to Schur functions \cite{KV}
(see \cite{Mac} for help):
\begin{eqnarray}\label{beskshur}
{}^p\tau_r^s(M,{\bf t},{\bf t}^*) ={}_pF_s\left.\left(a_1+M, \dots
,a_p+M\atop b_1+M,\dots ,b_s+M\right|t_1,t_2,\dots\right)=
\nonumber\\
\sum_{{\lambda}}\frac{(a_1+M)_{{\lambda}}\cdots
(a_p+M)_{{\lambda}}} {(b_1+M)_{{\lambda}}\cdots
(b_s+M)_{{\lambda}}} \frac{s_{{\lambda}}({\bf t})}{H_{{\lambda}}}.
\end{eqnarray}
In the last formula $H_{{\lambda}}$ is the following hook product
(compare with (\ref{hp})):
\begin{eqnarray}
H_{{\lambda}}=\prod_{(i,j)\in {\lambda}} h_{ij},  \qquad
h_{ij}=(n_i+n'_j-i-j+1).
\end{eqnarray}
We obtain ordinary hypergeometric function of one variable of type
\begin{equation}\label{onevar}
{}_{p-1}F_s(a_2 \pm 1,\dots,a_p\pm 1;b_1\pm 1,\dots,b_s\pm 1;\pm t_1t^*_1)=
\tau_r(\pm 1,{\bf t},{\bf T},{\bf t}^* ) ,
\end{equation}
 if we take $a_1=0$, ${\bf t}=(t_1,0,0,\dots)$,
${\bf t}^* =(t^*_1,0,0,\dots)$.
\\

Now we take ${\bf t}={\bf t}({\bf x}^N)$ the formula
(\ref{beskshur}) turns out to be
\begin{eqnarray}\label{tauzon}
{}^p\tau_r^s(M, {\bf t}({\bf x}^N), {\bf t^*})=
{}_pF_s\left.\left(a_1+M,\dots ,a_p+M\atop b_1+M,\dots ,
b_s+M\right| {\bf x}^N\right)=
\nonumber\\
\sum_{{\lambda}\atop l({\lambda})\le
N}\frac{(a_1+M)_{{\lambda}}\cdots (a_p+M)_{{\lambda}}}
{(b_1+M)_{{\lambda}}\cdots (b_s+M)_{{\lambda}}}
\frac{s_{{\lambda}}({\bf x}^N)}{H_{{\lambda}}}.\qquad \qquad
\qquad \qquad \qquad
\end{eqnarray}
We got the hypergeometric function (\ref{hZ}) related to zonal
 polynomials for
the symmetric space $GL(N,C)/U(N)$ \cite{V}.
Here $x_i=z_i^{-1},i=1,...,N$ are the eigenvalues of the matrix
 ${\bf X}$, and for zonal spherical polynomials there is the following
matrix integral representation
\begin{equation}\label{zonmatint}
Z_{\lambda}({\bf X})= Z_{\lambda}({\bf
I}_N)\int_{U(N,C)}\Delta^{\lambda} \left( U^*{\bf X}U\right)d_*U ,
\end{equation}
where $\Delta^{\lambda}\left({\bf X}\right)=\Delta^{n_1-n_2}_1
\Delta^{n_2-n_3}_2\cdots \Delta^{n_N}_N$ and
$\Delta_1,\dots\Delta_N$ are main minors of the matrix ${\bf X}$,
$d_*U$ is the invariant measure on $U(N,C)$,
see \cite{V}.\\

Due to (\ref{tau+xi}) hypergeometric function (\ref{beskshur}) has
the representation as follows
\begin{equation}\label{tau+xizonI}
{}_pF_s\left.\left(a_1+M,\dots ,a_p+M\atop b_1+M,\dots ,
b_s+M\right| {\bf x}^N\right)= \frac{1}{\Delta}
\exp(x_1r(D_1))\cdots \exp(x_Nr(D_N))\cdot \Delta,
\end{equation}
where $D_i=x_i\frac{\partial}{\partial x_i}$ and
\begin{equation}
\label{Delta} \Delta = \frac{\prod_{i<j}
(x_i-x_j)}{(x_1\cdots x_N)^{N-M-1}}
\end{equation}
(See also (\ref{zonrb}) for different representation)

Now let us take ${\bf t}^*=(t_1^*,0,0,\dots)$. Then we get the
same function (\ref{tauzon}), but with each $x_i$ changed by
$-t_1^*x_i$. Then one consider the equation (\ref{linur+}) with
$k=1$. We put $D_i=x_i\partial_{x_i}$ and get
\begin{equation}\label{linurtauzonEx}
\left(\sum_{i=1}^N D_i
-\frac{1}{{\Delta}}\left(\sum_{i=1}^N x_i r(D_i)\right) {\Delta}
\right) {}_pF_s\left.\left(a_1+M,\dots ,a_p+M\atop b_1+M,\dots ,
b_s+M\right| {\bf x}^N\right)=0
\end{equation}
where $r$ is given
by (\ref{op1}), and $\Delta $ see in (\ref{Delta}).

Taking $N=1$ we obtain the ordinary hypergeometric function of one
variable  $x=x_1$, which is (compare with (\ref{onevar})):
\begin{equation}\label{tau+xizonIordinary}
 {}_pF_s\left(a_1+M,\dots ,a_p+M;
b_1+M,\dots , b_s+M; x \right)= x^{-M} e^{xr(D)}\cdot x^M,\quad
D=x\frac{d}{dx}
\end{equation}
The ordinary hypergeometric series
satisfies the known hypergeometric equation (we put $M=0$) written
in the following form:
\begin{equation}\label{ord-eq}
\left(\partial_{x}-{}_pr_s(D)\right)
{}_pF_s(a_1,\dots,a_p;b_1,\dots ,b_s; x)=0, \quad D:=x\partial_{x}
.
\end{equation}
This relation helps us to understand the meaning
of function $r$.
\\
It is known that the series (\ref{tauzon}) diverges if $p>s+1$
(until any of $a_i+M$ is non positive integer). In case $p=s+1$ it
converges in certain domain in the vicinity of ${\bf x}^N={\bf
0}$. For $p<s+1$ the series (\ref{tauzon}) converges for all ${\bf
x}^N$. These known facts (see \cite{KV}) can be also obtained with
the help of the determinant representation (\ref{det1}) and
properties
of  (\ref{ordinary}).\\

Now we have the following representations:
\begin{eqnarray}\label{pFsint1}
{}_{p+1}F_s\left.\left(a, a_1,\dots ,a_p\atop b_1,\dots ,
b_s\right| x_1,\dots ,x_N \right)=
\frac{1}{\prod_{i<j}^N (x_i-x_j)}
\int_0^\infty \cdots \int_0^\infty \quad
\\ \nonumber
(\alpha_1\cdots\alpha_N)^{a-N}e^{-\alpha_1-\cdots -\alpha_N}
{}_pF_s\left.\left(a_1,\dots ,a_p\atop b_1,\dots ,
b_s\right|\alpha_1x_1,\dots,\alpha_Nx_N \right)
\prod_{i<j}^N (\alpha_ix_i-\alpha_jx_j)d\alpha_1\cdots d\alpha_N
\end{eqnarray}
Hankel type of representation
\begin{eqnarray}\label{pFsint2}
{}_pF_{s+1}\left.\left(a_1,\dots ,a_p\atop b, b_1,\dots ,
b_s\right| x_1,\dots ,x_N \right)=
\frac{1}{\prod_{i<j}^N (x_i-x_j)}
\frac{1}{(2\pi \sqrt{-1})^N}
\int_{C} \cdots \int_{C}\quad
 \\ \nonumber
(\alpha_1\cdots\alpha_N)^{1-N-b}e^{\alpha_1+\cdots +\alpha_N}
{}_pF_s\left.\left(a_1,\dots ,a_p\atop b_1,\dots ,
b_s\right|\frac{x_1}{\alpha_1},\dots,\frac{x_N}{\alpha_N} \right)
\prod_{i<j}^N (\alpha_ix_i-\alpha_jx_j)d\alpha_1\cdots d\alpha_N
\end{eqnarray}
where $C$ starts at $-\infty$ on the real axis, circles the origin in the
counterclockwise direction and returns to the starting point.

Beta type representation:
\begin{eqnarray}\label{pFsint3}
{}_{p+1}F_{s+1}\left.\left(a, a_1,\dots ,a_p\atop b, b_1,\dots ,
b_s\right| x_1,\dots ,x_N \right)=
\frac{1}{\prod_{i<j}^N (x_i-x_j)}\left(\frac{1}{\Gamma (b-a)}\right)^N
\int_0^1 \cdots \int_0^1 \quad
\\ \nonumber
{}_pF_s\left.\left(a_1,\dots ,a_p\atop b_1,\dots ,
b_s\right|\alpha_1x_1,\dots,\alpha_Nx_N \right)
\prod_{i<j}^N (\alpha_ix_i-\alpha_jx_j)
\prod_{i=1}^N \alpha_i^{a-N}(1-\alpha_i)^{b-a-1}
d\alpha_i
\end{eqnarray}

{\bf Example 2}. Let us fix a number $N$. In order to get a
hypergeometric function of two sets of variables ${\bf x}^N,{\bf
y}^N$ we put
\begin{equation}\label{rtauzon2}
 {}_pr_s(n)=
\frac{\prod_{i=1}^p (a_i+n)}{\prod_{i=1}^s(b_i+n)}
\frac{1}{N-M+n},
\end{equation}
Notice that $r$ explicitly depends on $N$.

Taking ${\bf t}={\bf t}({\bf x}^N)$ and ${\bf t}^*={\bf t^*}({\bf
y}^N)$, and using (\ref{PochSchur}) (see III of \cite{Mac}) (see
III of \cite{Mac}) we obtain  the formula (\ref{hZ})
\begin{eqnarray}\label{zon2}
\l M| e^{H({\bf t}({\bf x}^N))}e^{-A({\bf t^*}({\bf y}^N))}|M \r=
{}_p{\mathcal{F}}_s\left.\left(a_1+M,\dots ,a_p+M\atop b_1+M,\dots
,
b_s+M\right|{\bf x}^N,{\bf y}^N\right)=\nonumber\\
\sum_{{\lambda}\atop l({\lambda})\le N } \frac{s_{{\lambda}}({\bf
x}^N)s_{{\lambda}}({\bf y}^N)} {(N)_{{\lambda}}}
\frac{(a_1+M)_{{\lambda}}\cdots (a_p+M)_{{\lambda}}}
{(b_1+M)_{{\lambda}}\cdots (b_s+M)_{{\lambda}}}.
\end{eqnarray}

Due to (\ref{reprNN*}) hypergeometric function (\ref{zon2}) has
the representation as follows
\begin{equation}\label{tau+xizon2}
{}_p{\mathcal{F}}_s\left.\left(a_1+M,\dots ,a_p+M\atop b_1+M,\dots
, b_s+M\right|{\bf x}^N,{\bf y}^N\right)= \frac{1}{\Delta}
\prod_{i,j=1}^N(1-y_jx_ir(D_{x_i}))^{-1}\cdot \Delta
\end{equation}
Here and below the inverse operator
$(1-y_jx_ir(D_{x_i}))^{-1}$ is a formal series
$1+y_jx_ir(D_{x_i})+\cdots $.

For example when $N=1$ we get the ordinary hypergeometric function
of one variable
$xy=x_1y_1$':
\begin{equation}
{}_pF_s\left(a_1+M,\dots ,a_p+M ;
b_1+M,\dots ,b_s+M; xy\right)=
x^{-M}\left(1-xyr(D_x)\right)^{-1}\cdot x^M
\end{equation}
Looking
at (\ref{tau+xizonI}) one derives the following linear equation
\begin{equation}\label{linurtauzon2}
\left(D_{y_j}-\sum_{i=1}^N\frac{1}{1-y_jx_ir(D_{x_i})}
+N\right)\cdot \left(\Delta
{}_p{\mathcal{F}}_s\left.\left(a_1+M,\dots ,a_p+M\atop b_1+M,\dots
, b_s+M\right|{\bf x}^N,{\bf y}^N\right)\right)=0
\end{equation}
\\
{\bf Example 3} The $q$-generalization of the {\em Example 3} is
Milne's hypergeometric function of the single set of variables. To
get it we choose
\begin{equation}\label{rq}
{}_pr_s^{(q)}(n)=
\frac{\prod_{i=1}^p (1-q^{a_i+n})} {\prod_{i=1}^s(1-q^{b_i+n})} .
\end{equation}
Taking ${\bf t}={\bf t}({\bf x}^N), {\bf t}^*={\bf t^*}({\bf
y}_{(\infty)})$ and putting
\begin{equation}\label{chvy}
y_k=q^{k-1},\quad k=1,2,... ,\quad t^*_m=\sum_{k=1}^{+\infty}
\frac{y_k^m}{m}= \frac{1}{m(1-q^m)},\quad m=1,2,\dots
\end{equation}
we get Milne's hypergeometric function (\ref{vh}):
\begin{eqnarray}
\l M|e^{H({\bf t})}e^{-A({\bf t}^*)}|M \r=
{}_p\Phi_s\left.\left(a_1+M,\dots ,a_p+M\atop b_1+M,\dots ,b_s+M\right|q,
{\bf x}^N\right)=\nonumber\\
\sum_{{\lambda}\atop l({\lambda})\le N } \frac{(q^{a_1+M};
q)_{{\lambda}}\cdots (q^{a_p+M};q)_{{\lambda}}}
{(q^{b_1+M};q)_{{\lambda}}\cdots (q^{b_s+M};q)_{{\lambda}}}
\frac{q^{n({\lambda})}}{H_{{\lambda}}(q)}s_{{\lambda}}({\bf x}^N)
. \label{qtau}
\end{eqnarray}
According to (\ref{reprNN*}),(\ref{chvy}) we have the following
representation
\begin{eqnarray}\label{reprMilne1}
{}_p\Phi_s\left.\left(a_1+M,\dots ,a_p+M\atop b_1+M,\dots
,b_s+M\right|q, {\bf x}^N\right)= \frac{1}{\Delta(x)}
\prod_{i=1}^N\frac{1}{\left(x_ir(D_{x_i});q\right)_\infty}\cdot \Delta(x)\\
\nonumber
=\frac{1}{\Delta} \exp_q(x_1r(D_1))\cdots \exp_q(x_Nr(D_N))\cdot \Delta
\end{eqnarray}
(the notations $(b;q)_\infty$, $ \exp_q(\alpha)$ see in
(\ref{qPochh}), (\ref{qexp}), $\Delta$ is the same as in
(\ref{Delta})). (See also (\ref{Milnerb}) for different
representation).

For this hypergeometric functions we have the linear equation
(\ref{linurtauzonq}) (where we choose $r$ due to (\ref{rq})),
which is $q$-difference equation. For instance for the case
$N=1,M=0$ we get (compare it with (\ref{ord-eq}))
\begin{equation}\label{q-eq}
\left(\frac {1}{x}
\left(1-q^{D}\right)-{}_p r_s^{(q)}(D)\right) {}_p\Phi_s(a_1,\dots
,a_p;b_1,\dots ,b_s;q,x)=0, \quad D:=x\partial_{x} ,
\end{equation}
where ${}_p r_s^{(q)}(D)$ is defined by (\ref{rq}).
Milne-s function itself for $N=1$ is the ordinary basic
hypergeometric function:
\begin{eqnarray}\label{qssN=1}
{}_p\Phi_s\left.\left(a_1+M,\dots ,a_p+M\atop b_1+M,\dots ,b_s+M\right|q,x\right)
=\sum_{n=0}^{+\infty}\frac{(q^{a_1+M};q)_{n}
\cdots (q^{a_p+M};q)_{n}}{(q^{b_1+M};q)_{n}\cdots (q^{b_s+M};q)_{n}}
\frac{x^{n}}{(q;q)_n}, \quad x=x_1
\end{eqnarray}

Let us take $M=0$.
We have the following integral representation
\begin{eqnarray}\label{pFsint1q}
{}_{p+1}\Phi_s\left.\left(a,a_1,\dots ,a_p\atop b_1,\dots ,b_s\right|q,
x_1,\dots ,x_N \right)=
\frac{(-1)^Nq^{\frac{N-N^2}{2}-Na}}{(1-q)^{\frac{3N-N^2}{2}}}
\frac{1}{\prod_{i<j}^N (x_i-x_j)}
\int_0^\infty\cdots \int_0^\infty
\qquad \qquad \\  \nonumber
\prod_{i<j}^N (\alpha_i x_i-\alpha_j x_j)
{}_{p}\Phi_s\left.\left(a_1,\dots ,a_p\atop b_1,\dots ,b_s\right|q,
\frac{\alpha_1 (1-q)x_1}{q},\dots \frac{\alpha_N (1-q)x_N}{q}\right)
\prod_{i=1}^N \frac{\alpha_i^{a-1}}{\exp_q(\alpha_i)} d_q\alpha_i
\end{eqnarray}
Also we have the Beta-type representation:
\begin{eqnarray}\label{pFsint2q}
{}_{p+1}\Phi_{s+1}\left.\left(a,a_1,\dots ,a_p\atop b,b_1,\dots ,b_s\right|q,
x_1,\dots ,x_N \right)=
\left(\frac {1}{\Gamma_q(b-a)}\right)^N
\frac{1}{\prod_{i<j}^N (x_i-x_j)}
\int_0^1\cdots \int_0^1
\qquad \\  \nonumber
\prod_{i<j}^N (\alpha_i x_i-\alpha_j x_j)
{}_{p}\Phi_s\left.\left(a_1,\dots ,a_p\atop b_1,\dots ,b_s\right|q,
\alpha_1 x_1,\dots \alpha_N x_N\right)
\prod_{i=1}^N
 {\alpha_i}^{a-1}\frac{(\alpha_i q;q)_\infty }
{(\alpha_i q^{b-a};q)_\infty }
d_q\alpha_i
\end{eqnarray}

{\bf Example 4}.
 To obtain Milne's hypergeometric function of two
sets of variables ${\bf x}^N,{\bf y}^N$, we take
 ${\bf t}={\bf t}({\bf x}^N)$  and
${\bf t}^*={\bf t^*}({\bf y}^N)$. This choice restricts the sum
over partitions  ${\lambda}$ with $l({\lambda})\le N$. We put
\begin{equation}\label{rMilne2}
 {}_pr_s^{(q)}(n)=
\frac{\prod_{i=1}^p (1-q^{a_i+n})}{\prod_{i=1}^s(1-q^{b_i+n})}
\frac{1}{1-q^{N-M+n}},
\end{equation}
\begin{eqnarray}
 e^{-T_n}= \frac{1}{(1-q)^n\Gamma_q(n+N-M+1)}
\frac{\prod_{i=1}^p (1-q)^n \Gamma_q (a_i+n+1)}
{\prod_{i=1}^s (1-q)^n\Gamma_q(b_i+n+1)},\\
\Gamma_q(a)=(1-q)^{1-a}\frac{(q;q)_{\infty}}{(q^a,q)_{\infty}},
\qquad (q^a,q)_n=(1-q)^n\frac{\Gamma_q(a+n)}{\Gamma_q(a)}.
\end{eqnarray}
Here $\Gamma_q(a)$ is a $q$-deformed Gamma-function
\begin{equation}
\Gamma_q(a)=(1-q)^{1-a}\frac{(q;q)_{\infty}}{(q^a,q)_{\infty}},
\qquad (q^a,q)_n=(1-q)^n\frac{\Gamma_q(a+n)}{\Gamma_q(a)}.
\end{equation}
Using (\ref{PochSchur}) (see III of \cite{Mac})
we obtain the Milne's formula
(\ref{vh2})
\begin{eqnarray}\label{Milne2}
\tau_r(M,{\bf t}({\bf x}^N), {\bf t^*}({\bf y}^N))=
{}_p\Phi_s\left.\left(a_1+M,\dots ,a_p+M\atop b_1+M,\dots
,b_s+M\right|q,
{\bf x}^N,{\bf y}^N\right)=\nonumber\\
\sum_{{\lambda}\atop l({\lambda})\le N }
\frac{q^{n({\lambda})}}{H_{{\lambda}}(q)}\frac{s_{{\lambda}}({\bf
x}^N) s_{{\lambda}}({\bf y}^N)}{s_{{\lambda}}(1,q,\dots ,q^{N-1})}
\frac{(q^{a_1+M}; q)_{{\lambda}}\cdots (q^{a_p+M};q)_{{\lambda}}}
{(q^{b_1+M};q)_{{\lambda}}\cdots (q^{b_s+M};q)_{{\lambda}}}.
\label{qtauN}
\end{eqnarray}
This is the KP tau-function (but not the TL one because (\ref{rMilne2})
depends on TL variable $M$).

We have the same type of representation (\ref{tau+xizon2}) and the same
form of linear equation (\ref{linurtauzon2}), however we shall use
(\ref{rMilne2}) for $r$ in formulae (\ref{tau+xizon2}),(\ref{linurtauzon2}).

To receive the basic hypergeometric function of one set of variables
 we must put indeterminates ${\bf y}^N$ in (\ref{qtau}) as $y_i=q^{i-1}, i=(1,\dots ,N)$.
 Thus we have
\begin{eqnarray}\label{qssN}
{}_p\Phi_s\left.\left(a_1+M,\dots ,a_p+M\atop b_1+M,\dots
,b_s+M\right|q,{\bf x}^N\right) =\nonumber\\ \sum_{{\lambda}\atop
l({\lambda})\le N }\frac {(q^{a_1+M};q)_{{\lambda}} \cdots
(q^{a_p+M};q)_{{\lambda}}}
                            {(q^{b_1+M};q)_{{\lambda}}\cdots (q^{b_s+M};q)_{{\lambda}}}
\frac{q^{n({\lambda})}}{H_{{\lambda}}(q)}s_{{\lambda}}({\bf x}^N)
.
\end{eqnarray}
And for $N=1$ we have the ordinary $q$-deformed hypergeometric
function:
\begin{eqnarray}\label{qssN=1}
{}_p\Phi_s\left.\left(a_1+M,\dots ,a_p+M\atop b_1+M,\dots
,b_s+M\right|q,x,y\right)
=\sum_{n=0}^{+\infty}\frac{(q^{a_1+M};q)_{n} \cdots
(q^{a_p+M};q)_{n}}{(q^{b_1+M};q)_{n}\cdots (q^{b_s+M};q)_{n}}
\frac{(xy)^{n}}{(q;q)_n}, ~ xy=x_1y_1 \qquad
\end{eqnarray}

\subsection{Baker-Akhiezer functions and the elements of
Sato Grassmannian related to the tau function of hypergeometric
type}

This subsection is based on \cite{nl64}. Let us write down the
expression for Baker-Akhiezer functions (\ref{baker}) in terms of
Hirota-Miwa variables (\ref{HMxy}):
\begin{equation}\label{bak}
 w_{\infty}(M,-{\bf
t}({\bf x}^N),{\bf t^*},\frac{1}{z})= \frac{\tau_r(M,-{\bf t}({\bf
x}^{N+1}),{\bf t^*})} {\tau_r(M,-{\bf t}({\bf x}^N),{\bf t^*})}
\prod_{i=1}^N(1-\frac{x_i}{z}),\quad {\bf x}^{N+1}=(x_1,\dots
,x_N,z) ,
\end{equation}
\begin{equation}\label{bak*}
w_{\infty}^*(M,{\bf t}({\bf x}^N),{\bf t^*},\frac{1}{z})=
-\frac{\tau_r(M,{\bf t({\bf x}^N)+[z]},{\bf t^*})}{\tau_r(M,{\bf
x}^N,{\bf t^*})}
\prod_{i=1}^N\frac{1}{(1-\frac{x_i}{z})}\frac{dz}{z},\quad {\bf
[z]}=(z,\frac{z^2}{2},\dots).
\end{equation}

We see that the variables $x_k,k=1,\dots,N$ are zeroes of
$w_{\infty}(z)$ and poles of
$w_{\infty}^*(z)$.\\
\bremark
The associated linear problems for Baker-Akhiezer functions are read as
\begin{equation}\label{Loperator}
\left(\partial_{t_1}-\partial_{t_1}\phi_n\right)
w(n,{\bf t},{\bf t}^*,z)
=w(n+1,{\bf t},{\bf t}^*,z),
\end{equation}
\begin{equation}\label{Aoperator}
\partial_{t_1^*}
w(n,{\bf t},{\bf t}^*,z)=r(n)e^{\phi_{n-1}-\phi_{n}} w(n-1,{\bf
t},{\bf t}^*,z).
\end{equation}
where $w$ is either $w_\infty $ or
$w_0$. The compatibility of these equations gives rise to the
equation (\ref{rToda}). Taking into account the second
eq.(\ref{rToda}), equations (\ref{Loperator}), (\ref{Aoperator})
may be also viewed as the recurrent equations for the
tau-functions which depend on different number of variables ${\bf
x}^N$.
\eremark
Let us write down a plane of Baker-Akhiezer
functions (\ref{baker}), which characterizes elements of Sato
Grassmannian related to the tau-function (\ref{tau3})
$\tau_r(M,{\bf t},{\bf t}^*)$. We take $x_k=0,k=1,\dots,N$ in
(\ref{bak}) and obtain:
\begin{equation}\label{SatoG}
w_\infty
(n,{\bf 0},{\bf t}^*,z)=z^n(1+\sum_{m=1}^\infty r(n)r(n-1)\cdots
r(n-m+1)h_m(-{\bf t}^*)z^{-m}) , \quad n=M,M+1,M+2,... .
\end{equation}
The dual plane is
\begin{equation}\label{dSatoG}
 w_\infty^*
(n,{\bf 0},{\bf t}^*,z)=z^{-n}(1+\sum_{m=1}^\infty
r(n)r(n+1)\cdots r(n+m-1) h_m({\bf t}^*)z^{-m})dz , \quad
n=M,M+1,M+2,... .
\end{equation}
About these formulae see also (\ref{SatoGr}),(\ref{dSatoGr}).\\
We see that when $r$ has zeroes, then the element of the
Grassmannian defining the solution corresponds to
 the finite-dimensional Grassmannian.

If $r(0)=0$ one can construct a plane
\begin{equation}\label{plane*}
\{w_\infty^* (n,{\bf 0},{\bf t}^*,1/z),n=1,\dots,M\}
\end{equation}
which is a spanned by a finite number of vectors. The space
spanned by these vectors and in addition by the vectors
(\ref{SatoG}) gives the half infinite plane generated by
$\{z^n,n\ge 0\}$.

\subsection{Rational $r$ and their $q$-deformations}

{\bf Lemma 1} If
\begin{equation}\label{rrat}
r(n)=\frac{\prod_{i=1}^p(n+a_i)}{\prod_{i=1}^s(n+b_i)}
\end{equation}
then
\begin{equation}\label{rlambdarat}
r_\lambda(n)=\left(s_{{\lambda}}({\gamma
}(+\infty))\right)^{s-p}\frac{\prod_{k=1}^p{s_{{\lambda}}({\gamma}(a_k+n))}}
{\prod_{k=1}^s{s_{{\lambda}}({\gamma}(b_k+n))}}
\end{equation}
where
\begin{equation}\label{gammainftya}
\gamma(\infty)=(1,0,0,\dots),\quad \gamma(a)=\left(\frac{a}{1},
\frac{a}{2},\frac{a}{3},\dots \right)
\end{equation}
and $s_\lambda(\gamma)$ are Schur functions as functions of
variables $\gamma$ according to (\ref{ph}) and (\ref{Schurh}).

{\bf Lemma 2} If
\begin{equation}\label{rratq}
r(n)=\frac{\prod_{i=1}^p(1+q^{n+a_i})}{\prod_{i=1}^s(1+q^{n+b_i})}
\end{equation}
then
\begin{equation}\label{rlambdarat}
r_\lambda(n)=\left(s_{{\lambda}}({\gamma}(+\infty,q))\right)^{s-p}
\frac{\prod_{k=1}^p{s_{{\lambda}}({\gamma}(a_k+n,q))}}
{\prod_{k=1}^s{s_{{\lambda}}({\gamma}(b_k+n,q))}}
\end{equation}
where
\begin{equation}\label{gammainftyaq}
\gamma_m(a,q)=\frac{1-(q^a)^{m}}{m(1-q^m)}, \quad
\gamma(+\infty,q)=\frac{1}{m(1-q^m)},\quad m=1,2,\dots
\end{equation}

 For the proof we use the following relations (see
\cite{Mac}):
\begin{equation}\label{PochSchur}
(q^a;q)_{\lambda}=\prod_{(i,j)\in {\lambda}} (1-q^{a+j-i})=
\frac{s_{{\lambda}}({\gamma}(a,q))}{s_{{\lambda}}
({\gamma}(+\infty,q))},\quad (a)_{\lambda}=\prod_{(i,j)\in
{\lambda}}(a+j-i)=\frac{s_{{\lambda}}({\gamma}(a))}
{s_{{\lambda}}({\gamma}(+\infty)} ,
\end{equation}

Now we see that for arbitrary complex $b$ and for $r$ of
(\ref{rrat}) we have
\begin{equation}\label{sym=miwa}
<e^{\gamma_1},e^{\sum_{m=1}^\infty
m\gamma_mt_m^*}>_{r,n}=<e^{(b+n)\sum_{m=1}^\infty
m\gamma_m},e^{\sum_{m=1}^\infty m\gamma_mt_m^*}>_{r_b,n}, \quad
r_b(n)=\frac{r(n)}{b+n}.
\end{equation}
For $r$ of (\ref{rratq}) we have
\begin{equation}\label{qsym=miwa}
<e^{\sum_{m=1}^\infty \frac{\gamma_m}{1-q^m}},e^{\sum_{m=1}^\infty
m\gamma_mt_m^*}>_{r,n}=<e^{\sum_{m=1}^\infty
\frac{\gamma_m(1-q^{b+n+m})}{1-q^m}},e^{\sum_{m=1}^\infty
m\gamma_mt_m^*}>_{r_b,n}, \quad r_b(n)=\frac{r(n)}{1-q^{b+n}} .
\end{equation}

  Let
us note that parameters $\gamma_m(a,q)$ and $\gamma_m(a)$ are
chosen via 'generalized Hirota-Miwa transforms'
(\ref{gammainftyaq}),(\ref{gammainftya})  with 'multiplicity $a$'.
Remember that $|q|<1$. Due to
\begin{eqnarray}\label{t(a)}
 s_{{\lambda}}({\gamma}(+\infty,q))=\lim_{a\to +\infty}
s_{{\lambda}}({\gamma}(a,q))=\frac{q^{n({\lambda})}}
{H_{{\lambda}}(q)} ,\\
 s_{{\lambda}}({\gamma}(+\infty))=\lim_{a\to +\infty}
s_{{\lambda}}\left(\frac{\gamma_1(a)}{a},
\frac{\gamma_2(a)}{a^2},\dots\right)= \lim_{a\to
+\infty}\frac{1}{a^{|{\lambda}|}}s_{{\lambda}}({\gamma}(a))=
\frac{1}{H_{{\lambda}}} .
\end{eqnarray}
This allows to rewrite the series (\ref{tauzon}) and (\ref{qtau})
 only in terms of Schur functions, see
\cite{nl64}.

For functions ${}_pF_s$ of (\ref{tauzon}),(\ref{tau+xizonI}) and
${}_p\Phi_s$ of (\ref{qtau}),(\ref{reprMilne1}) we use the results
of subsection 5.4 to derive the following representations
\begin{eqnarray}
{}_pF_s\left.\left(a_1+M,\dots ,a_p+M\atop b_1+M,\dots ,
b_s+M\right| {\bf x}^N\right)= \frac{1}{\Delta(x)}
e^{x_1r(D_1)+\cdots+x_Nr(D_N)}\cdot \Delta(x) \label{zonrb1}\\=
\frac{1}{\Delta(x)} \prod_{i=1}^N
\left(1-x_i\frac{r(D_i)}{b+D_i}\right)^{-b}\cdot
\Delta(x)\label{zonrb}
\end{eqnarray}
\begin{eqnarray}
{}_p\Phi_s\left.\left(a_1+M,\dots ,a_p+M\atop b_1+M,\dots
,b_s+M\right|q, {\bf x}^N\right)= \frac{1}{\Delta(x)}
\prod_{i=1}^N\frac{1}{(x_ir(D_i);q)_\infty}\cdot
\Delta(x)\label{Milnerb1}\\= \frac{1}{\Delta(x)}
\prod_{i=1}^N\frac{\left(x_i{q^br(D_i)}{(1-q^{b+D_i})^{-1}};q\right)_\infty}
{(x_ir(D_i);q)_\infty}\cdot \Delta(x)\label{Milnerb}
\end{eqnarray}
where $D_i=x_i\frac{\partial}{\partial x_i}$, $\Delta(x)$ is defined in
(\ref{reprNN*}), the notation $(b;q)_\infty $ see in (\ref{qPochh}).

\subsection{Gauss factorization problem,
additional symmetries, string equations and $\Psi DO$ on the
circle \cite{nl64}} Let us describe relevant string equations
following Takasaki and Takebe \cite{TT},\cite{T}. We shall also
consider this topic in a more detailed paper.

Let us introduce infinite matrices to describe KP and TL
flows and symmetries, see \cite{UT}. Zakharov-Shabat dressing
matrices are $K$ and $\bar{K}$. $K$ is a lower triangular matrix with unit
main diagonal: $(K)_{ii}=1$. $\bar{K}$ is an upper triangular matrix.
The matrices $K,\bar{K}$ depend on parameters
$M,{\bf t},{\bf T},{\bf t}^*$.
The matrices $(\Lambda)_{ik}=\delta_{i,k-1}$,
$(\bar{\Lambda})_{ik}=\delta_{i,k+1}$.
For each value of
${\bf t},{\bf T},{\bf t}^*$ and $M \in Z$ they solve Gauss (Riemann-Hilbert)
factorization problem for infinite matrices:
\begin{eqnarray}
\bar{K}=KG(M,{\bf t},{\bf T},{\bf t}^*),\quad G(M,{\bf t},{\bf T},{\bf t}^*)=
\exp \left( \xi({\bf t},\Lambda)\right)
\Lambda^{M}G({\bf 0},{\bf T},{\bf 0})\bar{\Lambda}^{M}
\exp \left(\xi({\bf t}^*,\bar{\Lambda})\right) .\quad
\end{eqnarray}
We put
$\log (\bar{K}_{ii})=\phi_{i+M}$, and a set of fields
$\phi_i({\bf t},{\bf t}^*), ( -\infty <i<+\infty) $
solves the hierarchy of higher two-dimensional TL equations.

Take
$L=K\Lambda K^{-1}$, $\bar{L}=\bar{K}\bar{\Lambda}{\bar{K}}^{-1}$,
and
$(\Delta)_{ik}=i\delta_{i,k}$,
$\widehat{M}=K\Delta K^{-1}+M+\sum nt_n L^n$,
$\widehat{\bar{M}}=\bar{K}\Delta {\bar{K}}^{-1}+M+\sum nt^*_n{\bar{L}}^n$.
Then the KP additional symmetries \cite{TT},\cite{T},\cite{D'},\cite{ASM},
\cite{OW} and higher TL flows \cite{UT} are written as
\begin{equation}
\partial_{{\bf t}^*_n}K=
-\left(\left(r(\widehat{M})L^{-1}\right)^n\right)_-K,\quad
\partial_{{\bf t}^*_n}\bar{K}=
\left(\left(r(\widehat{M})L^{-1}\right)^n\right)_+\bar{K},\quad
\end{equation}
\begin{equation}
\partial_{t^*_n}K=-\left(\bar{L}^n\right)_-K,\quad
\partial_{t^*_n}\bar{K}=\left(\bar{L}^n\right)_+\bar{K}.\
\end{equation}
Then the string equations are
\begin{equation}\label{string1}
\bar{L}L=r(\widehat{M}),
\end{equation}
\begin{equation}\label{string2}
\widehat{\bar{M}}=\widehat{M}.
\end{equation}
The first equation is a manifestation of the fact that the group
time ${\bf t}^*_1$ of the additional symmetry of KP can be
identified with the Toda lattice time $t^*_1$. In terms of
tau-function we have the equation (\ref{stringvert1}) in terms of
vertex operator action \cite{D'},\cite{ASM}, or the
equations (\ref{linur+}) in case the tau-function is written in Hirota-Miwa variables.\\
 The second string equation (\ref{string2}) is related to the symmetry of
our tau-functions with respect to ${\bf t} \leftrightarrow {\bf
t}^*$.

When
\begin{equation}\label{rlin}
r(M)=M+a,
\end{equation}
the equations (\ref{string1}),(\ref{string2}) describe $c=1$ string , see
\cite{NTT},\cite{T}. In this case we easily get the relation
\begin{equation}\label{barLL}
[{\bar L},L]=1 .
\end{equation}
The string equations in the form of Takasaki
 allows us to notice the similarity to the different problem.
The dispersionless limit of (\ref{barLL})
(and also of (\ref{string2}), (\ref{string1}), where $r(M)=M^n$, and of
(\ref{rlin})) will be written as
\begin{equation}
{\bar \lambda}\lambda=\mu^n,\quad n \in Z ,
\end{equation}
\begin{equation}
{\bar \mu}=\mu .
\end{equation}
The case when ${\bar \lambda}$ and ${\bar \mu}$ are complex conjugate of
 $\lambda$ and of $\mu$ respectively, is of interest. These string equations
(mainly the case $n=1$) were
recently investigated to solve the so-called Laplacian
growth problem, see \cite{MWZ}. We are grateful to A.Zabrodin for
the discussion on this problem. For the dispersionless limit of the KP and TL
hierarchies see \cite{TT}.

In case the function $r(n)$ has zeroes (described by divisor
${\bf m}=(M_1,...)$: $r(M_k)=0$), one needs to
produce the replacement:
\begin{equation}
\Lambda \to \Lambda({\bf m}),\quad
\bar{\Lambda} \to \bar{\Lambda}({\bf m}) ,
\end{equation}
where new matrices $\Lambda({\bf m})$,$\bar{\Lambda}({\bf m})$ are defined
as
\begin{equation}
(\Lambda({\bf m}))_{i,j}=\delta_{i,j-1}, j\neq M_k,\quad
(\Lambda({\bf m}))_{i,j}=0, j= M_k ,
\end{equation}
\begin{equation}
(\bar{\Lambda}({\bf m}))_{i,j}=\delta_{i,j+1}, i\neq M_k,\quad
(\bar{\Lambda}({\bf m}))_{i,j}=0, i= M_k .
\end{equation}
This modification describes the open
TL equation (\ref{deltaToda}):
\begin{equation}\label{deltaToda'}
\partial_{t_1}\partial_{{\bf t}^*_1}\varphi_n=
\delta(n)e^{\varphi_{n-1}-\varphi_{n}}-
\delta(n+1)e^{\varphi_{n}-\varphi_{n+1}} .
\end{equation}
The set of fields $\phi_\infty,\dots ,\phi_{M_1},\phi_{M_1+1}\dots$
consists of the following
 parts due to the conditions
\begin{equation}\label{Todaspliting}
\sum_{n=-\infty}^{M_s}\phi_n=0,\quad
\sum_{n=M_k+1}^{M_{k+1}}\phi_n=0,\quad
\sum_{n=\infty}^{M_1+1}\phi_n=0,
\end{equation}
which result from $\tau(M_k,{\bf t},{\bf t}^*)=1$.

\bremark
The matrix $r(\widehat{M})$ contains $(\widehat{M}-b_i)$
in the denominator. The matrix $(\widehat{M}-b_i)^{-1}$
is $K(\Delta -b_i)^{-1}K^{-1}(1+O({\bf t}))$ (compare
the consideration of the inverse operators with \cite{OW}).
\eremark

The KP tau-function (\ref{tauhyp}) can be obtained as follows.
\begin{equation}
G(M,{\bf t},{\bf T},{\bf t}^*)= G({\bf 0},{\bf T},{\bf 0})U(M,{\bf
t},{\bf t}^*),\quad U(M,{\bf t},{\bf t}^*)=U^{+}({\bf
t})U^{-}(M,{\bf t}^* ).
\end{equation}
\begin{equation}
U^{+}({\bf t})=\exp \left( \xi({\bf t},\Lambda)\right),\quad
U^{-}(M,{\bf t}^* )= \exp \left(\xi({\bf t}^*,{\bar \Lambda}
r\left(\Delta+M\right))\right),
\end{equation}
The matrix $G({\bf
0},{\bf T},{\bf 0})$ is related to the transformation of the
eq.(\ref{Toda}) to the eq.(\ref{rToda}). By taking the projection
\cite{UT} $U \mapsto U_{--}$ for non positive values of matrix
indices we obtain a determinant representation of the tau-function
(\ref{tauhyp}):
\begin{equation} \label{2-c}
\tau_r(M,{\bf t},{\bf t}^*)= \frac {\det U_{--}(M,{\bf t},{\bf
t}^*)} { \det \left(U^{+}_{--}({\bf t}) \right) \det
\left(U^{-}_{--}(M,{\bf t}^* ) \right)} =\det U_{--}(M,{\bf
t},{\bf t}^*) ,
\end{equation}
 since both determinants in the
denominator are equal to one. Formula (\ref{2-c}) is also a
Segal-Wilson formula for ${GL}(\infty)$ 2-cocycle \cite{SW}
$C_M\left( U^{+}(-{\bf t}),U^{-}(-{\bf t}^* )\right)$. Choosing
the function $r$ as in {\em Section 3.2}
we obtain hypergeometric functions listed in the {\em Introduction}.\\
\bremark
 Therefore the hypergeometric functions which were
considered above have the meaning of $GL(\infty)$ two-cocycle on
the two multi-parametric group elements $U^{+}({\bf t})$ and
$U^{-}({M,{\bf t}^*})$. Both elements $U^{+}({\bf t})$ and
$U^{-}(M,{\bf t}^* )$  can be considered as elements of group of
pseudo-differential operators on the circle. The corresponding Lie
algebras consist of the multiplication operators $\{z^n;n \in
N_0\}$ and of the pseudo-differential operators $\{\left( \frac 1z
r(z\frac {d}{d z}+M) \right)^n;n \in N_0\}$. Two sets of group
times ${\bf t}$ and ${\bf t}^*$ play the role of indeterminates of
the hypergeometric functions (\ref{tau}). Formulas (\ref{tauhyp})
and (\ref{tauTLAAtauKP}) mean the expansion of ${GL}(\infty)$
group 2-cocycle in terms of corresponding Lie algebra 2-cocycle
\begin{equation}
c_M (z,\frac 1z r(D))=r(M),\quad c_M (\tilde{r}(D)z,\frac 1z
r(D+M))=\tilde{r}(M)r(M).
\end{equation}
 Japanese cocycle is
cohomological to Khesin-Kravchenko cocycle \cite{KhK} for the
$\Psi DO$ on the circle:
\begin{equation}
 c_M \sim c_0 \sim \omega_M ,
\end{equation}
 which is
\begin{equation}
\omega_M(A,B)=\frac{1}{2\pi \sqrt{-1}}\oint res_\partial
A[\log (D+M),B]dz, \quad A,B \in \Psi DO .
\end{equation}
 For the group cocycle we have
\begin{equation}\label{CM}
 C_M
\left(e^{-\sum z^nt_n},e^{-\sum (z^{-1}r(D))^n{\bf t}^*_n}\right)
=\tau_r(M,{\bf t},{\bf t}^* ),
\end{equation}
where we imply that the order of $\Psi$DO $r(D)$ is 1 or less.
About properties of $e^{-\sum (z^{-1}r(D))^n{\bf t}^*_n}$ see
(\ref{expPDO}),(\ref{1DGGF}). \eremark \bremark It is interesting
to note that in case of hypergeometric functions ${}_pF_s$
(\ref{hZ1}) the order of $r$ is $p-s$ (see {\em Example 3}), and
the condition $p-s \leq 1$ is the condition of the convergence of
this hypergeometric series, see \cite{KV}. Namely the radius of
convergence is finite in case $p-s=1$, it is infinite when $p-s<1$
and it is zero for $p-s>1$ (this is true for the case when no one
of $a_k$ in (\ref{hZ1}) is nonnegative integer). \eremark \bremark
The set of functions $\{ w(n,z), n=M,M+1,M+2,\dots \}$, where
\begin{equation}\label{SatoGr}
w(n,z)= \exp \left( -\sum_{m=0}^\infty t^*_m\left(\frac 1z
r(D)\right)^m\right) \cdot z^n ,
\end{equation}
 may be identified
with Sato Grassmannian (\ref{SatoG}) related to the cocycle
(\ref{CM}). The dual Grassmannian (\ref{dSatoG}) is the set of one
forms $\{ w^*(n,z),n=M,M+1,M+2,\dots \}$,
\begin{equation}\label{dSatoGr}
 w^*(n,z)=\exp \left( \sum_{m=0}^\infty t^*_m\left(\frac 1z r(-D)\right)^m
\right)\cdot z^{-n} dz .
\end{equation}
\eremark

\section{Appendices B}

\subsection{A scalar product in the theory of symmetric
functions \cite{Mac}}

We consider a ring of symmetric functions of variables
$x_i,i=1,2,\dots,N$ , where $N$ may be infinity. Let us remember
the notion of scalar product. Different choices of scalar product
give rise to different systems of bi-orthogonal symmetric
polynomial functions (like Schur functions, Macdonald functions
and so on).

Actually we consider two sets of variables, the first one is ${\bf
x}=(x_1,x_2,\dots ,x_N)$, the second is ${\bf p}=(p_1,p_2,\dots)$.
These two sets are related via
\begin{equation}\label{xp}
p_m=\sum_{i=1}^N x_i^m,\quad m=1,2,\dots
\end{equation}

If one takes
\begin{equation}\label{u,v}
e^{\sum_{m=1}^\infty \frac 1m v_m^{-1}p_mp^*_m}=\sum_\lambda
P_\lambda({\bf p})Q_\lambda({\bf p}^*)=\sum_\lambda
\frac{p_\lambda p_\lambda^* }{z_\lambda ^{(v)}} ,
\end{equation}
then for any partitions $\mu,\lambda$
\begin{equation}\label{biort}
\l  P_\mu,Q_\lambda \r  _v=\delta_{\mu,\lambda}
\end{equation}
with respect to the scalar product
\begin{equation}\label{pport}
\l p_\mu,p_\lambda\r  _v=z_\lambda^{(v)}  \delta_{\mu,\lambda}
\end{equation}
When $v=1$ then
\begin{equation}\label{zv}
z_\lambda^{(v)}=z_\lambda ,\quad z_\lambda=\prod_i i^{m_i}m_i!
\end{equation}
where $m_i=m_i(\lambda)$ is the number of parts of $\lambda$ equal
to $i$
\begin{equation}\label{powersums}
p_\mu=\prod_i p_{\mu_i},\quad p_{\mu_i}=\sum_k x_k^{\mu_i},\quad
mt_m=p_m
\end{equation}
If one takes
\begin{equation}\label{vm}
v_m=\frac{1-t^m}{1-q^m}
\end{equation}
he gets  Macdonald's polynomials $Q_\lambda({\bf
p})=Q_\lambda({\bf p};q,t),P_\lambda({\bf p})=P_\lambda({\bf
p};q,t)$. In this case
\begin{equation}\label{zMacdonald}
z_\lambda^{(v)}=z_\lambda(q,t)=z_\lambda
\prod_{i=1}^{l(\lambda)}\frac{1-q^{\lambda_i}}{1-t^{\lambda_i}},\quad
z_\lambda=\prod_i i^{m_i}m_i!
\end{equation}
where $l(\lambda)$ is the length of $\lambda$ and
$m_i=m_i(\lambda)$ is the number of parts of $\lambda$ equal to
$i$.

\subsection{A deformation of the scalar product and series of
hypergeometric type}

For different choice of function $v$ we choose numbers
$a^{(v)},b^{(v)}$ and a function $r^{(v)}$ which is a function of
one variable $M$, say. We define
\begin{equation}\label{defr}
r_\lambda^{(v)}(M)=\prod_{i,j \in \lambda}
r^{(v)}(M+a^{(v)}i+b^{(v)}j)
\end{equation}
as a product over all boxes of the Young diagram. In what follows
we shall omit the superscript $(v)$.

Now let us deform the scalar product (\ref{biort}) in a following
way
\begin{equation}\label{rort}
\l P_\mu,Q_\lambda\r  _{v,r,M}=r_\lambda (M) \delta_{\mu,\lambda}
\end{equation}

To get series of hypergeometric type we consider
\begin{equation}\label{hyptype}
\l e^{\sum_{m=1}^\infty \sum_{k=1}^N v_m^{-1}x_k^m
\frac{p_m}{m}},e^{\sum_{m=1}^\infty \sum_{k=1}^N v_m^{-1}y_k^m
\frac{p_m}{m}}\r  _{v,r,M}
\end{equation}
Using (\ref{u,v})
\begin{equation}\label{hyptype}
\l e^{\sum_{m=1}^\infty \sum_{k=1}^N v_m^{-1}x_k^m
\frac{p_m}{m}},e^{\sum_{m=1}^\infty \sum_{k=1}^N v_m^{-1}y_k^m
\frac{p_m}{m}}\r  _{v,r,M}=\sum_\lambda r_\lambda
(M)P_\lambda({\bf x}^N)Q_\lambda({\bf y}^N)
\end{equation}
In this paper we shall consider two cases of (\ref{vm}). First one
is
\begin{equation}\label{v1}
v_n=1
\end{equation}
for all $n$. The second one is
\begin{equation}\label{v2}
v_{2n}=0, \quad v_{2n+1}=2
\end{equation}
The first case corresponds to Schur function series. The second
one corresponds to projective Schur function series. These are the
cases when the scalar product (\ref{rort}) can be presented as a
vacuum expectation of certain fermionic fields, see {\em Section
4} and \cite{prSch}.

\subsection{Scalar product of tau functions. A conjecture.}

Before we considered tau functions of hypergeometric type
(\ref{tauhyp}). In this subsection we shall consider more general
TL tau functions.
 Let $\tau_1({\bf t},{\bf t}^*), \tau_2({\bf
t},{\bf t}^*)$ be TL tau functions of the form
\begin{equation}\label{t1XYt2}
\tau_1({\bf t},{\bf t}^*)=e^{X({\bf t})}\cdot e^{\sum_{m=1}^\infty
mt_mt_m^*},\quad \tau_2({\bf t},{\bf t}^*)=e^{Y({\bf t}^*)}\cdot
e^{\sum_{m=1}^\infty mt_mt_m^*}
\end{equation}
where the operators $X({\bf t})$ and $Y({\bf t}^*)$ are linear
combinations of vertex operators
 (\ref{vertex}) and (\ref{vertex0}) respectively:
\begin{equation}\label{xVV*}
X_i({\bf t})=\int  f(z,z')V_\infty({\bf t},z)V_\infty^*({\bf
t},z')dzdz'
\end{equation}
\begin{equation}\label{yVV*}
Y_i({\bf t}^*)=\int  g(z,z')V_0({\bf t}^*,z)V_\infty^*({\bf
t}^*,z')dzdz'
\end{equation}

If the TL tau functions (\ref{t1XYt2}) have the following forms
\cite{Tinit},\cite{TI},\cite{TII}
\begin{equation}\label{XKss}
\tau_1({\bf t},{\bf
t}^*)=\sum_{\lambda,\mu}K_{\lambda\mu}s_\lambda({\bf t})s_\mu({\bf
t}^*),\quad \tau_2({\bf t},{\bf
t}^*)=\sum_{\lambda,\mu}M_{\lambda\mu}s_\lambda({\bf t})s_\mu({\bf
t}^*)
\end{equation}
then their (standard) scalar product (\ref{ss}) is obviously equal
to
\begin{equation}\label{standardtaugen}
<\tau_1({\bf t},\gamma),\tau_2(\gamma,{\bf
t}^*)>=\sum_{\lambda,\mu}N_{\lambda\mu}s_\lambda({\bf
t})s_\mu({\bf t}^*),\quad N_{\lambda\mu}=
\sum_{\nu}K_{\lambda\nu}M_{\nu\mu}
\end{equation}
It is not immediately clear that what we got is a tau function
again (without solving Hirota equations). However on the other
hand we have
\begin{equation}\label{scaltaugenvert}
<\tau_1({\bf t},\gamma),\tau_2(\gamma,{\bf t}^*)>=e^{X({\bf
t})}\cdot e^{Y({\bf t}^*)}\cdot <e^{\sum_{m=1}^\infty
mt_m\gamma_m},e^{\sum_{m=1}^\infty m\gamma_mt_m^*}>=e^{X({\bf
t})}\cdot e^{Y({\bf t}^*)}\cdot e^{\sum_{m=1}^\infty mt_mt_m^*}
\end{equation}
The last relation shows that we get tau function again according
to general results of \cite{DJKM}. (It was shown there that the
action of vertex operators $e^X$ to KP tau function gives rise to
a new KP tau function. TL tau function is the tau function of the
pair of KP hierarchies related to the sets of times ${\bf t}$ and
 ${\bf t}^*$, see \cite{UT}).

 Due to (\ref{standscalreal}) it means that if the following integral over complex planes
$\gamma_m,m=1,2,\dots $
\begin{equation}\label{t1vact2}
\int \tau_1({\bf t},\gamma)e^{-\sum_{m=1}^\infty m|\gamma_m|^2}
\tau_2(-{\bar \gamma},{\bf t}^*)\prod_{m=1}^\infty \frac{md^2
\gamma_m}{\pi}
\end{equation}
exists, then it is a new TL tau function. Here the variables
${\bar \gamma}=({\bar \gamma}_1,{\bar \gamma}_1,\dots )$ are
complex conjugated to $\gamma =(\gamma_1,\gamma_2,\dots )$.

The proof can be done also directly as follows
\begin{equation}\label{t1vact2-1}
\int \tau_1({\bf t},\gamma)e^{-\sum_{m=1}^\infty m|\gamma_m|^2}
\tau_2(-{\bar \gamma},{\bf t}^*)\prod_{m=1}^\infty \frac{md^2
\gamma_m}{\pi}=
\end{equation}
\begin{equation}\label{t1vact2-2}
e^{X({\bf t})}\cdot e^{Y({\bf t}^*)}\cdot \int
e^{\sum_{m=1}^\infty
mt_m\gamma_m-m|\gamma_m|^2-m{\bar\gamma}_mt_m^*}\prod_{m=1}^\infty
\frac{md^2 \gamma_m}{\pi}=
\end{equation}
\begin{equation}\label{t1vact2-3}
e^{X({\bf t})}\cdot e^{Y({\bf t}^*)}\cdot e^{\sum_{m=1}^\infty
mt_mt_m^*}
\end{equation}
which is tau function.

\quad

Let us try to consider more general situation. Let $\tau_0({\bf
t},{\bf t}^*)$ be a TL tau function with the following properties.
First, $\tau_0({\bf t},{\bf t}^*)=\tau_0({\bf t}^*,{\bf t})$.
Second, each integral over complex plane $\gamma_m,m=1,2,\dots $

\begin{equation}\label{tauintegrgammam}
\int \tau_0(\gamma,-{\bar \gamma})d^2 \gamma_m=a_m
\end{equation}
where ${\bar \gamma}_m$ is complex conjugated of $\gamma_m$, is
convergent and non-vanishing. Then
\begin{equation}\label{tauintegr}
\int \tau_0(\gamma,-{\bar \gamma})\prod_{m=1}^\infty \frac{d^2
\gamma_m}{a_m}=1
\end{equation}

Given $\tau_0$, we introduce the following scalar product
\begin{equation}\label{scalprodtau0}
<f,g>_0= \int f(\gamma)\tau_0(\gamma,-{\bar \gamma})g(-{\bar
\gamma}) \prod_{m=1}^\infty \frac{d^2 \gamma_m}{a_m}
\end{equation}

Let us present a conjecture that scalar product of TL tau
functions is TL tau function again.

{\bf Conjecture.} Let $\tau_1({\bf t},{\bf t}^*), \tau_2({\bf
t},{\bf t}^*)$ be TL tau functions.
 If the integral over complex planes
$\gamma_m,m=1,2,\dots $
\begin{equation}\label{conjucture}
\int \tau_1({\bf t},\gamma)\tau_0(\gamma,-{\bar \gamma})
\tau_2(-{\bar \gamma},{\bf t}^*)\prod_{m=1}^\infty \frac{d^2
\gamma_m}{a_m},
\end{equation}
exists, then it is a new TL tau function.

\subsection{Integral representation of scalar products}

For the scalar product (\ref{rort}) one can present the following
integral realization.

In case $r=1$
\begin{equation}\label{irealortrone}
\l f,g\r  =\int \cdots \int f({\bf t})g({\bar {\bf t}})
e^{-\sum_{m=1}^\infty m|t_m|^2}\prod_{m=1}^\infty \frac {
d^2t_m}{a_m}
\end{equation}
where $mt_m$ are power sum functions, $m{\bar t}_m$ are complex
conjugated variables, and integration is going over complex planes
$(t_m,{\bar t}_m)$ . Normalization factors $a_m$ depend on $m$:
\begin{equation}\label{normak}
a_m =\frac{\pi} {m}
\end{equation}

In case $r(n)=n$ there is a representation
\begin{equation}\label{irealortrzero}
\l f,g\r_{r,n}  =\int \cdots \int f({\bf z})g({\bar {\bf z}})
e^{-\sum_{k=1}^N |z_k|^2}|\Delta (z)|^2\prod_{k=1}^N \frac {
d^2z_k }{b_k}
\end{equation}
where ${\bar {\bf z}}=({\bar z}_1,\dots,{\bar z}_N)$ are complex
conjugated to ${\bf z}=(z_1,\dots,z_N)$, normalizing constants are
\begin{equation}\label{normbk}
b_k=\pi
\end{equation}
$\Delta$ is Vandermond determinant
\begin{equation}\label{normbk}
\Delta(z)=\prod_{i<k}^N(z_k-z_i)
\end{equation}
and one puts $\Delta (z)=1$ for $N=1$. The integration is going
over complex planes $(z_k,{\bar z}_k)$.

For the scalar product (\ref{pport}) we present the representation

\begin{equation}\label{irealortrone}
\l f,g\r  _v =\int \cdots \int f({\bf t})g({\bar {\bf t}})
e^{-\sum_{m=1}^\infty \frac{m}{v_m}t_m{\bar
t}_m}\prod_{m=1}^\infty \frac { dt_md{\bar t}_m}{a_m}
\end{equation}
where $mt_m$ are power sum functions, $m{\bar t}_m$ are complex
conjugated variables, and integration is going over complex planes
$(t_m,{\bar t}_m)$ . Normalization factors $a_m$ depend on $m$:
\begin{equation}\label{normak}
a_m =\frac{2\pi v_m}{m}\sqrt{-1}
\end{equation}

In particular for Macdonald's polynomials $Q_\lambda({\bf
p};q,t),P_\lambda({\bf p};q,t)$, see (\ref{vm}), we get

\begin{equation}\label{Macort}
\int \cdots \int Q_\lambda({\bf p};q,t)P_\lambda({\bar{\bf
p}};q,t) e^{-\sum_{m=1}^\infty {\bf p}_m{\bar {\bf p}}_m
\frac{1-q^m} {m(1-t^m)}}\prod_{m=1}^\infty \frac { dp_md{\bar
p}_m}{a_m}=\delta_{\mu,\lambda}
\end{equation}
\begin{equation}\label{normak}
a_m ={2\pi}\frac{1-q^m} {m(1-t^m)}\sqrt{-1}
\end{equation}

\section{Appendix C. Hypergeometric functions}

{\bf Ordinary hypergeometric functions} First let us remember that
generalized hypergeometric function of one variable $x$ is defined
as
\begin{equation}\label{ordinary}
{}_pF_s\left(a_1,\dots ,a_p; b_1,\dots ,b_s; x \right)
=\sum_{n=0}^\infty\frac {(a_1)_n \cdots (a_p)_n}
             {(b_1)_n \cdots (b_s)_n}
\frac{x^n}{n!} .
\end{equation}
Here $(a)_n$ is Pochhammer's symbol:
\begin{equation}
(a)_n=\frac {\Gamma(a+n)}{\Gamma(a)}=a(a+1)\cdots (a+n-1) .
\end{equation}
Given number $q$, $|q|<1$, the so-called basic hypergeometric
series of one variable is defined as
\begin{equation}\label{qordinary}
{}_p\Phi_s\left(a_1,\dots ,a_p; b_1,\dots ,b_s; q, x \right)
=\sum_{n=0}^\infty \frac {(q^{a_1};q)_n \cdots (q^{a_p};q)_n}
             {(q^{b_1};q)_n \cdots (q^{b_s};q)_n}
\frac{x^n}{(q;q)_n} .
\end{equation}
Here $(q^a,q)_n$ is $q$-deformed  Pochhammer's symbol:
\begin{equation}\label{qPochh}
(b;q)_0=1,\quad (b;q)_n=(1-b)(1-bq^{1})\cdots(1-bq^{n-1}) .
\end{equation}
Both series converge for all $x$ in case $p<s+1$. In case $p=s+1$
they converge for $|x|<1$. We refer these well-known
hypergeometric functions as ordinary hypergeometric functions.

{\bf The multiple hypergeometric series related to Schur
polynomials \cite{V},\cite{KV},\cite{Milne}.} There are several
well-known different multi-variable generalizations of
hypergeometric series of one variable \cite{V, KV}. If one
replaces the sum over $n$ to a sum over partitions
 ${\lambda}=(n_1,n_2,\dots)$, and replaces the single
variable $x$ to a Hermitian matrix $X$, he get one of
generalizations of hypergeometric series which is called
hypergeometric function of matrix argument ${\bf X}$ with indices
$\bf{a}$ and $\bf{b}$ \cite{GR},\cite{V}:
\begin{eqnarray}\label{hZ1z}
{}_pF_s\left.\left(a_1,\dots ,a_p\atop b_1,\dots
,b_s\right|{{\bf{X}}} \right) =\sum_{{\lambda}\atop
l({\lambda})\le N} \frac {(a_1)_{{\lambda}} \cdots
(a_p)_{{\lambda}}}
                     {(b_1)_{{\lambda}} \cdots (b_s)_{{\lambda}}}
\frac{Z_{\lambda}({\bf X})} {|{\lambda}|!} .
\end{eqnarray}
Here the sum is over all different partitions
${\lambda}=\left(n_1,n_2,\dots,n_k\right)$, where $n_1\ge n_2\ge
\cdots \ge n_k$, $k\le |{\lambda}|$, $|{\lambda}|=n_1+\cdots+n_r$
and whose length $l({\lambda})=k\le N$ ($n_k\neq 0 )$. ${\bf X}$
is a Hermitian $N\times N$ matrix, and $Z_{\lambda}({\bf X})$ are
zonal spherical polynomial for the symmetric spaces of the
following types: $GL(N,R)/SO(N)$, $GL(N,C)/U(N)$, or
$GL(N,H)/Sp(N)$
 see \cite{KV}. The definition of symbol $(a)_{\lambda}$ depends
on the choice of the symmetric space:
\begin{equation}
(a)_{\lambda}=(a)_{n_1}(a-\frac{1}{\alpha})_{n_2}
\cdots(a-\frac{k-1)}{\alpha})_{n_k},\quad (a)_{\bf 0}=1,
\end{equation}
where $\alpha=2$, $\alpha=1$ and $\alpha =\frac12$ for the
symmetric spaces $GL(N,R)/SO(N)$, $GL(N,C)/U(N)$, and
$GL(N,H)/Sp(N)$ respectively. The function (\ref{hZ1z}) actually
depends on the eigenvalues of matrix $X$ which are
 ${\bf x}^N=(x_1,\dots,x_N)$ .
In what follows we consider only the case of  $GL(N,C)/U(N)$
symmetric space. Then  zonal spherical polynomial
$Z_{\lambda}({\bf X})$ is proportional to the Schur function
$s_{\lambda}(x_1,x_2,...,x_N)$ corresponding to a partition
${\lambda}$ \cite{Mac}, $s_{\lambda}(x_1,x_2,...,x_N)$ is a
symmetric function of variables $x_k$.

For this choice of the symmetric space the hypergeometric function
can be rewritten as follows
\begin{eqnarray}\label{hZ1}
{}_pF_s\left.\left(a_1,\dots ,a_p\atop b_1,\dots
,b_s\right|{{\bf{X}}} \right) = \sum_{{\lambda}\atop
l({\lambda})\le N} \frac {(a_1)_{{\lambda}} \cdots
(a_p)_{{\lambda}}}
                     {(b_1)_{{\lambda}} \cdots (b_s)_{{\lambda}}}
\frac{s_{\lambda}({\bf x}^N)} {H_{\lambda}} .
\end{eqnarray}
where $H_{\lambda}$ is 'hook product':
\begin{equation}\label{hookprod}
H_{{\lambda}}=\prod_{(i,j)\in {\lambda}} h_{ij} , \quad
h_{ij}=(n_i+n'_j-i-j+1)
\end{equation}
and
\begin{equation}\label{Poch}
(a)_{\lambda}=(a)_{n_1}(a-1)_{n_2} \cdots(a-k+1)_{n_k}, \quad
(a)_{\bf 0}=1
\end{equation}
Taking $N=1$ we get (\ref{ordinary}).

For this hypergeometric functions we suggested different
representations like (\ref{zonrb1}),(\ref{zonrb}) or like
(\ref{pFsint1}),(\ref{pFsint2}),(\ref{pFsint3}).

Let $|q|<1$. The multiple {\em basic} hypergeometric series
related to Schur polynomials were suggested by L.G.Macdonald and
studied by S.Milne \cite{Milne}. These series are as follows
\begin{eqnarray}\label{vh}
{}_p\Phi_s\left(a_1,\dots ,a_p;b_1,\dots ,b_s;q,{\bf x}^N\right)
=\sum_{{\lambda}\atop l({\lambda})\le N}\frac
{(q^{a_1};q)_{{\lambda}} \cdots (q^{a_p};q)_{{\lambda}}}
{(q^{b_1};q)_{{\lambda}}\cdots (q^{b_s};q)_{{\lambda}}}
\frac{q^{n({\lambda})}}{H_{{\lambda}}(q)} s_{{\lambda}} \left(
{\bf x}^N \right) ,
\end{eqnarray}
where the sum is over all different partitions
${\lambda}=\left(n_1,n_2,\dots,n_k\right)$, where $n_1\ge n_2\ge
\cdots \ge n_k \ge 0$, $k\le |{\lambda}|$,
$|{\lambda}|=n_1+\cdots+n_k$ and whose length $l({\lambda})=k\le
N$. Schur polynomial  $s_{{\lambda}} \left( {\bf x}^N \right)$,
with $N \ge l({\lambda})$, is a symmetric function of variables
${\bf x}^N$ and defined as follows \cite{Mac}:
\begin{equation}\label{Schur}
s_{{\lambda}}({\bf
x}^N)=\frac{a_{{\lambda}+\delta}}{a_{\delta}},\quad
a_{{\lambda}}=\det (x_i^{n_j})_{1\le i,j\le N},\quad \delta= (N-1,
N-2,\dots,1,0).
\end{equation}
Coefficient $(q^c;q)_{\lambda}$ associated with partition
${\lambda}$
 is expressed in terms of the $q$-deformed Pochhammer's Symbols
$(q^c;q)_{n}$ (\ref{qPochh}):
\begin{equation}
(q^c;q)_{\lambda}=(q^c ;q)_{n_1}(q^{c-1};q)_{n_2}\cdots
(q^{c-k+1};q)_{n_k}.
\end{equation}
The multiple $q^{n({\lambda})}$ defined on the partition
${\lambda}$:
\begin{equation}\label{n(n)}
q^{n({\lambda})}=q^{\sum_{i=1}^k (i-1)n_i} ,
\end{equation}
and $q$-deformed 'hook polynomial' $H_{{\lambda}}(q)$ is
\begin{eqnarray}\label{hp}
H_{{\lambda}}(q)=\prod_{(i,j)\in {\lambda}}
\left(1-q^{h_{ij}}\right) , \quad h_{ij}=(n_i+n'_j-i-j+1) ,
\end{eqnarray}
where ${\lambda}'$ is the conjugated partition (for the definition
see \cite{Mac}).

For this hypergeometric functions we suggested different
representations like (\ref{Milnerb1}),(\ref{Milnerb}) or like
(\ref{pFsint1q}),(\ref{pFsint2q}).

For $N=1$ we get (\ref{qordinary}). \\

Let us note that in the limit $q \to 1$ series (\ref{vh})
reduces to (\ref{hZ1}), see \cite{KV}.\\

{\bf Hypergeometric series of double set of arguments
\cite{V},\cite{KV},\cite{Milne}}

Another generalization of hypergeometric series is so-called
hypergeometric function of two matrix arguments ${\bf X},{\bf Y}$
with indices $\bf{a}$ and $\bf{b}$:
\begin{eqnarray}\label{hZz}
{_p{\mathcal{F}}_s}\left(a_1,\dots ,a_p;b_1,\dots ,b_s;{{\bf{X}}},
{{\bf{Y}}}\right) =\sum_{{\lambda}}\frac {(a_1)_{{\lambda}} \cdots
(a_p)_{{\lambda}}}
                     {(b_1)_{{\lambda}} \cdots (b_s)_{{\lambda}}}
\frac{Z_{{\lambda}}({\bf X})Z_{{\lambda}}({\bf Y})}
{|{\lambda}|!Z_{{\lambda}}({\bf I}_N)} .
\end{eqnarray}
Here ${\bf X},{\bf Y}$ are Hermitian $N\times N$ matrices and
$Z_{{\lambda}}({\bf X}), Z_{{\lambda}}({\bf Y})$ are zonal
spherical polynomials for the symmetric spaces $GL(N,C)/U(N)$,
$GL(N,R)/SO(N)$ and $GL(N,H)/Sp(N)$
 see \cite{KV}. The notations are the same as in previous subsection.
 Again we shall consider only the case of
$GL(N,C)/U(N)$ symmetric spaces. In this case (\ref{hZz}) it may
be written as
\begin{eqnarray}\label{hZ}
{_p{\mathcal{F}}_s}\left(a_1,\dots ,a_p;b_1,\dots
,b_s;{{\bf{x}}^{(N)}}, {{\bf{y}}^{(N)}}\right)
=\sum_{{\lambda}}\frac {(a_1)_{{\lambda}} \cdots
(a_p)_{{\lambda}}}
                     {(b_1)_{{\lambda}} \cdots (b_s)_{{\lambda}}}
\frac{s_{{\lambda}}({\bf x}^N)s_{{\lambda}}({\bf y}^N)}
{H_{\lambda}s_{{\lambda}}({\bf 1}^{N})} .
\end{eqnarray}
We suggest a different representations like (\ref{tau+xizon2}) or
different integral representations.

The q-deformation of the hypergeometric function (\ref{hZ}) is as
follows
\[
{}_p\Phi_s\left.\left(a_1,\dots ,a_p\atop b_1,\dots ,b_s\right|q,
{\bf x}^N,{\bf y}^N\right)=
\]
\begin{equation}
\sum_{{\lambda}\atop l({\lambda})\le N}\frac
{(q^{a_1};q)_{{\lambda}}\cdots (q^{a_p};q)_{{\lambda}}}
                            {(q^{b_1};q)_{{\lambda}}\cdots (q^{b_s};q)_{{\lambda}}}
\frac{q^{n({\lambda})}}{H_{{\lambda}}(q)} \frac {s_{{\lambda}}
\left( {\bf x}^N \right)s_{{\lambda}} \left( {\bf y}^N \right)}
{s_{{\lambda}} \left(1,q,q^2,...,q^{N-1} \right)} \label{vh2}
\end{equation}
This is the multi-variable basic hypergeometric function of two
sets of variables which was also studied by S.Milne, see
\cite{Milne}, \cite{KV}.
\\

There are also hypergeometric functions related to Jack
polynomials $C_{\lambda}^{(d)}$ \cite{KV}:
\begin{eqnarray}\label{hJ}
{_p{\mathcal{F}}_s}^{(d)}\left(a_1,\dots ,a_p;b_1, \dots b_s;{\bf
x}^N,
{\bf y}^N \right)=\nonumber\\
\sum_{{\lambda}}\frac {(a_1)^{(d)}_{{\lambda}}\cdots
(a_p)^{(d)}_{{\lambda}}}
                    {(b_1)^{(d)}_{{\lambda}}\cdots (b_s)^{(d)}_{{\lambda}}}
\frac{C_{\lambda}^{(d)}({\bf x}^N) C_{\lambda}^{(d)}({\bf y}^N)}
{|{\lambda}|!C_{\lambda}^{(d)}(1^n)} ,
\end{eqnarray}
where
\begin{equation}\label{ad}
(a)^{(d)}_{{\lambda}}=\prod_{i=1}^{l({\lambda})}
\left(a-\frac{d}{2}(i-1)\right)_{n_i} .
\end{equation}
Here $(c)_k=c(c+1)\cdots (c+k-1)$. For the special value $d=2$ the
last expression (\ref{hJ}) coincides with (\ref{vh}), and reduces
to (\ref{hZ}) as $|q| \to 1$. The open problem is to get a
fermionic representation of (\ref{hJ}) for arbitrary value of the
parameter $d$.

\section{ Further generalization. Examples of Gelfand-Graev hypergeometric
functions}

\subsection{Scalar product and series in plane partitions}

Let us consider the scalar product (\ref{rortSchur})
\begin{equation}\label{rortSchurmulti}
\l s_\mu,s_\lambda \r_{r,n}=r_\lambda(n) \delta_{\mu,\lambda}
\end{equation}
For the simplicity of notations sometimes we shall write $\l,\r
_r$ instead of $\l,\r  _{r,n}$.

We introduce notations
\begin{equation}\label{multixi}
({\bf t},{\bf p})=\sum_{m=1}^\infty p_mt_m ,\quad ({\bf p},\gamma
_{i})=\sum_{m=1}^\infty p_m \gamma _{im},\quad ({\bf p},\gamma
^*_{i})=\sum_{m=1}^\infty p_m \gamma ^*_{im}
\end{equation}

{\bf Lemma}

\begin{equation}\label{matele^xi}
\l s_\lambda ,e^{({\bf p},\gamma )}s_\mu\r  _{r,n}=\l s_\mu
e^{({\bf p},\gamma )},s_\lambda\r =r_\lambda(n)s_{\lambda
/\mu}(\gamma )
\end{equation}
One sees that for $\gamma =0$ and $r=1$ the right hand side is
equal to $\delta_{\mu,\lambda}$.

{\bf Proof} One develops $e^{(p,\gamma _i)}$ using
\begin{equation}\label{orttp}
e^{(p,\gamma _i)}=\sum_\nu s_\nu({\tilde {\bf p} })s_\nu({\bf
\gamma }_i)
\end{equation}
and taking into account the orthogonality of Schur functions
(\ref{rortSchurmulti}) and the formulas (see section 5 of chapter
I of \cite{Mac})
\begin{equation}\label{MacskewSchur}
s_\nu s_\mu=\sum_\lambda c^\lambda_{\mu\nu}s_\lambda,\quad
s_{\lambda/\mu}=\sum_\nu c^\lambda_{\mu\nu} s_\nu
\end{equation}

Let us consider a set of scalar products (\ref{rortSchurmulti})
which differ by the choice of $r$. The scalar product of functions
of $p_i$ will be labelled by script $i$: we shall denote this
scalar product as $\l,\r  _{r^{(i)}}$.

{\bf Lemma}
\begin{equation}\label{expdevelopskew}
e^{({\bf p}_{i},\gamma _{i}+ {\tilde {\bf p}}_{i-1})}
=\sum_{\nu_1,\nu_2}s_{\nu_1}({\tilde {\bf p}}_{i})
s_{\nu_1/\nu_2}(\gamma _i)s_{\nu_2}({\tilde {\bf p}}_{i-1})
\end{equation}
and for given partition $\nu_{i-1}$ we have
\begin{equation}\label{expdevelopskewvac}
\l e^{({\bf p}_{i},\gamma _{i}+ {\tilde {\bf
p}}_{i-1})},s_{\nu_{i-1}}({\tilde {\bf p}}_{i-1})
\r_{r^{(i-1)},n_{i-1}} =\sum_{\nu_i}s_{\nu_i}({\tilde {\bf
p}}_{i}) s_{\nu_i/\nu_{i-1}}(\gamma _i)
r^{(i-1)}_{\nu_{i-1}}(n_{i-1})
\end{equation}
where
\begin{equation}\label{tildep}
{\tilde {\bf p}}_i=(\frac {p_{i1}}{1},\frac {p_{i2}}{2},\frac
{p_{i3}}{3},\dots )
\end{equation}
{\bf Proof} follows from the relations (see \cite{Mac})
\begin{equation}\label{prf}
e^{\sum_{n=1}^\infty (p_{in}\gamma _{in}+\frac 1n
p_{in}p_{i-1,n})}=\sum_{\nu_1}s_{\nu_1}(\gamma _i+{\tilde {\bf
p}}_{i-1})s_{\nu_1}({\tilde {\bf p}}_{i}),\quad s_{\nu_2}(\gamma
_i+{\tilde {\bf p}}_{i-1})=\sum_{\nu_2} s_{\nu_1/\nu_2}(\gamma
_i)s_{\nu_2}({\tilde {\bf p}}_{i-1})
\end{equation}
and definition (\ref{rortSchur}).

Let us consider a set of $\{ {\bf p}_i,i=0,1,2,\dots, k+1 \}$ and
a set of $\{ \gamma_i,i=1,2,\dots, k+1 \}$ with conditions
\begin{equation}\label{gammapboundary}
{\bf p}_0=0,\quad {\bf p}_{k+1}={\bf p}
\end{equation}

With the help of Lemma we consider scalar product of scalar
products  (compare with (\ref{tauscal})):

\begin{eqnarray}
 \l e^{ ({\bf p}_{k},\gamma _{k}+{\tilde {\bf p}}_{k-1})}
 \cdots  ,\l e^{
({\bf p}_{3},\gamma _{3}+ {\tilde {\bf p}}_{2})},\l e^{ ({\bf
p}_{2},\gamma _{2}+ {\tilde {\bf p}}_{1})} ,e^{ ({\bf
p}_{1},\gamma _{1}+{\tilde {\bf p}}_{0})} \r _{r^{(1)},n_1} \r
_{r^{(2)},n_2} \cdots \r _{r^{(k-1)},n_k}
\end{eqnarray}
\begin{eqnarray}\nonumber
=\sum_{\nu_{k}\ge \cdots \ge \nu_0}
r^{(k-1)}_{\nu_{k-1}}(n_{k-1})r^{(k-2)}_{\nu_{k-2}}(n_{k-2})\cdots
r^{(2)}_{\nu_{2}}(n_{2})r^{(1)}_{\nu_{1}}(n_1)\\
\label{p_k} s_{\nu_{k}}({\tilde {\bf p} }_{k})
s_{\nu_{k}/\nu_{k-1}}(\gamma _{k}) s_{\nu_{k-1} /\nu_{k-2}}(\gamma
_{k-1})\cdots s_{\nu_{2}/\nu_{1}}({\bf \gamma }_2)
s_{\nu_1/\nu_{0}}({\bf \gamma }_1)s_{\nu_{0}}({\tilde {\bf p}
}_{0})\\=I_k({\bf p},{\bf r}^{k-1},{\lambda}^{k-1},\gamma,{\bf
p}_0)
\end{eqnarray}
where ${\bf r}^{k-1}=(r^{(1)},r^{(2)},\dots
,r^{(k-1)}),{\lambda}_{k-1}=(n_1,\dots , n_{k-1})$,
$s_{\nu_{0}}({\tilde {\bf p}
}_{0}=0)=\delta_{0,\nu_0}$.\\
This is the sum over all {\em plane partitions} \cite{Mac} with
the largest part k: $\pi^{(k)} =(\nu_1\le \cdots \le \nu_k)$.

Now let us consider scalar product of different $I$ and with
respect to standard scalar product (\ref{rortSchurmulti}) .
Putting ${\bf p}_0=0$ we get a function of $\gamma =\{ \gamma
_{in},i=1,\dots ,k,~ n=1,2,\dots \}$, $\gamma ^* =\{ \gamma ^*
_{in},i=1,\dots ,l, n=1,2,\dots \}$, which will appear to be a tau
function
\begin{eqnarray} \label{scalII}
\tau_{{\tilde {\bf r}},{\bf
r}}({\lambda}_{l-1},{\lambda}^*_{k},\gamma,\gamma ^*)=\l I_k({\bf
p},{\tilde {\bf r}}_{k-1},\gamma,0),I_l({\bf p},{\bf
r}_{l-1},\gamma^*,0)\r_{r^{(k)},n_k}
\end{eqnarray}
where ${\tilde {\bf r}}=({\tilde r}^{(1)},{\tilde r}^{(2)},\dots
,{\tilde r}^{(k-1)}),{\bf r}=(r^{(1)},r^{(2)},\dots
,r^{(k-1)},r^{(k)})$ and ${\lambda}_{l-1}=(n_1,\dots
,n_{l-1}),{\lambda}^*_{k}=(n^*_1,\dots ,n^*_{k})$. We see that
(\ref{scalII}) is a sum over a pair of plane partitions.

\subsection{Integral representation for $\tau_{{\tilde {\bf r}},{\bf r}}$}

In case the function $r$ has zero the scalar product
(\ref{rortSchurmulti}) is degenerate one. For simplicity let us
take $r(0)=0,r(k)\neq 0, k \le n$. The scalar product
(\ref{rortSchurmulti}) is non generate on the subspace of
symmetric functions spanned by Schur functions $\{ s_\lambda
,l(\lambda)\le n \}$, $l(\lambda)$ is the length of partition
$\lambda$.
\begin{equation}\label{}
\l f,g \r_{r,n}=\int \cdots \int \prod_{i=1}^n d{\bar z}_i dz_i
\mu_r({\bar z}_iz_i)\Delta_n({\bar z})\Delta_n(z)f({\bar
z}^n)g(z^n)
\end{equation}
where ${\bar z}^n=({\bar z}_1,\dots ,{\bar z}_n),{z}^n=(z_1,\dots
,z_n)$ and
\begin{equation}\label{DD}
\Delta_n({\bar z})=\prod_{i<j}^n({\bar z}_i-{\bar z}_j), \quad
\Delta_n(z)=\prod_{i<j}^n(z_i-z_j)
\end{equation}
Thus get the following integral representation for (\ref{scalII})
\begin{equation}\label{tauintmultiscal}
\tau_{{\tilde {\bf r}},{\bf
r}}({\lambda}_{l-1},{\lambda}^*_{k},\gamma,\gamma ^*)=
\end{equation}
\begin{equation}\label{intmultiscal}
 \int \cdots
\int \prod_{j=1}^{l-1}\prod_{i=1}^{n_j}d{\bar
z}_{ji}dz_{ji}\mu_{{\tilde r}^{(j)}}({\bar z}_{ji}z_{ji})D_j
\delta(z_{ki}-z^*_{l-1,i})\delta({\bar z}_{ki}-{\bar z}^*_{l-1,i})
\prod_{j=1}^{k}\prod_{i=1}^{n^*_j}d{\bar
z}^*_{ji}dz^*_{ji}\mu_{r^{(j)}}({\bar z}^*_{ji}z^*_{ji})D^*_j
\end{equation}
\begin{equation}\label{D}
D_j=\Delta_{n_j}({\bar z}^*_j)\Delta_{n_j}(z^*_j) e^{{ \xi}_{j}},
\quad D^*_j=\Delta_{n^*_j}({\bar
z}^*_j)\Delta_{n_j}(z^*_j)
e^{{ \xi}^*_{j}}
\end{equation}
\begin{equation}\label{xiij}
{\xi}_{j}=\sum_{m=1}^\infty
\sum_{k=1}^{n_j}(z_{jk})^m\left(\gamma_{jm}+\frac 1m
\sum_{i=1}^{n_{j-1}}({\bar z}_{j-1,i})^m\right),j>1,\quad
{\xi}_{1}=\sum_{m=1}^\infty \sum_{k=1}^{n_1}(z_{1k})^m\gamma_{1m}
\end{equation}
\begin{equation}\label{xiij*}
{\xi}^*_{j}=\sum_{m=1}^\infty \sum_{k=1}^{n_j^*}({\bar
z}^*_{1k})^m\left(\gamma_{jm}^*+\frac 1m
\sum_{i=1}^{n_{j-1}^*}({z}^*_{j-1,i})^m\right),j>1,\quad
{\xi}^*_{1}=\sum_{m=1}^\infty \sum_{k=1}^{n_1}({\bar
z}^*_{1k})^m\gamma_{1m}^*
\end{equation}

\subsection{Multi-matrix integrals}

In case ${\tilde r}^{(j)}(n)=n+a_j, r^{(j)}(n)=n+a^*_j$ we get
that the tau function (\ref{tauintmultiscal}) is equal to the
following multi matrix integral
\begin{equation}\label{mmi}
\int \cdots \int \prod_{j=1}^{l-1}dM_jd{\bar M}_je^{Tr M_j{\bar
M}_j} e^{V_j} \delta(M_{k}-M^*_{l-1})\delta({\bar M}_{ki}-{\bar
M}^*_{l-1,i}) \prod_{j=1}^kdM_j^*d{\bar M}^*_je^{Tr M_j^*{\bar
M}_j^*}e^{V_j^*}
\end{equation}
\begin{equation}\label{VMM}
V_{j}=\sum_{m=1}^\infty Tr (M_j)^m\left(\gamma_{jm}+ \frac 1m Tr
({\bar M}_{j-1})^m\right),j>1,\quad V_{1}=\sum_{m=1}^\infty Tr
(M_1)^m \gamma_{1m}
\end{equation}
\begin{equation}\label{VMM*}
 V_{j}^*=\sum_{m=1}^\infty Tr
({\bar M}_j^*)^m\left(\gamma_{jm}^*+ \frac 1m Tr (
M_{j-1}^*)^m\right),j>1,\quad V_{1}^*=\sum_{m=1}^\infty Tr ({\bar
M}_1^*)^m \gamma_{1m}^*
\end{equation}
where $M_j,{\bar M}_j$ has a size $n_j=M+a_j$ and $M_j^*,{\bar
M}_j^*$ has a size $n_j^*=M+a^*_j$.

\subsection{Fermionic representation \cite{nl64}}

Due to (\ref{vacscal_rn}) scalar product (\ref{scalII}) is equal
to a vacuum expectation value. Let us write down it explicitly
using \cite{nl64}.

 Formula
(\ref{tau3}) is related to 'Gauss decomposition' of operators
inside vacuums $\langle M|\dots|M\rangle $ into diagonal operator
$e^{H_0({\bf T})}$ and upper triangular operator $e^{H({\bf t})}$
and lower triangular operator $e^{-H^*({\bf t}^*)}$ the last two
have the Toeplitz form. Now let us consider more general
two-dimensional Toda chain tau function
\begin{equation}\label{general}
\tau=\langle M|e^{H({\bf t})}ge^{-A({\bf t}^*)}|M\rangle ,
\end{equation}
where  we decompose $g$ in the following way:
\begin{equation}\label{general1}
g(\gamma,\gamma ^*)=e^{\tilde{A}_1(\gamma_1)} \cdots
e^{\tilde{A}_k(\gamma_k)}e^{-A_l(\gamma ^*_l)} \cdots
e^{-A_1(\gamma ^*_1)},
\end{equation}
where
\begin{equation}
\tilde{\gamma ^*_i}=(\gamma_{i1},\gamma_{i2}, \dots),\quad \gamma
^*_i=(\gamma ^*_{i1},\gamma ^*_{i2},\dots )
\end{equation}
\begin{equation}
\tilde{A}_k(\gamma_k)=\sum_{m=1}^\infty
\tilde{A}_{km}\gamma_{km},\quad {A}_k({\gamma
^*}_k)=\sum_{m=1}^\infty {A}_{km}{\gamma ^*}_{km}
\end{equation}
Here each of $A_k(\gamma ^*_i)$ has a form as in (\ref{tdopsim})
and corresponds to operator $r^{(k)}(D)$, while each of
$\tilde{A}_k(\tilde{\gamma ^* }_k)$ has a form of (\ref{tdopsim'})
and corresponds to operator $\tilde{r}^{(k)}(D)$:
\begin{equation}\label{tdopsimmulti}
\tilde{A}_{km}=-\frac{1}{2\pi \sqrt{-1}}
\oint \psi^*(z) \left({\tilde {r}}^{(k)}(D)z\right)^m\psi(z),
\quad m=1,2,\dots
\end{equation}
\begin{equation}\label{dopsimmulti}
A_{km}=\frac{1}{2\pi \sqrt {-1}}
\oint \psi^*(z) \left(\frac{1}{z}r^{(k)}(D)\right)^m \psi(z),
\quad m=1,2,\dots  ,
\end{equation}

 Collections of variables
$\gamma=\{\gamma_{in}\}, \gamma ^*=\{\gamma ^*_{in}\}$ play the
role of coordinates for some wide enough class of Clifford group
elements $g$. This tau function is related to rather involved
generalization of the hypergeometric functions we considered
above. Tau function (\ref{general}),(\ref{general1}) may be
considered as the result of applying of the additional symmetries
to the vacuum
tau function, which is 1, see {\em Appendix ``The vertex operator action''}.\\
Let us calculate this tau function. First of all we introduce a
set consisting of $m+1$ partitions:
\begin{equation}
 ({\lambda^1},\dots ,{\lambda^m}, {\lambda^{m+1}}={\lambda}),\quad 0={\lambda^0}\le {\lambda^1}\le {\lambda^2}\le \cdots\le {\lambda^m}\le {\lambda^{m+1}}={\lambda},
\end{equation}
see \cite{Mac} for the notation $\le$ for
the partitions. The corresponding set
\begin{equation}
\Theta_{{\lambda}}^m=({\lambda^1},\theta^1,\dots
,\theta^{m}),\quad \theta^i= {\lambda^{i+1}}-{\lambda^{i}},\quad
i=1,\dots ,m
\end{equation}
depends on the partition ${\lambda}$ and the number $m+1$ of the
partitions. We take as $s_{\Theta}({\bf t}^*,\gamma ^*)$  the
product which is relevant to the set $\Theta^m_{{\lambda}}$ and
depending on the set of variables $\mu_{i}=\{\mu_{ij}\}$
($i=(1,\dots , m+1)$, $j=(1,2\dots)$)
\begin{eqnarray}
s_{\Theta^m_{{\lambda}}}({\bf
\mu})=s_{{\lambda^1}}(\mu_1)s_{\theta^1}(\mu_2)\cdots
s_{\theta^m}(\mu_{m+1}).
\end{eqnarray}
Here $s_{\theta^i}$ is a skew Schur function (see \cite{Mac}).
Further we define function $r_{\Theta^m_{{\lambda}}}(M)$:
\begin{equation}
r_{\Theta^m_{{\lambda}}}(M)=r_{{\lambda^1}}(M)r^{1}_{\theta^1}(M)\cdots
r^{m}_{\theta^m}(M),
\end{equation}
where the function $r^i_{\theta^i}(M)$ , a skew analogy of
$r_{\lambda}(M)$ of (\ref{rlambda}), is
\begin{equation}\label{rskew}
r_{\theta^i}(M)=\prod_{j=1}^s r(n^{(i)}_{j}-j+1+M)\cdots
r(n^{(i+1)}_{j}-j+M),
\end{equation}
where ${\lambda^{i+1}}=(n^{(i+1)}_{1},\dots ,n^{(i+1)}_{s})$. If
the function $r^i(m)$  has no poles and zeroes at integer points
then the relation
\begin{equation}
r^{i}_{\theta^i}(M)=\frac{r^{i}_{{\lambda^{i+1}}}(M)}{r^{i}_{{\lambda^{i}}}(M)},\quad
i=1,\dots , m
\end{equation}
is correct.
To calculate the tau function we need the
{\em Lemma}\\
{\bf Lemma 3} Let partitions ${\lambda}=(i_1,\dots
,i_s|j_1-1,\dots ,j_s-1)$ and ${\bf \tilde{n}}=(\tilde{i}_1,\dots
,\tilde{i}_r|\tilde{j}_1-1,\dots ,\tilde{j}_r-1)$ satisfy the
relation ${\lambda}\ge{\bf \tilde{n}}$. The following  is valid:
\begin{eqnarray}\label{skewScurvac}
\l 0|\psi^*_{\tilde{i}_1}\cdots\psi^*_{\tilde{i}_r}\psi_{-\tilde{j}_r}\cdots
\psi_{-\tilde{j}_1}e^{A^i(\g_i)}\psi^*_{-j_1}\cdots
\psi^*_{-j_s}\psi_{i_s}\cdots\psi_{i_1}|0\r=\nonumber\\
=(-1)^{\tilde{j}_1+\cdots+\tilde{j}_r+j_1+\cdots+j_s}s_{\theta}(\g_i)r_{\theta}(0),\qquad
\theta={\lambda}-\tilde{{\lambda}}.
\end{eqnarray}
\begin{eqnarray}\label{tildeskewScurvac}
\l 0|\psi^*_{i_1}\cdots\psi^*_{i_s}\psi_{-j_s}\cdots \psi_{-j_1}
e^{{\tilde A}^i(\gamma ^*_i)} \psi^*_{-{\tilde j}_1}\cdots
\psi^*_{-{\tilde j}_r}\psi_{{\tilde i}_r}\cdots\psi_{{\tilde i}_1}|0\r=\nonumber\\
=(-1)^{\tilde{j}_1+\cdots+\tilde{j}_r+j_1+\cdots+j_s}s_{\theta}(\g_i)r_{\theta}(0),\qquad
\theta={\lambda}-\tilde{{\lambda}}.
\end{eqnarray}
{\bf Proof:} One  proof is achieved by a development of
$,e^A=1+A+\cdots,e^{\tilde A}=1+{\tilde A}+\cdots$ and a direct
evaluation of vacuum expectations
(\ref{skewScurvac}),(\ref{tildeskewScurvac}).
(see Example 22 in Sec 5 of \cite{Mac} for help).\\
The second proof of (\ref{skewScurvac}),(\ref{tildeskewScurvac})
is achieved using the developments (\ref{AskewSchurdec}),
(\ref{tildeAskewSchurdec}) respectively.\\

Then we obtain the generalization of
{\em Proposition 1}:\\
\bprop
\begin{equation}\label{gen}
\tau_M({\bf t},{\bf t}^* ;\g , \g ^*)=\sum_{{\lambda}}
\sum_{\Theta^k_{{\lambda}}} \sum_{\Theta^l_{{\lambda}}}
\tilde{r}_{\Theta^k_{{\lambda}}}(M) r_{\Theta^l_{{\lambda}}}(M)
s_{\Theta^k_{{\lambda}}}({\bf t},{\g})
s_{\Theta^l_{{\lambda}}}({\bf t}^*,\g ^*) ,
\end{equation}
where $\tilde{r}_{\Theta^k_{{\lambda}}}(M)$ and
$r_{\Theta^l_{{\lambda}}}(M)$ are given by (\ref{rskew}). \eprop

With the help of this series one can obtain different hypergeometric
functions.

In the end of the subsection we put $\tilde {\gamma ^*}=0$. For
the case  ${\bf t}={\bf t}({\bf x}^N)$, all $\gamma=0$
 one can obtain the analog of (\ref{psixir})
\begin{equation}\label{psixirmulti}
e^{A({\bf t}^*)}g^{-1}(\gamma=0,\gamma^*)\psi(z)
g(\gamma=0,\gamma^*)e^{-A({\bf t}^*)}=
e^{-\xi_{r_l}(\gamma^*_l,z^{-1})}\cdots
e^{-\xi_{r_1}(\gamma^*_1,z^{-1})} e^{-\xi_r({\bf t}^*,z^{-1})}
\psi(z)
\end{equation}
\begin{equation}\label{psi*xirmulti}
e^{A({\bf t}^*)}g^{-1}(\gamma=0,\gamma^*)\psi^*(z)
g(\gamma=0,\gamma^*)e^{-A({\bf t}^*)}=
e^{\xi_{{r'}_l}(\gamma^*_l,z^{-1})}\cdots
e^{\xi_{{r'}_1}(\gamma^*_1,z^{-1})} e^{\xi_{r'}({\bf t}^*,z^{-1})}
\psi^*(z)
\end{equation}
where
\begin{equation}\label{xirA}
\xi_{r_k}({\bf t}^*,z^{-1})=
\sum_{m=1}^{+\infty}t_m\left(\frac 1z r_k(D)\right)^m,
\quad D=z\frac{d}{dz},\quad {r'}_k(D)=r_k(-D)
\end{equation}
In (\ref{psixirmulti}),(\ref{psi*xirmulti}) $\xi_r$
are operators which act on $z$ variable of $\psi,\psi^*$.

In cases ${\bf t}={\bf t}({\bf x}^N)$ (using (\ref{mm*r+}) and
(\ref{psi*xir})) and ${\bf t}=-{\bf t}({\bf x}^N)$ (using
(\ref{mm*r-}) and (\ref{psixir})) one gets the following
representations
\bprop
\begin{equation}\label{tau+ximulti}
\tau(M,{\bf t}({\bf x}^N),{\bf t}^*;\gamma=0,\gamma^*)=
\Delta^{-1} e^{{\eta_l}'}\cdots e^{{\eta_1}'}e^{\eta '}
\Delta,\quad \Delta= \frac{\prod_{i<j} (x_i-x_j)}{(x_1\cdots
x_N)^{N-1-M}}
\end{equation}
\begin{equation}\label{tau-ximulti}
 \tau(M,-{\bf t}({\bf
x}^N),{\bf t}^*;\gamma=0,\gamma^*)= \Delta^{-1}e^{-\eta_l}\cdots
e^{-\eta_1}e^{-\eta} \Delta,\quad \Delta= \frac{\prod_{i<j}
(x_i-x_j)}{(x_1\cdots x_N)^{N-1+M}}
\end{equation}
where
\begin{equation}\label{etamulti}
\eta=\sum_{i=1}^N\xi_{r}({\bf t}^*,x_i),\quad
{\eta_k}=\sum_{i=1}^N\xi_{r_k}({\bf t}^*,x_i)
\end{equation}
\begin{equation}\label{eta'multi}
{\eta}'=\sum_{i=1}^N\xi_{r'}({\bf t}^*,x_i),\quad
{\eta_k}'=\sum_{i=1}^N\xi_{{r'}_k}({\bf t}^*,x_i)
\end{equation}
and (cf (\ref{xir}))
\begin{equation}\label{xirxmulti}
\xi_{{r'}_k}({\bf t}^*,x_i)=
\sum_{m=1}^{+\infty}t_m\left(x_ir_k(D_i)\right)^m,\quad
\xi_{r_k}({\bf
t}^*,x_i)=\sum_{m=1}^{+\infty}t_m\left(x_ir_k(-D_i)\right)^m,
\quad D_i=x_i\frac{\partial}{\partial x_i}
\end{equation}
\eprop
Let us note that $[\xi_{{r'}_k}({\bf t}^*,x_i),\xi_{{r'}_k}({\bf
t}^*,x_j)]=0$ and $[\xi_{r_k}({\bf t}^*,x_i),\xi_{r_k}({\bf
t}^*,x_j)]=0$ for all $k,i,j$, while in general
$[\eta_k,\eta_n]\neq 0$.

(\ref{tau-ximulti}) may be also obtained from (\ref{tau+ximulti})
with the help of (\ref{tausim}).

Looking at (\ref{tau+ximulti}) one easily derives a set of linear
equations, which are differential equations with respect to
variables $\gamma^*$ and ${\bf t}^*$:
\begin{equation}\label{linur+multigamma}
\left(\frac{\partial}{\partial \gamma^*_{km}}- e^{{\eta_l}'}\cdots
e^{{\eta_{k-1}}'}  \cdot \left( \sum_{i=1}^N x_ir_k(D_i)\right)^m
\cdot e^{-{\eta_{k-1}}'}\cdots e^{-{\eta_l}'} \right)\left(
\Delta\tau(M,{\bf t}({\bf x}^N),{\bf
t}^*;\gamma=0,\gamma^*)\right) =0
\end{equation}
\begin{equation}\label{linur+multit*}
\left(\frac{\partial}{\partial t^*_m}- e^{{\eta_l}'}\cdots
e^{{\eta_{1}}'}  \cdot \left( \sum_{i=1}^N x_ir_k(D_i)\right)^m
\cdot e^{-{\eta_{1}}'}\cdots e^{-{\eta_l}'} \right)\left(
\Delta\tau(M,{\bf t}({\bf x}^N),{\bf
t}^*;\gamma=0,\gamma^*)\right) =0
\end{equation}
where $\Delta$ is the same as in (\ref{tau+ximulti}).

\bprop For the tau function $\tau_r(M,{\bf t},{\bf t}^*)$ of
(\ref{tauhyp1})
 and the vertex operators $\Omega_{\tilde{r}^1}^{(\infty)},
\Omega_{r^l}^{(0)} $ defined by (\ref{Omega}) we have
\begin{equation}
 e^{-\Omega_{\tilde{r}^1}^{(\infty)}(\g_1)} \cdots
 e^{-\Omega_{\tilde{r}^k}^{(\infty)}(\g_k)}\cdots
e^{\Omega_{r^l}^{(\infty)}(\g_l^*)} \cdots
 e^{\Omega_{r^1}^{(\infty)}(\g_1^*)}\cdot
\tau_r(M,{\bf t},{\bf t}^*)= \tau (M,{\bf t},{\bf t}^*;\g,\g^*) ,
\end{equation}
\begin{equation}
 e^{\Omega_{r^1}^{(0)}(\g_1^*)}\cdots
 e^{\Omega_{r^l}^{(0)}(\g_l^*)} \cdots
 e^{-\Omega_{\tilde{r}^k}^{(0)}(\g_k)}\cdots
 e^{-\Omega_{\tilde{r}^1}^{(0)}(\g_1)}\cdot
\tau_r(M,{\bf t},{\bf t}^*)= \tau (M,{\bf t},{\bf t}^*;\g,\g^*) .
\end{equation}
\eprop

\subsection{The example of Gelfand,Graev and Retakh
hypergeometric series \cite{nl64}}

Below we also put $\gamma=0$. Let us consider the  tau function:
\begin{equation}
\tau(M,\tilde{\beta},\beta;0,\g^*)= \l
M|e^{\tilde{A}(\tilde{\beta})}e^{-A_l(\gamma^*_l)} \cdots
e^{-A_1(\gamma^*_1)}e^{-A(\beta)}|M\r .
\end{equation}
We put
\begin{equation}
 \tilde{\beta}=(x, \frac{x^2}{2},\frac{x^3}{3},\dots),
 \quad \beta=(y_1,0,0,\dots ),\quad \g^*_i
 =(y_{i+1},0,0,\dots )\quad i=(1,\dots,l).
\end{equation}
We obtain the series
\begin{eqnarray}\label{Horn}
\tau(M, x, y_1,\dots,y_{l+1})=\sum_{n_1,\dots,
n_{l+1}=0}^{+\infty} \tilde{r}_{(n_1+\cdots+n_{l+1})}(M)
r_{\Theta^l_{{\lambda}}}(M)\frac{(xy_1)^{n_1}\cdots
(xy_{l+1})^{n_{l+1}}}{n_1!\cdots n_{l+1}!}=\qquad \qquad \\
\sum_{n_1,\dots,n_{l+1}\in Z}c(n_1,\dots,n_{l+1}) (xy_1)^{n_1}
\cdots (xy_{l+1})^{n_{l+1}}, c(n_1,\dots,n_{l+1})=
\frac{\tilde{r}_{(n_1+\cdots+n_{l+1})}(M)
r_{\Theta^l_{{\lambda}}}(M)}{\Gamma(n_1+1)\cdots
\Gamma(n_{l+1}+1)},\quad
\end{eqnarray}
where $\Theta^l_{{\lambda}}$ corresponds to the set of simple
partitions-rows
\begin{equation} {\lambda^1}=(n_1),{\lambda^2}=(n_1+n_2),\dots, {\lambda}_{l+1}=(n_1+\cdots+n_{l+1})
\end{equation}
When functions $b_i(n_1,\dots,n_{l+1})$ defined as
\begin{equation}
 b_i(n_1,\dots, n_{l+1})=
\frac{c(n_1,\dots, n_i+1,\dots, n_{l+1})}{c(n_1,\dots, n_{l+1})},
\quad i=1,\dots, l+1
\end{equation}
are rational functions of $(n_1,\dots, n_{l+1})$, then tau function
(\ref{Horn}) is a Horn hypergeometric series \cite{KV}. \\
Above series for the special choice of functions $r^i(D)$ can be
deduced from the Gelfand, Graev and Retakh series \cite{GGR}
defined on the special lattice and corresponding to the special
 set of parameters. Let us take the  rational functions $r^i(D)$:
\begin{eqnarray}
r^i(D)=\frac{\prod_{j=1}^{p^{(i)}}(D+a^{(i)}_j)}{\prod_{m=1}^{s^{(i)}}(D+b^{(i)}_m)},\quad
(i=0,\dots , l),\qquad
r^0(D)=r(D)\\
\tilde{r}(D)=\frac{\prod_{j=1}^{p^{(l+1)}}(D+a^{(l+1)}_j)}
{\prod_{m=1}^{s^{(l+1)}}(D+b^{(l+1)}_m)}
\end{eqnarray}
Let define $N=p^{(0)}+s^{(0)}+2\sum_{j=1}^l (p^{(j)}+2s^{(j)})+p^{(l+1)}+s^{(l+1)}+l+1$
and consider complex space $C^N$. In this space we consider the $l+1$-dimensional basis $B$
and the vector $\upsilon$ consisting of parameters.
\begin{eqnarray}
p_0=s_0=0,\quad p_i=p^{(i-1)}+s^{(i)},\quad
s_i=s^{(i-1)}+p^{(i)},\quad
i=(1,\dots , l)\nonumber\\
p_{l+1}=p^{(l)}+p^{(l+1)},\quad s_{l+1}=s^{(l)}+s^{(l+1)},\quad
N=\sum_{j=1}^{l+1} (p_j+s_j)+l+1
\end{eqnarray}
\begin{eqnarray}\label{f_i}
{\bf f}^i=-({\bf e}_{p_0+s_0+\cdots +p_{i-1}+s_{i-1}+1}+\cdots
+{\bf e}_{p_1+s_1+
\cdots +p_{i-1}+s_{i-1}+p_i})+\nonumber\\
+({\bf e}_{p_1+s_1+\cdots +p_{i-1}+s_{i-1}+p_i+1}+\cdots +{\bf
e}_{p_1+s_1+\cdots +p_{i-1}+s_{i-1}+p_i+s_i}), \quad i=1,\dots
,l+1
\end{eqnarray}
where ${\bf e}_{i}=\underbrace{(0,\dots , 0,\stackrel{i}{\hat{1}}, 0,\dots)}_{\mbox{N}}$.
The lattice $B\in C^N$ is generated by the vector basis of dimension $l+1$:
\begin{equation} {\bf b}^i={\bf f}^i+\cdots +{\bf f}^{l+1}+{\bf e}_{N-l-1+i},\qquad i=1,\dots ,l+1
\end{equation}
Vector $\upsilon\in C^N$ is defined as follows (compare with (\ref{f_i})):
\begin{eqnarray}
\upsilon^i =-(a_1^{(i-1)}{\bf e}_{p_0+s_0+\cdots
+p_{i-1}+s_{i-1}+1}+\cdots
+a^{(i-1)}_{p^{(i-1)}}{\bf e}_{p_0+s_0+\cdots +p_{i-1}+s_{i-1}+p^{(i-1)}}+\nonumber\\
+b_1^{(i)}{\bf e}_{p_0+s_0+\cdots
+p_{i-1}+s_{i-1}+p^{(i-1)}+1}+\cdots
+b^{(i)}_{s^{(i)}}{\bf e}_{p_0+s_0+\cdots +p_{i-1}+s_{i-1}+p_i})+\nonumber\\
+((b^{(i-1)}_1-1){\bf e}_{p_0+s_0+\cdots
+p_{i-1}+s_{i-1}+p_i+1}+\cdots + (b^{(i-1)}_{s^{(i-1)}}-1){\bf
e}_{p_0+s_0+\cdots +p_{i-1}+s_{i-1}+p_i+s^{(i-1)}}+
\nonumber\\
+(a^{(i)}_1-1){\bf e}_{p_0+s_0+\cdots
+p_{i-1}+s_{i-1}+p_i+s^{(i-1)}+1}+\cdots +
(a^{(i)}_{s^{(i)}}-1){\bf e}_{p_0+s_0+\cdots
+p_{i-1}+s_{i-1}+p_i+s_i})
\end{eqnarray}
for $i=(1,\dots l)$, and
\begin{eqnarray}
\upsilon^{l+1} =-(a_1^{(l)}{\bf e}_{p_0+s_0+\cdots
+p_{l}+s_{l}+1}+\cdots
+a^{(l)}_{p^{(l)}}{\bf e}_{p_0+s_0+\cdots +p_{l}+s_{l}+p^{(l)}}+\nonumber\\
+a_1^{(l+1)}{\bf e}_{p_0+s_0+\cdots +p_{l}+s_{l}+p^{(l)}+1}+\cdots
+a^{(l+1)}_{s^{(l+1)}}{\bf e}_{p_0+s_0+\cdots +p_{l}+s_{l}+p_{l+1}})+\nonumber\\
+((b^{(l)}_1-1){\bf e}_{p_0+s_0+\cdots
+p_{l}+s_{l}+p_{l+1}+1}+\cdots + (b^{(l)}_{s^{(l)}}-1){\bf
e}_{p_0+s_0+\cdots +p_{l}+s_{l}+p_{l+1}+s^{(l)}}+
\nonumber\\
+(b^{(l+1)}_1-1){\bf e}_{p_0+s_0+\cdots
+p_{l}+s_{l}+p_{l+1}+s^{(l)}+1}+\cdots +
(b^{(l+1)}_{s^{(l+1)}}-1){\bf e}_{p_0+s_0+\cdots
+p_{l}+s_{l}+p_{l+1}+s_{l+1}})
\end{eqnarray}
Vector $\upsilon$ is:
\begin{equation}
\upsilon=\upsilon^1+\cdots +\upsilon^{l+1}
\end{equation}
Now we can write down Gelfand, Graev and Retakh hypergeometric series corresponding to the
lattice $B$ and vector $\upsilon$:
\begin{eqnarray}
F_B(\upsilon;z)=\sum_{{\bf b}\in B}\prod_{j=1}^N \frac{z_j^{\upsilon_j+b_j}}
{\Gamma(\upsilon_j+b_j+1)}+\nonumber\\
\sum_{n_1,\dots , n_{l+1}\in Z}\prod_{j=1}^N
\frac{z_j^{\upsilon_j+n_1b^1_j+\cdots +n_{l+1}b^{l+1}_j}}
{\Gamma(\upsilon_j+n_1b^1_j+\cdots +n_{l+1}b^{l+1}_j+1)}
\end{eqnarray}
Let us compare this series with tau function (\ref{Horn}):
\begin{equation}
F_B(\upsilon;{\bf z})=c_1(a,b)g_1({\bf z})\cdots c_{l+1}(a,b)g_{l+1}({\bf z})
\tau(M, x, y_1,\dots,y_{l+1})
\end{equation}
where
\begin{eqnarray}
c_i^{-1}(a,b) =\Gamma(1-a_1^{(i-1)})\cdots\Gamma(1-a^{(i-1)}_{p^{(i-1)}})
\Gamma(1-b_1^{(i)})\cdots\Gamma(1-b^{(i)}_{s^{(i)}})\times\nonumber\\
\times\Gamma(b^{(i-1)}_1)\cdots\Gamma(b^{(i-1)}_{s^{(i-1)}})
\Gamma(a^{(i)}_1)\cdots\Gamma(a^{(i)}_{s^{(i)}}),\quad i=1,\dots ,l
\end{eqnarray}

\begin{eqnarray}
c_{l+1}^{-1}(a,b) =\Gamma(1-a_1^{(l)})\cdots\Gamma(1-a^{(l)}_{p^{(l)}})\Gamma(1-a_1^{(l+1)})
\cdots\Gamma(1-a^{(l+1)}_{s^{(l+1)}})\times\nonumber\\
\times\Gamma(b^{(l)}_1)\cdots\Gamma(b^{(l)}_{s^{(l)}})\Gamma(b^{(l+1)}_1)
\cdots\Gamma(b^{(l+1)}_{s^{(l+1)}})
\end{eqnarray}
\begin{eqnarray}
\frac{z_{p_0+s_0+\cdots +p_{i-1}+s_{i-1}+p_i+1}\cdots
z_{p_1+s_1+\cdots + p_{i-1}+s_{i-1}+p_i+s_i}}{(-z_{p_0+s_0+\cdots
+p_{i-1}+s_{i-1}+1})\cdots (-z_{p_0+s_0+\cdots
+p_{i-1}+s_{i-1}+p_i})}=1, \quad i=2,\dots ,l
\end{eqnarray}
\begin{eqnarray}
\frac{z_{p_1+1}\cdots z_{p_1+s_1}z_{N-l}}{(-z_{1})\cdots
(-z_{p_1})}=y_1
\end{eqnarray}
\begin{equation} y_i=z_{N-l-1+i},\quad i=2,\dots ,l+1
\end{equation}
\begin{eqnarray}
\frac{z_{p_1+s_1+\cdots +p_{l}+s_{l}+p_{l+1}+1}\cdots
z_{p_1+s_1+\cdots
+p_{l}+s_{l}+p_{l+1}+s_{l+1}}}{(-z_{p_1+s_1+\cdots
+p_{l}+s_{l}+1})\cdots (-z_{p_1+s_1+\cdots
+p_{l}+s_{l}+p_{l+1}})}=x
\end{eqnarray}

\begin{eqnarray}
g_i({\bf z})={z}^{\left(-a_1^{(i-1)}\right)}_{p_0+s_0+\cdots
+p_{i-1}+s_{i-1}+1}\cdots
{z}^{\left(-a^{(i-1)}_{p^{(i-1)}}\right)}_{p_0+s_0+\cdots
+p_{i-1}+s_{i-1}+p^{(i-1)}}
\times\nonumber\\
\times {z}^{\left(-b_1^{(i)}\right)}_{p_0+s_0+\cdots
+p_{i-1}+s_{i-1}+p^{(i-1)}+1}\cdots
{z}^{\left(-b^{(i)}_{s^{(i)}}\right)}_{p_0+s_0+\cdots +p_{i-1}+s_{i-1}+p_i}\times\nonumber\\
\times {z}^{\left(b^{(i-1)}_1-1\right)}_{p_0+s_0+\cdots
+p_{i-1}+s_{i-1}+p_i+1}\cdots
{z}^{\left(b^{(i-1)}_{s^{(i-1)}}-1\right)}_{p_0+s_0+\cdots
+p_{i-1}+s_{i-1}+p_i+s^{(i-1)}}\times
\nonumber\\
\times {z}^{\left(a^{(i)}_1-1\right)}_{p_0+s_0+\cdots
+p_{i-1}+s_{i-1}+p_i+s^{(i-1)}+1}\cdots
{z}^{\left(a^{(i)}_{s^{(i)}}-1\right)}_{p_0+s_0+\cdots
+p_{i-1}+s_{i-1}+p_i+s_i-1}
\end{eqnarray}
for $i=(1,\dots l)$, and
\begin{eqnarray}
g_{l+1}({\bf z}) ={z}^{\left(--a_1^{(l)}\right)}_{p_1+s_1+\cdots
+p_{l}+s_{l}+1}\cdots
{z}^{\left(-a^{(l)}_{p^{(l)}}\right)}_{p_1+s_1+\cdots +p_{l}+s_{l}+p^{(l)}}\times\nonumber\\
\times {z}^{\left(-a_1^{(l+1)}\right)}_{p_1+s_1+\cdots
+p_{l}+s_{l}+p^{(l)}+1}\cdots
{z}^{\left(-a^{(l+1)}_{s^{(l+1)}}\right)}_{p_1+s_1+\cdots +p_{l}+s_{l}+p_{l+1}}\times\nonumber\\
\times {z}^{\left(b^{(l)}_1-1\right)}_{p_1+s_1+\cdots
+p_{l}+s_{l}+p_{l+1}+1}\cdots
{z}^{\left(b^{(l)}_{s^{(l)}}-1\right)}_{p_1+s_1+\cdots
+p_{l}+s_{l}+p_{l+1}+s^{(l)}}\times
\nonumber\\
\times {z}^{\left(b^{(l+1)}_1-1\right)}_{p_1+s_1+\cdots
+p_{l}+s_{l}+p_{l+1}+s^{(l)}+1}\cdots
{z}^{\left(b^{(l+1)}_{s^{(l+1)}}-1\right)}_{p_1+s_1+\cdots
+p_{l}+s_{l}+p_{l+1}+s_{l+1}-1}
\end{eqnarray}

\section{Acknowledgements}

I thank J.Harnad  and T.Shiota for the helpful numerous
discussions. I thank S.Milne for sending \cite{Milne}, L.A.Dickey
for sending \cite{DK}, I thank also A.Zabrodin and M.Mineev for
the discussions of the paper \cite{MWZ}. I thank grant RFBR
02-02-17382a.

\end{document}